\documentclass[prb,aps,twocolumn,floatfix,superscriptaddress]{revtex4-1}
\usepackage{eurosym}
\usepackage{amsmath,amsfonts,amssymb,graphics,graphicx,color,bm}
\usepackage{hyperref}
\usepackage{braket}
\usepackage{accents}
\usepackage{tikz}

\begin{document}

\title{Staircase of crystal phases of hard-core bosons on the Kagome lattice}

\author{Daniel Huerga}
\affiliation{Institut f\"ur Theoretische Physik III, Universit\"at Stuttgart, Pfaffenwaldring 57, 70550 Stuttgart, Germany}
\author{Sylvain Capponi}
\affiliation{Laboratoire de Physique Th\'eorique, IRSAMC, Universit\'e de Toulouse, CNRS, UPS, France}
\author{Jorge Dukelsky}
\affiliation{Instituto de Estructura de la Materia, C.S.I.C., Serrano 123, E-28006 Madrid, Spain}
\author{Gerardo Ortiz}
\affiliation{Department of Physics, Indiana University, Bloomington, IN 47405, USA}

\begin{abstract}
We study the quantum phase diagram of a system of hard-core bosons on the Kagome lattice 
with nearest-neighbor repulsive interactions, for arbitrary densities, by means of the hierarchical 
mean field theory and exact diagonalization techniques. This system is isomorphic to the 
spin S=1/2 XXZ model in presence of an external magnetic field, a paradigmatic 
example of frustrated quantum magnetism.
In the {\it non-frustrated} 
regime, we find two crystal phases at densities 1/3 and 2/3 that melt into a 
superfluid phase when increasing the hopping amplitude, in semi-quantitative 
agreement with quantum Monte Carlo computations. 
In the {\it frustrated} regime and away from half-filling, we find a series of plateaux
with densities commensurate with powers of 1/3. The broader density plateaux (at densities 
1/3 and 2/3) are remnants of the classical degeneracy in the Ising limit. For densities near half-filling, this 
staircase of crystal phases melts into a superfluid, which
displays finite chiral currents when computed with clusters having
an odd number of sites. Both the staircase of crystal phases and the superfluid phase 
prevail in the non-interacting limit, suggesting that the lowest dispersionless 
single-particle band may be at the root of this phenomenon. 
\end{abstract}


\maketitle

\section{INTRODUCTION}\label{sec:introduction}
%
Frustrated systems are distinguished by the 
subtle interplay between crystal geometry and interaction among its microscopic 
components~\cite{Lacroix}. This 
interplay may lead to stabilization of a plethora of competing thermodynamic phases, some of which are 
characterized by exotic orders. Particularly interesting is the response of Mott --or any other strongly 
interacting-- insulating materials to applied magnetic fields at low temperatures, since very often 
the system transitions between commensurate and incommensurate phases, each signaled 
by the presence of a magnetization plateau as the field changes~\cite{Takigawa2013}. 

These so-called (classical or quantum) frustrated magnets are effectively described by spin degrees 
of freedom distributed in a graph or lattice, such as Kagome, interacting through Heisenberg 
or Ising-type interaction terms \cite{Nishimori-Ortiz-2011}. Eventually plaquette or ring-exchange 
interactions may become relevant \cite{Roger1983}.  In Ising or easy axis models of the 
magnet, the system may display a complete 
\textit{devil staircase}~\cite{Bak1982}, i.e. a series of magnetization plateaux where wide plateaux alternate with 
quasi-infinite series of smaller ones, with characteristic correlation lengths large but not infinite. 
On the other hand, in Heisenberg models quantum fluctuations can induce melting of those small plateaux states
into a long-range ordered phase, leading to an incomplete staircase. To assess this melting phenomenon is hard, 
both from the theoretical and experimental standpoints. From the theoretical side, large system sizes 
are needed, while high NMR resolution and very clean samples are experimentally desired~\cite{Bak1982}.

Recent numerical studies of the Kagome Heisenberg antiferromagnetic model (KHAF) found 
a series of plateaux at magnetizations commensurate with nine units of the saturation 
magnetization \cite{Capponi2013,Nishimoto2013,Picot2016}.  
In particular, the crystal phase state just below saturation can be exactly represented by a localized 
resonant magnon over a background of fully polarized spins~\cite{Schulenburg2002}.  
At zero magnetic field, the KHAF is a paradigmatic example of a frustrated quantum magnet and 
a prominent candidate for hosting a translational invariant paramagnetic ground state, so-called 
\textit{quantum spin liquid}. The KHAF is conjectured to be realized in recently synthetized herbertsmithite, 
a layered compound that shows the absence of magnetic order for very low temperatures~\cite{Shores2005}, 
although the true nature of its excitations remains elusive due to the presence of 
impurities~\cite{Helton2007,Mendels2007}. Latest experiments on samples with $5$-$10\%$ concentration of 
impurities show a finite spin excitation gap~\cite{Fu2015}.  
Understanding the response to external magnetic fields will help to
unveil the physics of its excitations.

Motivated by experimental efforts to synthesize new magnetic materials, 
in this paper we investigate the quantum phase diagram, and nature of the low-lying excitations, 
of a general class of frustrated magnets that includes spin anisotropy. We consider  
a Kagome lattice with ${\cal N}$ sites (vertices) where on each vertex $j$ lies a quantum spin, 
S=1/2, described by the operator 
$S_j^\nu$ ($\nu=x,y,z$), interacting with its nearest-neighbor spin $i$, defining the link 
$\langle ij\rangle$. Its model Hamiltonian is given by
\begin{equation}
H= \sum_{\langle ij\rangle}\left[J\left(S_i^xS_j^x+S_i^yS_j^y\right) + \Delta S_i^zS_j^z\right]
-h\sum_j S^z_j
\label{eq:XXZ_ham},
\end{equation}
where $h$ represents the external magnetic field and the exchange interaction along the $z$ 
quantization axis is always taken to be antiferromagnetic $(\Delta\geq 0)$. Parameter values 
include ferromagnetic (FM), $J<0$, and antiferromagnet (AFM), $J>0$ regimes. 
For $\Delta=J$ and $h=0$, Eq. (\ref{eq:XXZ_ham}) reduces to the SU(2) 
symmetric Heisenberg model. For any other point in the phase diagram, this ``XXZ model'' possesses 
global U(1) symmetry corresponding to rotations of spin operators in the $xy$ plane. 
Since current ultracold atom technology may allow for a clean quantum simulation of this 
model~\cite{Jo2012}, it is appropriate and convenient to analyze its isomorphically equivalent  
hard-core boson model  with nearest-neighbor repulsive interactions, as will be shown below.  
%
%
\begin{figure}[t!]
\centering
\includegraphics[width=0.35\textwidth]{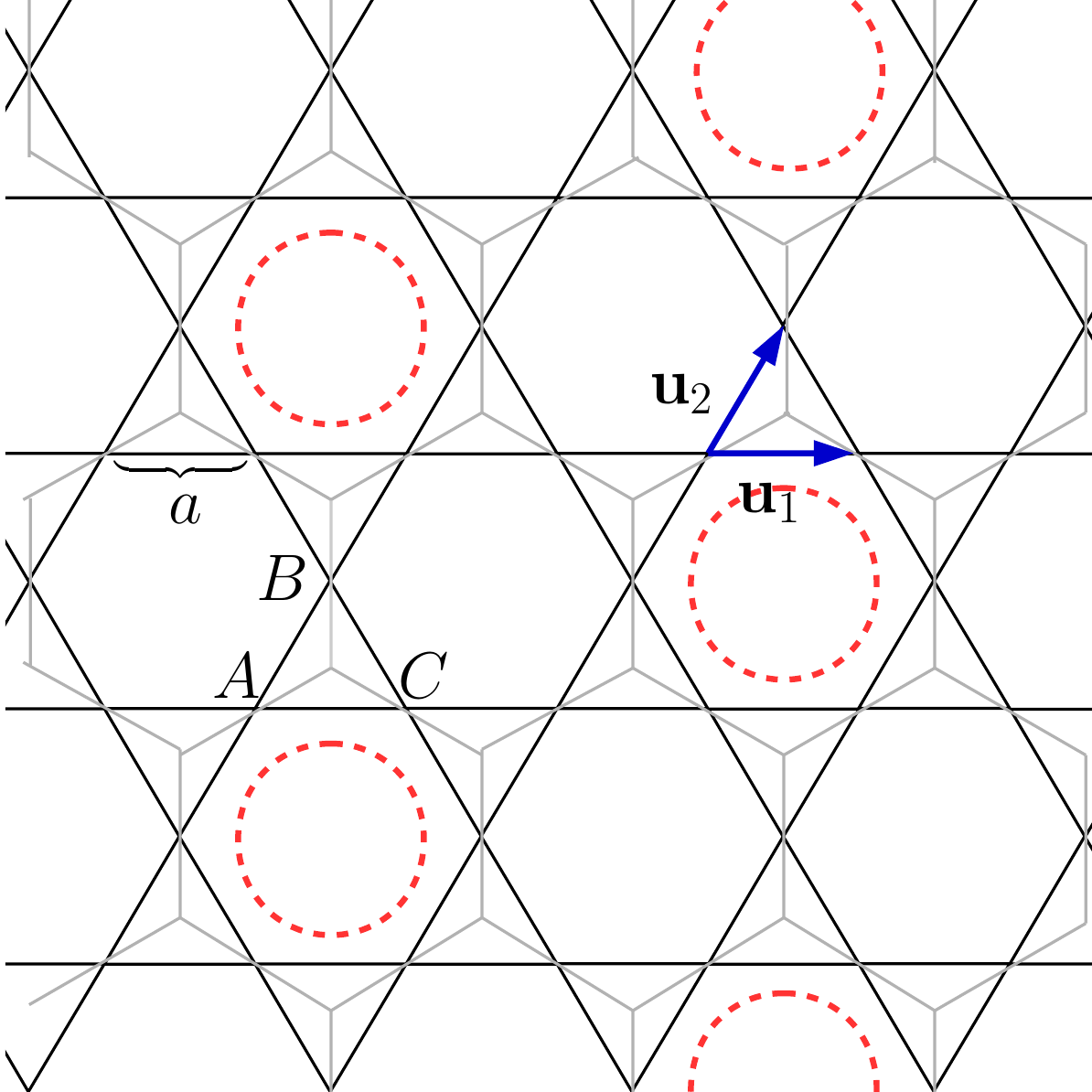}
\caption{\label{fig:intro} (Color online) Schematics of a Kagome lattice (black lines) and the 
underlying \textit{honeycomb} lattice defined by the centers of the triangular plaquettes (gray). Blue arrows 
mark the two basis vectors of the Kagome lattice, $\mathbf{u}_1$ and $\mathbf{u}_2$. 
Capital letters label the 3-sites basis $A,B,C$. Red dashed circles indicate the pattern of 
the localized resonant hole valence bond crystals (VBCs), as explained in the text.}
\end{figure}

Before proceeding with our main findings let us briefly summarize current knowledge 
about the XXZ model \eqref{eq:XXZ_ham}, which can be re-written in terms 
of the total spin of a corner-sharing triangle (see Fig. \ref{fig:intro}). Up to an 
irrelevant additive constant, the Hamiltonian takes the form
\begin{equation}
H=\frac{1}{2}\sum_{p} 
\left[ J\mathbf{S}_p^2 +(\Delta-J)(S_p^{z})^2 -hS_p^z\right],\label{projector_ham}
\end{equation}
where $\mathbf{S}_p=\sum_{j\in p}\mathbf{S}_j$ and $S^z_p=\sum_{j\in p}S^z_j$ are the total spin 
and its $z$ component of triangle $p$, respectively. In the  Ising limit, 
$J=0$, and $h=0$, the exact ground-state manifold is defined by all spin 
configurations with $S^z_p=\pm1/2$.  When applying a magnetic field ($h>0$), this degeneracy is reduced 
as the exact ground-state manifold is defined by those configurations with $S^z_p=+1/2$. 
The number of configurations coincides with all hard-core dimer coverings of the honeycomb 
lattice~\cite{Cabra2005}. A small but finite value of the XY anisotropy further lifts the degeneracy favoring 
valence bond crystal (VBC) states characterized by a fully packed pattern of localized three-magnon 
resonant states over a background of fully polarized spins, in a pattern represented  by 
dashed red lines in Fig. \ref{fig:intro}. This pattern was found to characterize the VBC$_{\rho}$ 
phases at $\rho=1/3$ and $2/3$ in the FM regime by quantum Monte Carlo (QMC) computations 
on the isomorphically equivalent hard-core boson model~\cite{Isakov2006}.

At precisely the SU(2) Heisenberg point (i.e. the KHAF), (\ref{projector_ham}) becomes a sum of local projectors 
over the subspace with $S_p=3/2$,
\begin{equation}
    H= J\sum_{p} P_{p}^{3/2},
\end{equation}
up to an irrelevant constant, implying that any state containing a singlet in a triangle satisfies 
locally the energy constraint. The ground-state manifold is thus defined by those many-body states minimizing 
the number of frustrated triangles. In this regime, as QMC computations are affected 
by the sign-problem,  
different analytical and numerical approaches have been used to unveil the nature of the ground state.  
Projection onto the short-range resonant valence-bond subspace~\cite{Mambrini2000,Poilblanc2010} 
has been used to study the proliferation of the low-lying singlet states found by exact diagonalization (ED)~\cite{Waldtmann1998}.
Series expansions around the dimer limit~\cite{Singh-H-07} found a VBC with a 36-site unit cell. 
Density matrix renormalization group (DMRG) computations find a translational invariant ground state 
with an excitation gap~\cite{Yan2011} and $\mathbb{Z}_2$ topological order~\cite{Depenbrock2012,Jiang2012}, 
while other methods have proposed different translational invariant gapped~\cite{Messio2012,Capponi2013a} or 
gapless ground states~\cite{Iqbal2013}. 

In presence of a magnetic field, the KHAF  has been studied by means of ED~\cite{Capponi2013}, 
DMRG~\cite{Nishimoto2013} and, more recently, by infinite 
projector entangled pair states (iPEPS)~\cite{Picot2016}. These works find  
a series of magnetization plateaux at values of the magnetization $M$ commensurate with nine 
(in units of the saturated magnetization), 
$M=1/9,1/3,5/9,7/9$, with $1/3$ representing the broader plateau. While the 
exact nature of the $M=1/9$ plateau state is still debated -- existing proposals argue for either a topological 
state~\cite{Nishimoto2013} or a more conventional VBC state~\cite{Picot2016} --
the wave functions for 
the other three plateaux states ($M=1/3,5/9,7/9$) can be approximately described by hexagon 
magnon states distributed over a background of fully polarized spins in the fully stacked pattern 
described above~\cite{Capponi2013,Nishimoto2013}. 
In particular, the one-resonant magnon state approximating 
the $M=7/9$ plateau is the exact ground state of model (\ref{eq:XXZ_ham}) previous to 
saturation~\cite{Schulenburg2002}. 

As anticipated,  we may rewrite the XXZ model (\ref{eq:XXZ_ham}) in terms of hard-core bosons by 
applying the Matsubara-Matsuda isomorphism to the SU(2) spin operators~\cite{Matsubara1956}. 
Explicitely, the ladder operators of the S=1/2 representation, $S_j^\pm=S_j^x\pm {\rm i}S_j^y$, 
are mapped to creation and 
annihilation hard-core boson operators, $S^+_j=a^\dag_j$ and $S^-_j=a_j$, and the 
Cartan to the number operator $n_j=a^\dag_j a^{\;}_j$, $S^z_j=n_j-1/2$, leading to 
\begin{equation}
H= t\sum_{\langle ij\rangle}\left( a^{\dag}_{i}a_j + \text{H.c.}\right) + 
V\sum_{\langle ij \rangle} n_i n_j - \mu\sum_j n_j +\text{C}, \label{hardcore_ham}
\end{equation}
where
\begin{eqnarray} \hspace*{-0.5cm}
t&=&J/2 \ , \ 
V=\Delta \ , \ 
\mu=h+2\Delta \ , \  \text{C}={\cal N}(\Delta+h)/2 .
\end{eqnarray}
The chemical 
potential $\mu$ controls the total hard-core bosons density,  
$\rho$, the nearest neighbour density-density 
interaction is repulsive $(V\geq 0)$ and frustrated, and the hopping amplitude $t$ is tuned from the non-frustrated 
$(t<0$, FM) to the frustrated ($t>0$, AFM) regimes.~\footnote{This terminology is often used since the 
model is amenable to sign-free QMC simulations but we should emphasize that the 
density-density interaction is always frustrated.}  

\begin{figure}[t!]
\centering
\includegraphics[width=0.45\textwidth]{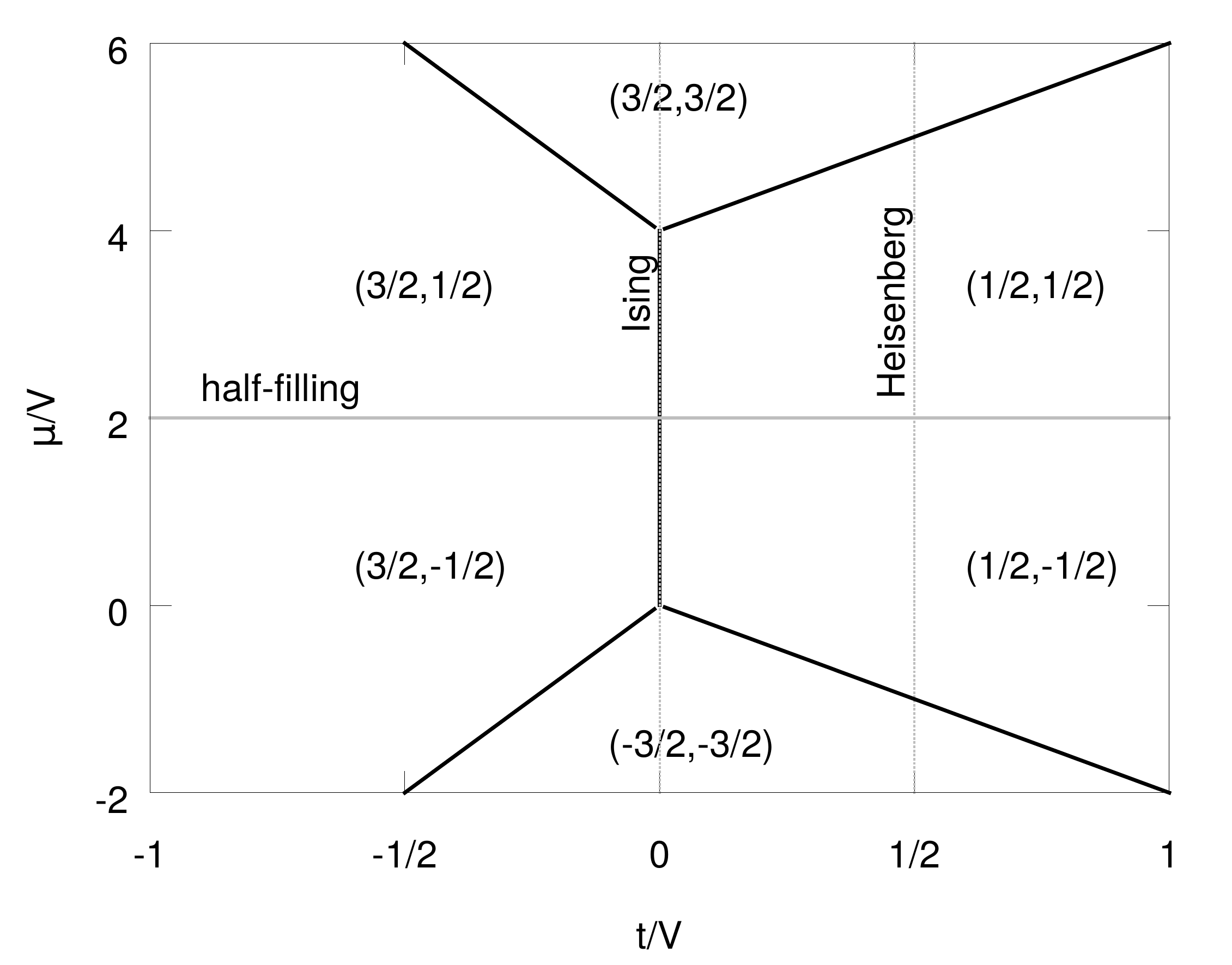}
\caption{\label{fig:diag_spin} (Color online) Parameter regimes of the hard-core boson model on the 
Kagome lattice (\ref{hardcore_ham}). Regions are classified by the total 
and third component of the spin of a triangle $(S_p,S^z_p)$ that locally minimize the equivalent 
XXZ Hamiltonian written in the honeycomb lattice defined by the centers of the triangles 
comprising the Kagome (\ref{projector_ham}). Solid black lines 
correspond to parameters where the Hamiltonian has an exact solution. Dotted gray lines 
correspond to $h=0$ (half-filling), Ising, and Heisenberg 
limits, as explained in the text.}
\end{figure}
The total spin of the corner-sharing triangles partitions the quantum phase diagram 
in terms of $(S_p,S_p^z)$ pairs which minimize locally the Hamiltonian, as shown in 
Fig. \ref{fig:diag_spin}. At vanishing external magnetic field ($\mu=2\Delta$) the spin version 
(\ref{eq:XXZ_ham}) corresponds to half-filling in its bosonic counterpart, where a particle-hole 
symmetry holds. In particular, the AFM XY limit  translates into a tight-binding model of 
hard-core bosons characterized by a dispersionless lowest-energy band. The Ising and Heisenberg 
limits (arbitrary $h$) are thus placed at $t/V=0$ and $t=V/2$, respectively. 
The global U(1) symmetry of the spin rotations in the $xy$ plane is translated into the conservation 
of the total number of bosons. The density of hard-core bosons is related 
to the relative magnetization by 
\begin{equation}
\rho=\frac{M+1}{2} ,
\end{equation}
once taken into account the three-site basis 
of the Kagome lattice. Thus, the magnetization plateaux encountered in 
previous works have a direct translation to the density of hard-core bosons,
\begin{equation}
\begin{array}{cccccc}
M&=& \{1/9, & 1/3, & 5/9, & 7/9\} \notag\\
& &\updownarrow &\updownarrow &\updownarrow &\updownarrow \notag\\
\rho &=& \{5/9, & 2/3, & 7/9, & 8/9\}. \notag
\end{array}
\end{equation}

In this work we aim at providing a unified description of the complete quantum phase diagram, including the 
Ising, Heisenberg, and the FM and AFM XY limits of the hard-core boson model (\ref{hardcore_ham}), 
or equivalently the XXZ model \eqref{eq:XXZ_ham}, by means of the hierarchical mean-field theory 
(HMFT)~\cite{Ortiz2003,Batista2004,Isaev2009,Isaev2012,Huerga2014} and ED. 
The HMFT is a versatile algebraic framework based on the use of clusters of the original degrees of freedom 
as the basic building blocks containing the short-range correlations which account for the main features 
of the phases present in the system under study. Following this idea, cluster states are represented as 
the action of a \textit{composite boson} (CB) over a new vacuum. As the relation between the original 
spin, or bosonic operators, and the new ones can be cast in a canonical form, we can rewrite the 
Hamiltonian of interest in terms of CBs and approach it by standard techniques, with the advantage 
that short-range quantum correlations are taken into account \emph{exactly} from the onset. 
To further support our findings, 
we will also perform ED calculations on finite clusters with periodic boundary conditions (PBC).

\begin{figure}[b!]
\centering
\includegraphics[width=0.45\textwidth]{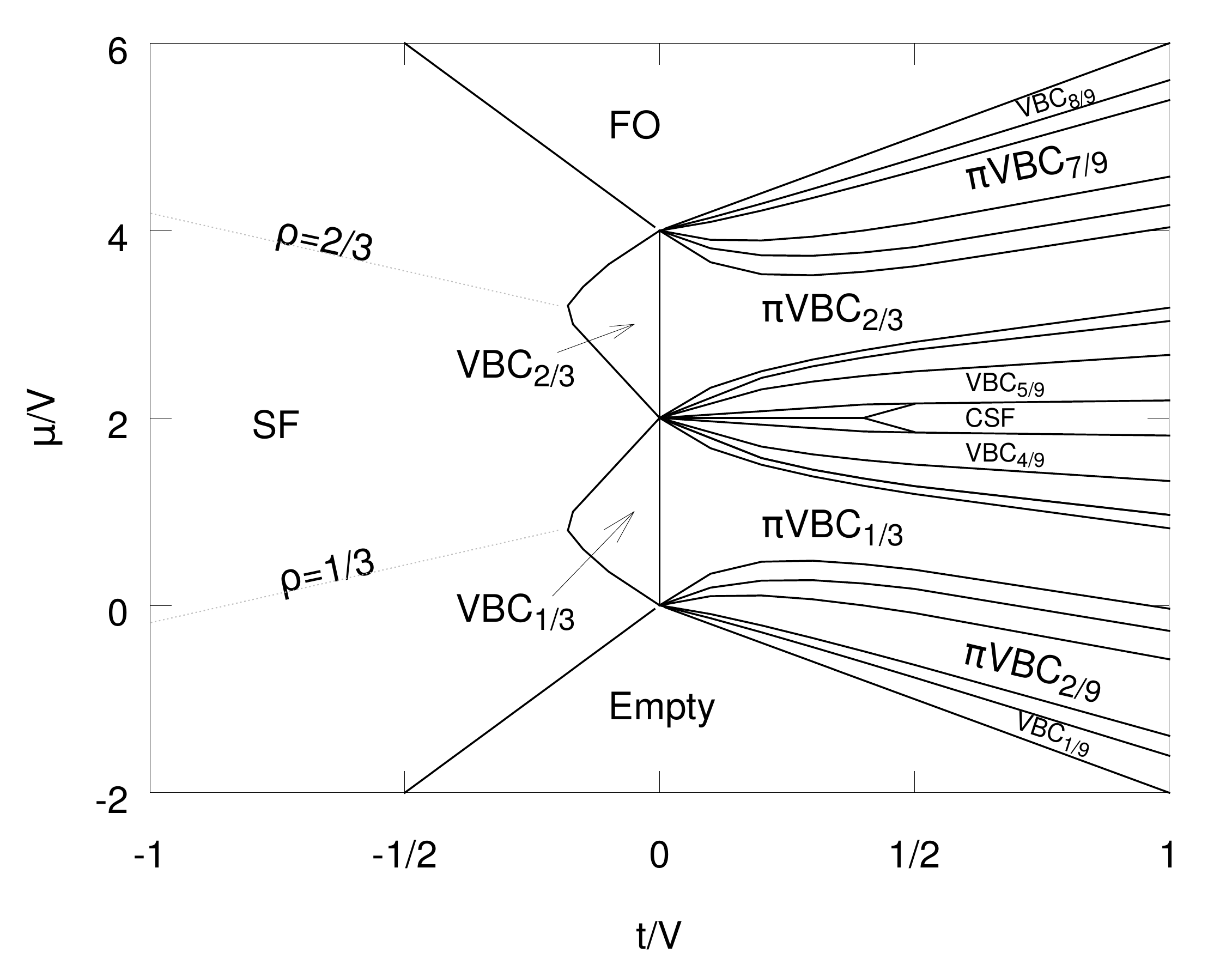}
\caption{\label{fig:intro_results} (Color online) Quantum phase diagram of the hard-core boson model 
(\ref{hardcore_ham}) as obtained within HMFT using 27-sites clusters (referred in the text as 27-HMFT). 
Phases are labeled in capital letters as: fully occupied (FO), superfluid (SF), chiral superfluid (CSF) and 
valence bond crystal of density $\rho$ (VBC$_\rho$). The prefix $\pi$ is used to distinguish those 
VBCs with real positive amplitudes in their wave functions, from those containing 
some negative ones. Non-labeled phases correspond to crystal phases 
with densities commensurable with the 27-sites, but not with the 9-sites cluster. Dotted gray lines 
correspond to constant density lines within the SF phase.}
\end{figure}

Our main findings are summarized in Fig.~\ref{fig:intro_results}. In the non-frustrated 
regime (FM),  we obtain two VBC$_\rho$ lobes of densities $\rho=1/3$ and $2/3$ that melt 
into a superfluid (SF) in semi-quantitative agreement with previous 
QMC computations~\cite{Isakov2006}. We have reproduced these QMC data by simulating the 
model using the Stochastic Series Expansion (SSE) algorithm~\cite{Sandvik1991,Syljuasen2002} 
of the ALPS library~\cite{ALPS2}. Moreover, the superfluid order parameter (i.e. stiffness) was 
computed following the steps of Ref. [\onlinecite{Rousseau2014}], where a precise method for computing 
this quantity was put forward for non-Bravais lattices, a case often overlooked in the literature. 
 
In the frustrated regime --or, equivalently, AFM-- and away from half-filling, we obtain the series of 
VBC$_\rho$ \textit{main plateaux} of densities $\rho$ commensurate with 1/9 as previously described. 
The prefix $\pi$ is used to distinguish the wave functions from the non-frustrated region. Around 
half-filling ($\mu/V\sim 2$) and hopping amplitudes $t>0.4V$, the staircase of crystal phases melts 
into a superfluid characterized by the breakdown of the global U(1) symmetry and onset of 
Bose-Einstein condensation (BEC). 
For clusters with an odd number of sites, this superfluid possesses non-vanishing chiral currents 
(CSF) and is doubly degenerate; each state displaying a current of opposite chirality.
For clusters with an even number of sites, e.g. a 18-sites cluster, the chirality vanishes. 
This is consistent with previous ED studies \cite{Waldtmann1998} performed at the 
SU(2) Heisenberg point, where the first S=1 excitation was found to have non-zero Chern 
number only in clusters with an odd number of sites. We find this non-trivial ``odd-even effect'' 
along the whole $t/V>0$ axis. This superfluid 
region considerably diminishes its size upon increasing the cluster size, indicating that it might perhaps  
only survive at $\mu= 2 V$ in the thermodynamic limit. Note that chiral phases have been found to be 
competing states at half-filling, meaning that they can be stabilized by longer-range 
interactions~\cite{Gong2014} or explicit chiral interactions~\cite{Bauer2014}, and also in the absence 
of SU(2) symmetry~\cite{He2014}.

For hopping amplitudes $0<t<0.4V$, the half-filling 
line defines a first order transition between VBCs with densities $\rho=13/27$ and 14/27. In-between 
plateaux, we obtain a series of narrower plateaux commensurate with 1/27 which diminish in 
width, discarding the possibility of the onset of a standard superfluid order characterized by 
BEC at momentum $\mathbf{k}=(k_x,k_y)$ consistent with the 27-sites cluster. 
Nonetheless, we cannot asses their stability 
in the thermodynamic limit, as they were obtained with a single coarse-graining. 
This is in partial disagreement with the findings 
of Ref.~\onlinecite{Picot2016} at the Heisenberg line, where a 9-sites unit cell was 
used and some regions between main plateaux where claimed to support U(1) breakdown, 
while others were not fully characterized.

All the quantum phases we found essentially prevail in the XY limit, consistent with similar 
conclusions obtained at zero magnetic field (half-filling) in Refs. 
[\onlinecite{LauchliMoessner}] and [\onlinecite{He2015}]. 
In particular, the VBC$_{8/9}$ is exact at precisely the phase boundary in the whole AFM regime, in agreement with Ref.~\onlinecite{Schulenburg2002}. 
Similarly, we found the $\rho=7/9$ plateau ($\pi$VBC$_{7/9}$) to be described by a pattern of fully 
stacked resonant localized-magnons from the Ising to the AFM XY regime, finding no evidence 
of additional degeneracies that would support either a VBC with further degeneracy -- as claimed 
in Ref. ~\onlinecite{Picot2016} -- or a topologically ordered state -- as claimed in Ref.~\onlinecite{Kumar2014}. 
From our ED and HMFT results, the $\rho=2/3$ plateau ($\pi$VBC$_{2/3}$) smoothly transitions from a
fully stacked localized resonant-magnon VBC (similar to the VBC$_{8/9}$ and $\pi$VBC$_{7/9}$) 
to a more complex VBC pattern 
in the AFM XY limit. 
Based on our analysis of the low-lying ED spectrum of finite clusters, 
we cannot exclude the possibility 
that, in this limit, the $\rho=2/3$ plateau becomes either a translational invariant gapped state~\cite{Kumar2014}, or a gapless BEC. 
Similarly, 
we find the $\rho=5/9$ plateau ($\pi$VBC$_{5/9}$) to be a complex VBC in the whole AFM regime.
Again, we cannot exclude the possibility it becomes a translational invariant 
three-fold degenerate gapped state in the XY limit~\cite{Nishimoto2013}. 

The outline of the paper is as follows. In Sec. \ref{sec:methods} we briefly describe the HMFT approach 
at zero temperature and 
some details about the ED computations, together with the order parameters and observables 
used to characterize the different quantum phases. In Sec. \ref{sec:phase_diag} we 
present the complete quantum phase diagram 
and provide a detailed analysis of the quantum correlations 
in the various crystal and superfluid phases 
encountered in both the frustrated and non-frustrated regimes. Finally,  Sec. \ref{sec:conclusion} 
concludes with a summary and outlook.

\begin{figure*}[t!]
\centering
\includegraphics[width=0.95\textwidth]{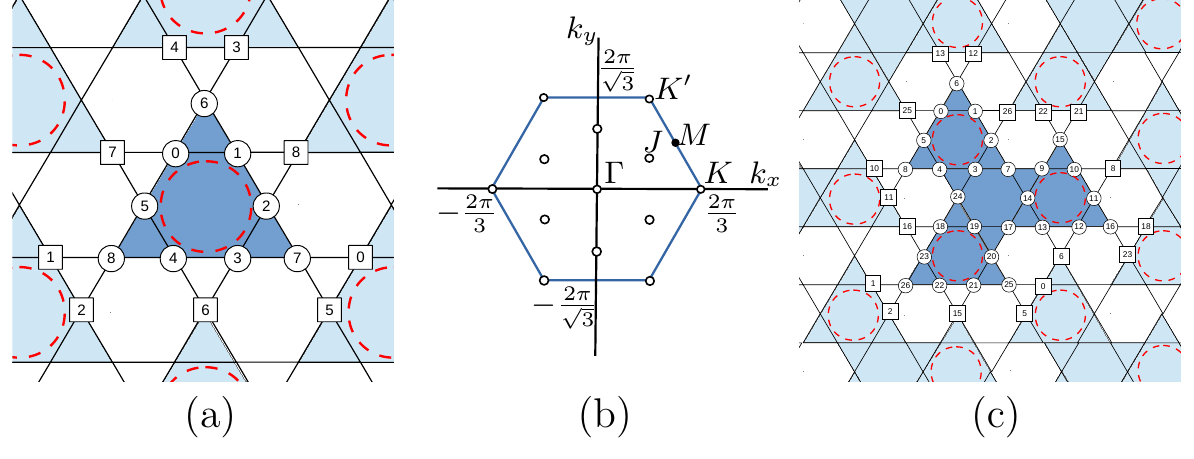}
\caption{\label{9site_embed} (Color online) (a) and (c): Schematic pictures showing the (a) 9-sites and (c) 27-sites 
cluster tilings used in the HMFT approach of this work. Both tilings contain the fully packed resonant-hole 
hexagon pattern of the exact VBC$_{8/9}$ state, represented by dashed red lines. 
Numbers label the lattice sites within the cluster. Circles represent intra-cluster sites, while squares represent those
sites where the auxiliary mean-fields are evaluated. 
(b) First Brillouin zone of the Kagome lattice. 
Empty circles mark the momenta $(\mathbf{k})$ consistent with the 9-site and 27-site tilings.
In particular, the 9-HMFT contains $\Gamma$, $K$ and $K'$ points, while the 27-HMFT contains 
in addition the $J$ point and its rotations. 
The $M$ point, identified with a solid circle, is contained within the 36-sites cluster used for ED.}
\end{figure*}

\section{Theoretical methods\label{sec:methods}}
%
In this section we provide a brief description of the two methods used in this work: HMFT  in the 
Gutzwiller approximation, and ED of finite clusters. From a technical standpoint, each method 
emphasizes a different aspect of the infinite-size (thermodynamic) limit solution 
and thus their combined use provide complementary information. 

The HMFT in the Gutzwiller approximation involves the iterative diagonalization of a finite cluster of size $N$ with 
open boundary conditions (OBC) and a set of self-consistently defined mean-fields acting 
on its boundaries. The CB Gutzwiller wave functions are {\it exact} ground states previous to 
saturation. In the rest of the phase diagram, while short-range correlations within the clusters are computed exactly, 
the mean-fields carry information about the thermodynamic limit allowing for potential 
breakdown of symmetries and the concurrent stabilization of long-range ordered phases. These CB Gutzwiller 
wave functions permit the description of quantum phases characterized by the onset of long-range 
superfluid order, signaled by 
the occurrence of BEC at various momenta, as well as crystal or superfluid 
phases characterized by chiral currents~\cite{Huerga2014,Greschner2015}. 

On the contrary, ED consists of a single diagonalization performed on a finite cluster with PBC. 
Since lattice symmetries and conservation of the total number of bosons (or, equivalently, magnetization) 
can be exploited to reduce the dimension of the Hilbert space, one can simulate 
cluster sizes beyond those used in HMFT. In the ED case, assessing the stability or 
absence of long-range order is determined by a finite-size scaling analysis.

\subsection{Hierarchical mean-field theory\label{sec:hmft}}
%
The HMFT is an approach based on the use of a cluster as the basic degree of freedom. In practice, 
the original lattice is tiled with clusters such that every site belongs to a unique cluster, preserving 
the original symmetries of the problem as much as possible. The quantum states of each cluster 
$\ket{\alpha}_\mathbf{R}$ are represented by the action of a bosonic (CB) operator over a vacuum $\ket{0}$, 
\begin{equation}
b^{\dag}_{\mathbf{R},\alpha}\ket{0}\equiv \ket{\alpha}_\mathbf{R},
\end{equation}
where $\alpha$ labels the quantum state of the cluster in a generic basis, and 
$\mathbf{R}$ its position in the cluster superlattice.  In this way, quantum correlations 
with a range smaller than the size of the cluster are taken into account exactly,  
and thus local information about different 
competing orders is unbiased. The mapping relating the original bosons to 
the new CBs is canonical whenever the latter satisfy the Schwinger constraint~\cite{Isaev2009,Huerga2013}, 
\begin{equation}
\sum_{\alpha} b^\dag_{\mathbf{R},\alpha}b^{\;}_{\mathbf{R},\alpha}=1,~~~~\forall ~\mathbf{R},
\end{equation}
which is equivalent to demand that any many-body state of the original problem can be written, 
in terms of CBs, as a lineal combination of product states with one CB per 
cluster. As the mapping is canonical, the original Hamiltonian can thus be rewritten 
in terms of these new CBs and solved by standard many-body techniques. 
The method is variational whenever the constraint is fulfilled exactly.
However, proposing a generic ansatz 
satisfying the Schwinger constraint exactly is a non-trivial 
task common to all slave-particle approaches.

In this work we use a homogeneous Gutzwiller wave function of CBs (CB Gutzwiller), 
which satisfies the Schwinger constraint exactly,
\begin{equation}
\ket{\Psi}= \prod_\mathbf{R}\ket{\Phi}_\mathbf{R},~~~~\ket{\Phi}_\mathbf{R}=
\sum_{\mathbf{n}} U_{\mathbf{n}} \ b^{\dag}_{\mathbf{R},\mathbf{n}}\ket{0},\label{CB_Gutz}
\end{equation}
where $\mathbf{n}$ refers to the quantum state in the occupation basis. The amplitudes 
$U_{\mathbf{n}}$ are determined by making stationary the expectation value of the Hamiltonian 
(\ref{hardcore_ham}). The resulting set of non-linear equations can be cast in a Hartree eigensystem 
form which is solved iteratively until self-consistency is reached. This homogeneous Gutzwiller 
approach is equivalent to perform ED on a finite $N$-sites cluster, with OBC and a set of self-consistent 
auxiliary fields acting on the boundaries of the cluster, where the Hartree or mean-field 
cluster Hamiltonian has the form
\begin{equation}
H_{\sf mf}= H^{\square} + H^{\times}(\{\psi,\eta\}),\label{hartree}
\end{equation}
with superscripts $\square$ and $\times$ referring to intra- and inter-cluster terms, respectively. 
The inter-cluster terms depend upon a set of self-consistently defined mean-fields $\{\psi,\eta\}$ that 
rely on  the particular form of the Hamiltonian and the 
tiling performed. In the present case,
\begin{eqnarray}
H^\square&=&t\sum_{\langle ij\rangle\in\square}\left( a^{\dag}_{i}a_j + \text{H.c.}\right)\notag\\
&&+ V\sum_{\langle ij \rangle\in\square} n_i n_j
- \mu\sum_{j\in\square} n_j,
\end{eqnarray}
with sums running over sites within the same cluster, and
\begin{equation}
H^\times=
t\sum_{\langle i,j\rangle}\left( a^{\dag}_{i}\psi_j + \text{H.c.}\right)
+ V\sum_{\langle i,j\rangle} n_i \eta_j ,
\end{equation}
where $\langle ij\rangle$ represent the inter-cluster links connecting a cluster site $i$ and a 
site $j$ belonging to the neighbouring cluster. The set of auxiliary fields are defined as
\begin{eqnarray}
\psi_j^\ast&=&\bra{\Phi} a_j^\dag\ket{\Phi} \ , \ 
\eta_j=\bra{\Phi} n_j\ket{\Phi},
\end{eqnarray}
where we have dropped the superlattice index $\mathbf{R}$, as in the homogeneous CB Gutzwiller 
wave function all clusters are equivalent. These auxiliary fields are evaluated on the boundaries 
of the embedding clusters.
In Fig.~\ref{9site_embed} (a) and (c) we show schematic pictures of the two main tilings used in 
this work, i.e. 9-sites and 27-sites clusters, both containing exactly the fully packed localised 
resonant-hole pattern of the exact VBC$_{8/9}$ previously described.

The CB Gutzwiller wave function (\ref{CB_Gutz}) allows systematic computation of order 
parameters and observables. In particular, as it satisfies the Schwinger constraint exactly, it 
provides a variational upper bound to the energy and phase boundaries  are established 
by monitoring any non-analytic behavior in its derivatives.  In addition, we computed the total 
density, the condensate density of hard-core bosons, signaling the breakdown of global U(1) 
symmetry and onset of BEC, the bond-currents signaling the breakdown of time-reversal 
symmetry, and the expectation value of the local hoppings on the links.

The total density is simply defined as the average value of the density in the cluster,
\begin{equation}
\rho=\frac{1}{N}\sum_{j\in\square} \left\langle n_j\right\rangle .
\end{equation}
%

From the macroscopic eigenvalues of the density matrix one can determine 
the condensate density ~\cite{Leggett2001}. Due to translational symmetry, the density 
matrix is diagonal in momentum space, 
\begin{equation}
\rho^{{\sf c}}_{\alpha,\beta}(\mathbf{k})= 
\frac{1}{{\cal N}_s}\langle a^{\dag}_{\mathbf{k},\alpha}a^{\;}_{\mathbf{k},\beta}\rangle,
\end{equation}
where ${\cal N}=3{\cal N}_s$ is the total number of sites of the lattice and $\mathbf{k}$ refers to 
a vector within the first Brillouin zone. The Kagome lattice is a triangular Bravais 
lattice with a 3-sites basis. Each point in the lattice is determined by the 
triangular lattice vector $\mathbf{r}$ and the two basis vectors  
$\mathbf{u}_1,\mathbf{u}_2$ (see Fig. \ref{fig:intro}). When computed with the CB 
Gutzwiller ansatz (\ref{CB_Gutz}), the density matrix becomes
\begin{eqnarray}
\rho^{{\sf c}}_{\alpha,\beta}(\mathbf{k})&=& 
\frac{9}{{\mathcal N}^2} \ 
\sum_{\mathbf{r}\in\square}\langle a^{\dag}_{\mathbf{r+u}_\alpha} \rangle 
\ e^{-{\rm i}\mathbf{k}(\mathbf{r+u}_\alpha)}\notag\\
&&\times\sum_{\mathbf{r'}\in\square}\langle a^{\;}_{\mathbf{r'+u}_\beta} \rangle 
\ e^{{\rm i}\mathbf{k}(\mathbf{r'+u}_\beta)}
,\label{eq:BEC_op_offd}
\end{eqnarray}
to order $1/{\cal N}_s$, where the sums run over the sites of a given species inside the cluster. 
For each $\mathbf{k}$-point we further diagonalize in the species subspace, and we can 
define the {\it total} condensate density at $\mathbf{k}$ as the sum
\begin{equation}
\rho^{\sf c}(\mathbf{k})= \sum_{\tilde{\alpha}} \rho_{\tilde{\alpha}\tilde{\alpha}}^{\sf c}(\mathbf{k}).
\end{equation}
where $\tilde{\alpha}$ refers to the diagonal basis in the species space. At each point in the Brillouin 
zone, two of the three eigenvalues of (\ref{eq:BEC_op_offd}) are null and the third is positive, 
in both   9-HMFT and 27-HMFT,

The bond-currents can be defined through the Heisenberg equation for the local number 
operator, ${\rm i}\partial n_j/\partial t=\left[n_j,H\right]$, in units of $\hbar$. By requiring the 
local density to be a conserved quantity, $\left[n_j,H\right]=0$ one can define local 
bond-currents satisfying $\sum_{\langle ij\rangle} \mathcal{J}_{ij}=0$ for a given site $i$,
\begin{equation}
\mathcal{J}_{ij}= \frac{\rm i}{2} \left( a^{\dag}_{i} a_{j}-\text{H.c.}\right).\label{eq:bond_current}
\end{equation}
This quantity is  the $z$-component of the vector spin chirality~\cite{Hassanieh2009}, 
$\mathbf{\kappa}_{ij}^z= (\mathbf{S}_{i}\times\mathbf{S}_{j})^z$, when written in terms of 
hard-core bosons. 
When taking the expectation value of the bond-current operator, one can distinguish between 
two cases, i.e. when the bond is contained within the cluster, 
\begin{equation}
\langle \mathcal{J}_{ij}\rangle^\square=\frac{\rm i}{2}\left(\langle a^\dag_i a_j\rangle- \text{c.c.}\right),
\end{equation}
and when the bond is connecting two different clusters after the tiling,
\begin{equation}
\langle \mathcal{J}_{ij}\rangle^\times=
\frac{\rm i}{2}\left(\psi^\ast_i \psi_j - \text{c.c.}\right),
\end{equation}
where $i\in\square$ and $j\in\square'\neq\square$. The occurrence of nonzero 
bond-currents is associated with complex-valued self-consistent auxiliary fields.

Similarly, the expectation value of the hopping operator,
\begin{equation}
    B_{ij}=\frac{1}{2}\left( a^\dag_i a_j +\text{H.c.}\right),\label{bond_op}
\end{equation}
depends on whether the bond lies within the cluster or connecting the two clusters.

\begin{figure}[t!]
\centering
\includegraphics[scale=0.6]{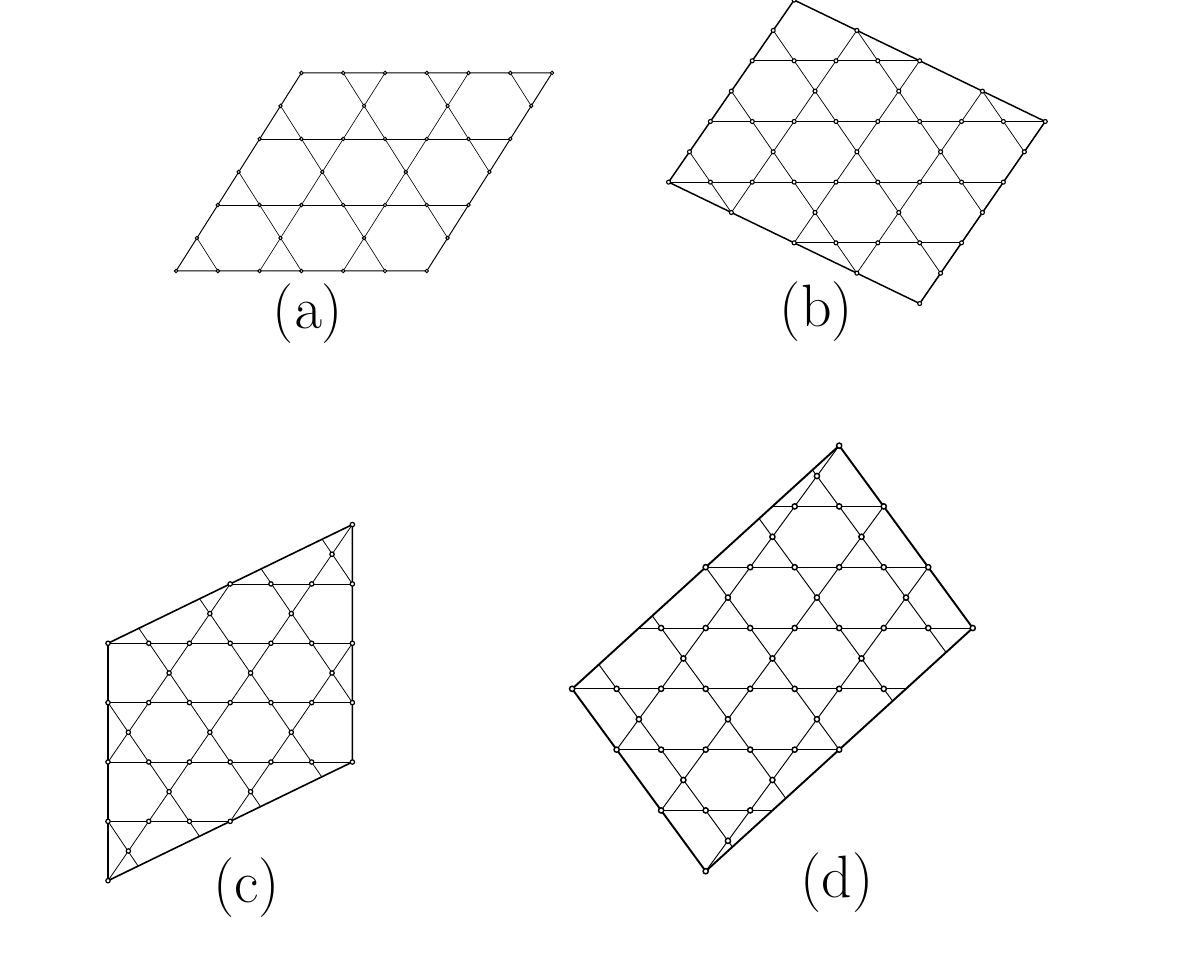}
\caption{\label{fig:clusters_ED} Clusters used for the ED computations in this work: 
(a) 27, (b) 36c, (c) 36, and (d) 45. }
\end{figure}
%
\begin{figure*}[htb]
\centering
\includegraphics[scale=0.55]{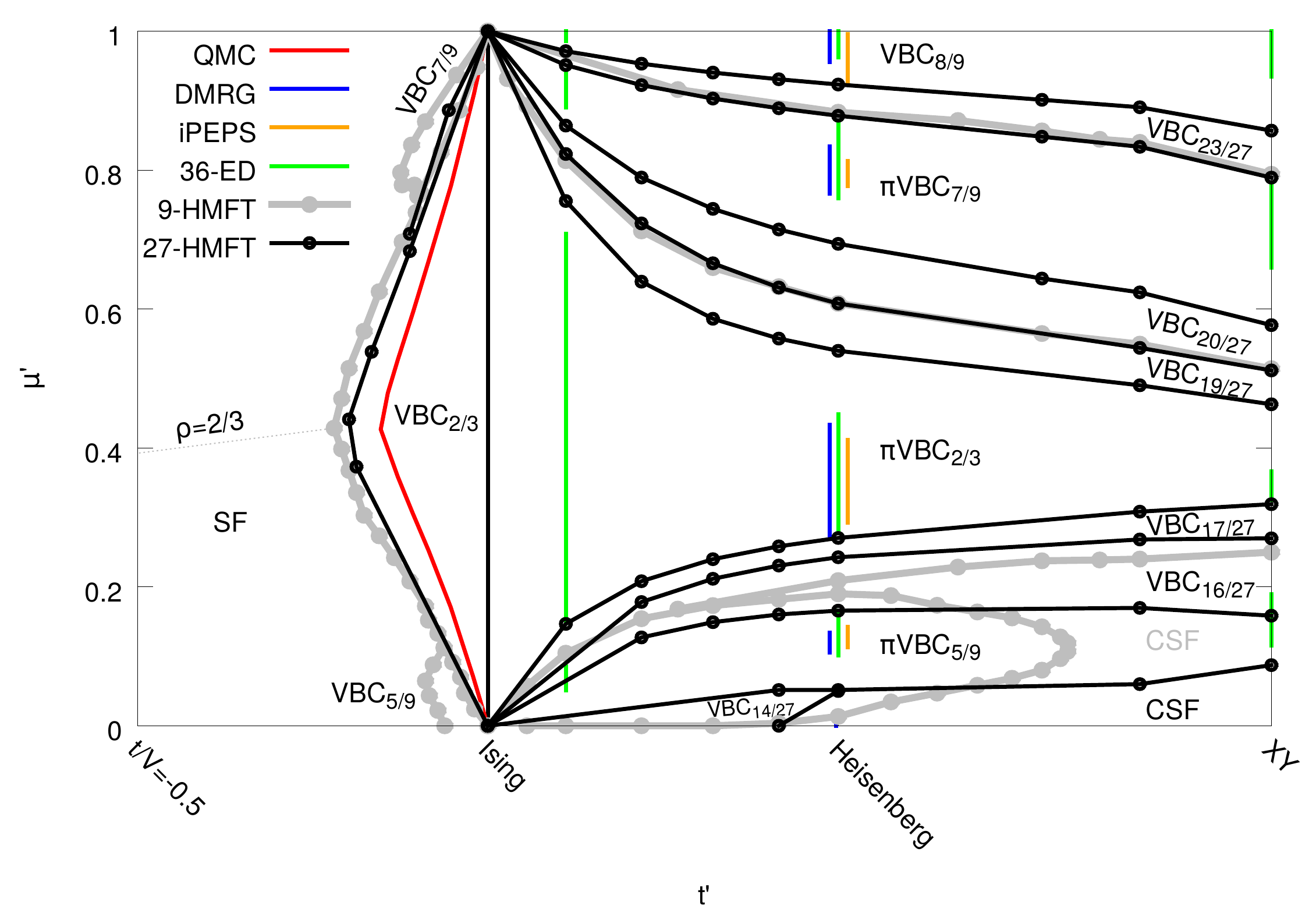}
\caption{\label{fig:9site_diag} (Color online) Quantum phase diagram of model (\ref{hardcore_ham}) 
computed with a 9-sites (gray) and 27-sites (black) CB Gutzwiller ansatz and ED. Also shown are 
QMC results from Ref.~\onlinecite{Isakov2006} (red), 
and DMRG and iPEPS plateaux widths at the Heisenberg line ($t/V=+0.5$) from 
Ref.~\onlinecite{Nishimoto2013} (blue) and Ref.~\onlinecite{Picot2016} (orange), respectively. 
The normalized chemical potential $(\mu')$ and hopping amplitude $(t')$ axis are defined 
 to fit all physical limits. Phases are labeled as superfluid (SF), chiral superfluid 
(CSF) and valence bond crystal of density $\rho$ (VBC$_{\rho}$). The prefix $\pi$  
distinguishes frustrated from non-frustrated regimes. The 
constant $\rho=2/3$ line within the SF phase (dashed) is computed 
with 9-HMFT. Due to particle-hole symmetry, the diagram is symmetric around the half-filling line 
$\mu'=0$. The trivial fully occupied and empty phases, correspond to $\mu'>1$ and $\mu'<-1$, respectively.}
\end{figure*}
\subsection{\label{sec:ed}Exact diagonalization}
%
We use a standard Lanczos algorithm to compute the density as a function of the 
chemical potential -- equivalent to the magnetization curve in the spin language -- for various cluster 
sizes. As discussed in more detail in Ref.~\onlinecite{Capponi2013}, in order to accomodate the 
exact VBC$_{8/9}$ state, we restrict ourselves to clusters containing the $K$ points 
of the Brillouin zone, i.e. clusters with 27, 36 and 45 sites, as shown in Fig. \ref{fig:clusters_ED}. 

We characterize the quantum phases by computing various correlation functions related to the 
observables previously defined. In particular, we compute the connected kinetic correlations,
\begin{equation}\label{eq:dimerdimer}
\langle B_{ij}B_{kl}\rangle_c=
\langle B_{ij}B_{kl}\rangle - \langle B_{ij}\rangle \langle B_{kl}\rangle,
\end{equation}
where $ij$ and $kl$ denote pairs of nearest-neighbor sites comprising the bonds. 
Analogously, we compute current correlations to detect time-reversal symmetry breaking 
and the onset of chirality,
\begin{equation}\label{eq:jj}
\langle \mathcal{J}_{ij}  \mathcal{J}_{kl} \rangle_c=
\langle \mathcal{J}_{ij}  \mathcal{J}_{kl} \rangle-
\langle \mathcal{J}_{ij} \rangle \langle \mathcal{J}_{kl} \rangle.
\end{equation}

We also compute the fidelity susceptibility $\chi_F$ to establish whether a quantum phase 
transition, not captured by the HMFT  approach,  takes place at fixed density 
when varying the density-density interaction strength $V$~\cite{Gu2010,Albuquerque2010}.  
This quantity detects phase transitions \emph{without any a priori 
knowledge of the order parameter}, when varying a driving 
parameter, by measuring the change in the ground-state 
wave function $| \psi_0\rangle$. In our case 
\begin{equation}\label{eq:fidelity}
|\langle \psi_0(V)|\psi_0(V+\delta V)\rangle| \simeq 1 - \frac{1}{2}\chi_F \ ( \delta V)^2.
\end{equation}

\section{Quantum Phase diagram\label{sec:phase_diag}}
%
The quantum phase diagram is obtained by computing the ground state energy and its 
derivatives, together with the observables and order parameters defined in the previous 
section. We assess the stability of the phases obtained with the 9-sites cluster CB Gutzwiller  
ansatz (9-HMFT) by performing a second coarse-graining with a 27-sites cluster (27-HMFT). 
In the first case, all $2^9=512$ cluster configurations are considered in the Hartree 
optimization (\ref{hartree}). In the 
second case, we introduce a cutoff in the number of states considered due to the extremely 
large Hilbert space dimension. 
We adjust this cutoff depending on the value of $\mu$ controlling the total density of the system. 
In particular, in-between the main plateaux we have used cluster configurations satisfying the 
condition $\rho_{\sf low}\le \rho\le \rho_{\sf ab}$, 
where $\rho_{\sf low}$ and $\rho_{\sf ab}$ refer to the densities of the immediate below ({\sf low}) 
and above ({\sf ab}) plateaux. Near half-filling, we keep those states with $12\le N_\square\le 15$, 
$N_\square=\rho N$, summing up to $74.884.320$ states and allowing for a density range 
$4/9\leq\rho\leq 5/9$.  A wider density range is available with increasing chemical potential, e.g. 
above the $\pi$VBC$_{2/3}$ plateau, where we have used $18\le N_\square\le 27$.
We have verified that it does not affect the final results, within 
error bars, by choosing different cutoffs for a given region in the phase diagram. 

Figure \ref{fig:9site_diag} displays the phase diagram obtained with 9-HMFT (gray lines) 
and 27-HMFT (black lines) together with the plateaux obtained with ED on 36-sites clusters 
(36-ED), in green (see Sec.~\ref{sec:ED}) for different hopping 
strengths, as well as QMC results~\cite{Isakov2006}, DMRG~\cite{Nishimoto2013}, and 
iPEPS~\cite{Picot2016} in the Heisenberg limit. We have normalized the hopping amplitude 
and the chemical potential in order to fit all parameter regimes within a single plot,
\begin{equation}   
t'=\frac{t}{\sqrt{t^2 + V^2}} \ , \ \mu'=
\left\{
\begin{array}{cc}
       \frac{\mu-2V}{2(V-2t)},& ~~~~t/V< 0, \\ & \\
       \frac{\mu-2V}{2(V+t)},& ~~~~t/V>0. \\
\end{array}
\right. 
\end{equation}
Note that the phase diagram is symmetric around the half-filling line $(\mu'=0)$ due 
to particle-hole symmetry, thus we only plot the $\rho\geq 1/2$ region. 

\begin{figure}[htb]
\centering
\includegraphics[width=0.4\textwidth]{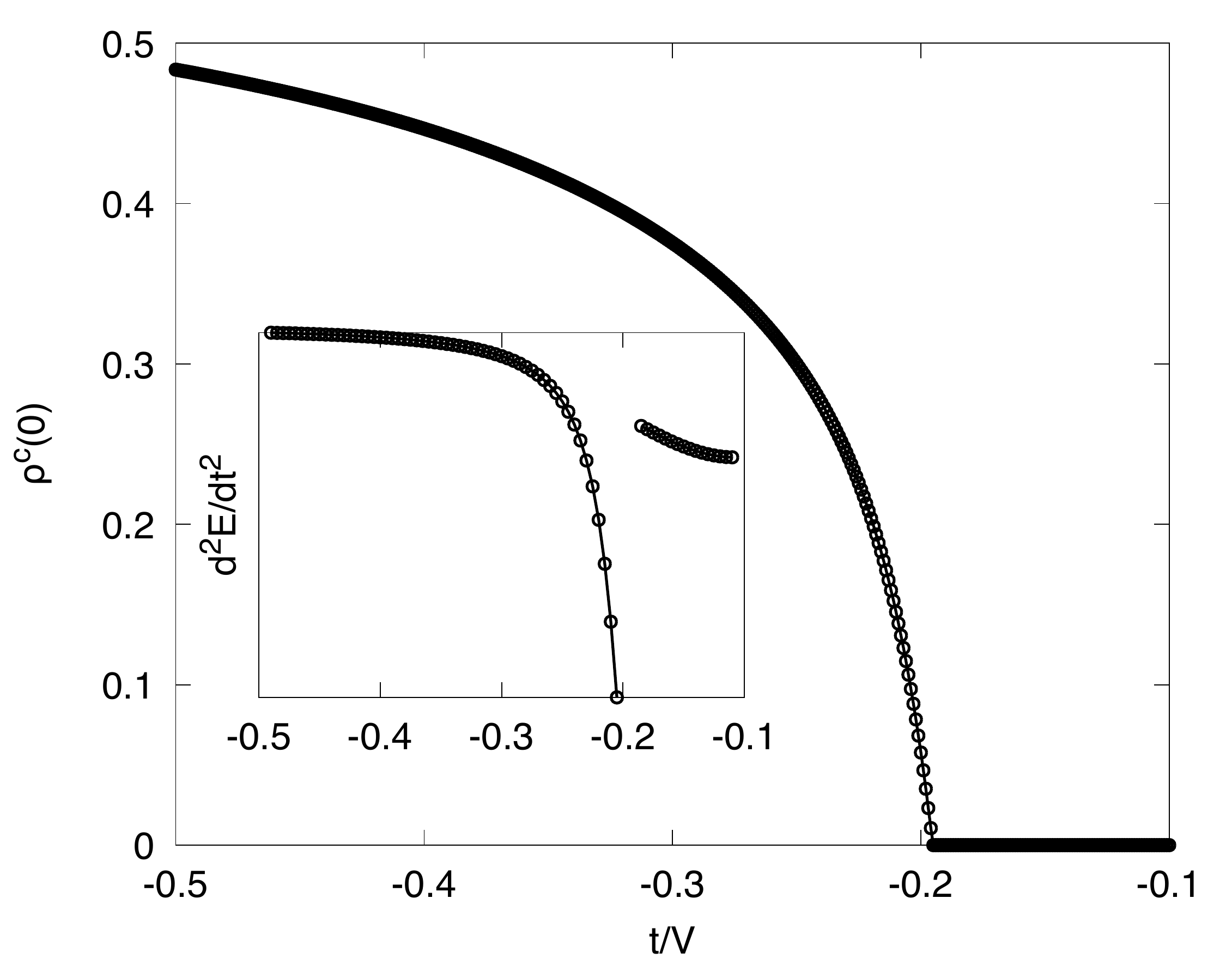}
\caption{\label{fig:sf_23} Condensate density computed with a 9-sites CB Gutzwiller along 
the $\rho=2/3$ density line, and across the SF-VBC$_{2/3}$ transition in the 
non-frustrated regime of the phase diagram (Fig. \ref{fig:9site_diag}). Inset: Second-order 
derivative of the ground-state energy along the same line.}
\end{figure}

\begin{figure*}[!htp]
\centering
\includegraphics[width=0.95\textwidth]{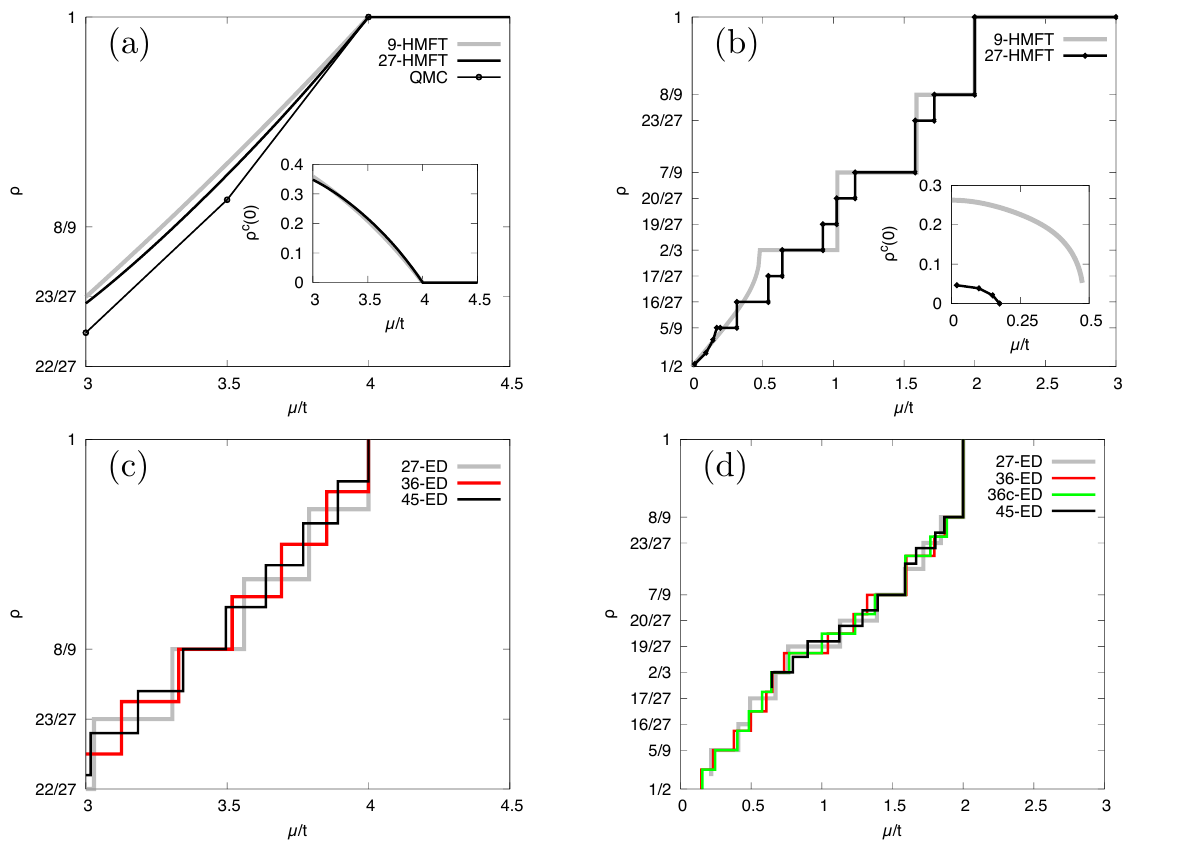}
\caption{\label{densXY}Total density and condensate density (inset) in the XY limit 
($V=0$) for the (a) non-frustrated regime $t<0$  and (b) frustrated regime $t>0$  computed with 9-HMFT and 27-HMFT. 
Total density for the (c) non-frustrated and (d) frustrated  regimes in the same XY limit obtained with 
ED.  In the frustrated regime, the widths of the ED plateaux do not have a monotonic dependence with 
increasing cluster size,  contrary to the non-frustrated regime.}
\end{figure*}

\subsection{Non-frustrated regime $(t<0)$}
In the non-frustrated region we find a superfluid (SF), the trivial fully occupied (FO), and a 
VBC$_{2/3}$ phase. We also observe two small lobes of density $\rho=5/9,7/9$ around 
$\mu'=0$ and $\mu'=1$, respectively, that shrink considerably upon increasing the cluster 
size, and presumably vanish in the thermodynamic limit. This is in agreement with previous 
analytical ~\cite{Jiang2006,Sengupta2006} and numerical~\cite{Isakov2006} studies 
where a stable SF phase for any $t<0$ at half-filling was found. 

The VBC$_{2/3}$ wave function obtained within the 9-HMFT approach is dominated by 
configurations containing three localized resonant holes, 
\begin{equation}
\ket{\Phi^{2/3}} = \alpha \left(
    \left\vert
    \begin{array}{c}
    \includegraphics[clip=true,scale=0.1]{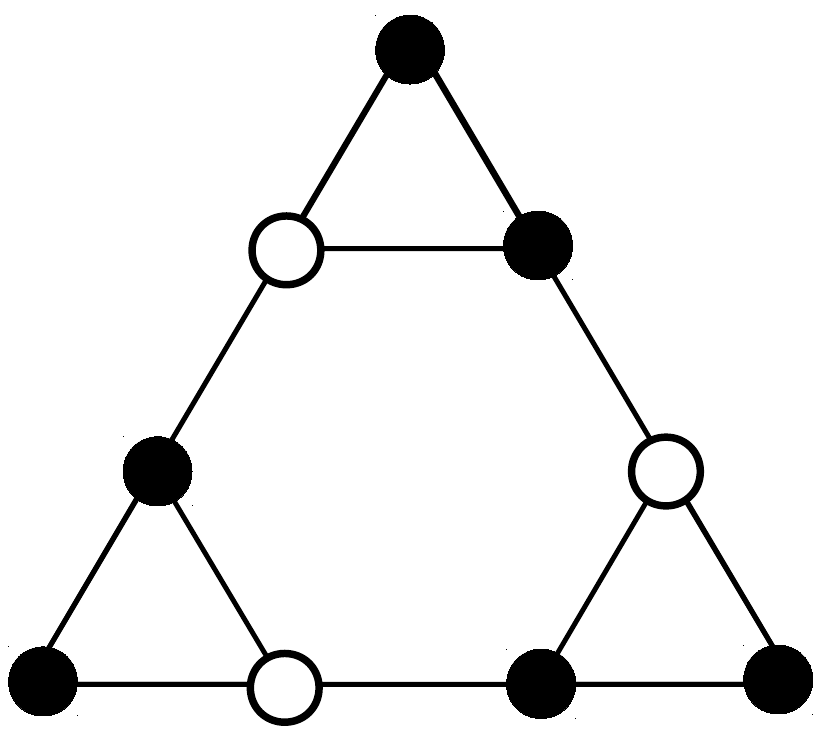}
    \end{array}
    \right\rangle+
    \left\vert
    \begin{array}{c}
    \includegraphics[clip=true,scale=0.1]{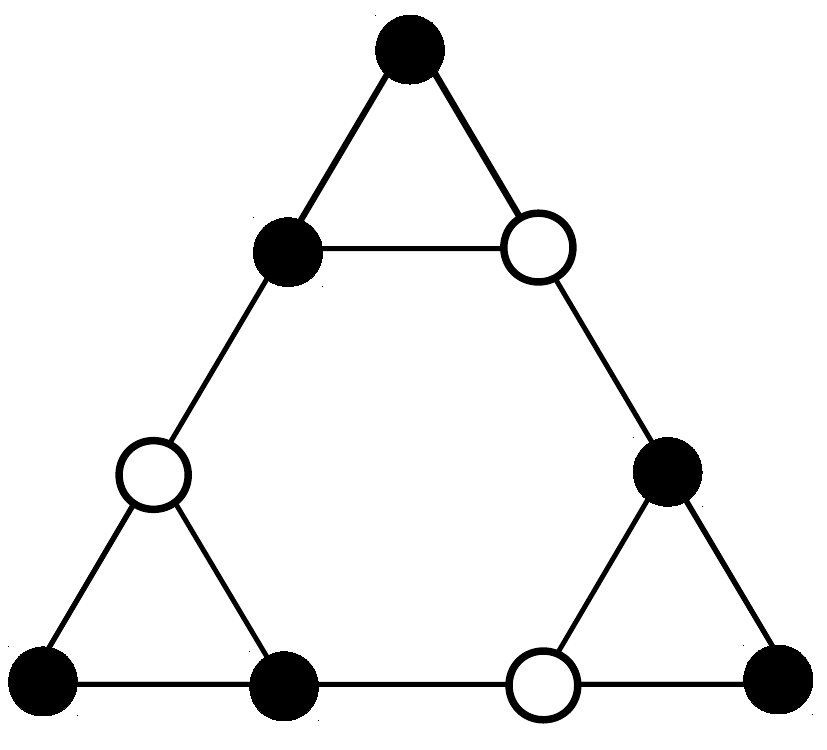}
    \end{array}
    \right\rangle    \right)
    +\ldots\ , \label{wf:23}
\end{equation}
where we refer to occupied (empty) sites as black (white) dots. For small hoppings, 
$-0.1V<t<0$, $\alpha^2\sim0.49$.

The SF phase is characterized by a finite condensate density at the $\Gamma$ ($\mathbf{k}=0$) 
point of 
the Brillouin zone. The phase transition from the SF to the VBC$_{2/3}$, at constant 
density $\rho=2/3$, is found to be second order (see Fig. \ref{fig:sf_23}), in qualitative 
agreement with previous QMC studies~\cite{Isakov2006,Damle2006}, where it was 
argued to be weakly first order after a refined scaling and histogram analysis~\cite{Isakov2006}. 

The phase boundary between the SF and FO phase can be exactly determined by equating 
the energy of the local Hamiltonian (\ref{projector_ham}) for $(S_p,S^z_p)=(3/2,1/2)$ and 
$(3/2,3/2)$, leading to
\begin{equation}
\mu_{\sf full}(t<0)=4(V-t) .  \label{eq:mufull_neg}
\end{equation}
It could be equally computed noticing that a unique delocalized hole over a background of 
fully occupied sites is indeed an exact eigenstate of the Hamiltonian (\ref{hardcore_ham}). 

\subsection{Frustrated regime $(t>0)$}
%
Similarly, in the frustrated regime the boundary to the FO state is determined by finding 
the degeneracy point of the local energies for the $(1/2,1/2)$ and $(3/2,3/2)$ local 
Hamiltonian (\ref{projector_ham}),
\begin{equation}
\mu_{\sf full}(t>0)=2(2V+t),\label{eq:mufull_pos}
\end{equation}
or by equating the exact energies of the VBC$_{8/9}$ and FO eigenstates of 
Hamiltonian (\ref{hardcore_ham}). Along this line, the ground state is macroscopically 
degenerate, due to the dispersionless nature of the non-interacting lowest-energy 
band.

Below this saturation line, we find a staircase of VBC$_{\rho}$ phases with fractional 
densities that dilutes into a CSF near half-filling, when diminishing 
the chemical potential. Similar to a devil staircase, smaller plateaux appear between 
the main ones at $\rho=5/9,2/3,7/9,8/9$, when increasing the cluster size from 9 to 27 sites. 
Moreover, these crystal phases prevail even in the absence of the density-density 
interaction, i.e. in the XY regime.

Note that the phase boundaries between main plateaux -- $\pi$VBC$_{2/3}$-$\pi$VBC$_{7/9}$ 
and $\pi$VBC$_{7/9}$-VBC$_{8/9}$ -- computed with 9-HMFT almost coincide with the boundaries of the 
intermediate plateaux obtained with 27-HMFT. Also, notice the agreement between the lower boundary 
of the $\pi$VBC$_{2/3}$ computed with 27-HMFT and the one obtained by 36-ED in all limits of the AFM regime.

When performing ED on different clusters we obtain various staircases, analogously 
to the FM regime. However, contrary to the FM regime, the \textit{main plateaux} in the AFM regime
do not decrease its width monotonically with increasing cluster size. This feature prevails in the 
XY regime, too.

In the following we give a detail analysis and comparison 
of the results obtained with both techniques in the AFM regime.

\subsubsection{HMFT approach}
When computing with 9-HMFT, the main plateaux with $\rho>5/9$ prevail up to the XY 
limit (see Fig.~\ref{densXY}),  and their transitions are all first order. The VBC$_{5/9}$ plateau 
exhibits a lobe surrounded by a doubly degenerate CSF, having 
its tip at $t=1.1V$. Upon increasing the cluster size to 27 sites, the main plateaux shrink 
leading to the appearance of other narrower plateaux characterized by densities commensurate 
with the cluster size. The VBC$_{5/9}$ in this case extends to the XY limit, while the CSF region 
shrinks to near half-filling. Around half-filling and for small values of the hopping amplitude, 
$t<0.4V$, the CSF disappears giving rise to an additional VBC with density $\rho=14/27$,  
rendering the half-filling line for this region to be a first order transition line to a $\rho=13/27$ state. 

These results are in partial disagreement with iPEPS results obtained in Ref.~\onlinecite{Picot2016}, 
where also a 9-sites unit cell was used. There, U(1) symmetry broken phases were found  
in regions between the main plateaux, although not all regions were fully characterized. We cannot 
rigorously discard the possibility that the intermediate phase between the main plateaux 
may have a characteristic length scale larger than the clusters used in this work, and thus be 
characterized by either BEC condensation in $\mathbf{k}$-points of the Brillouin zone different 
from the ones contained within the 9-sites and 27-sites clusters (see Fig. \ref{9site_embed}), or by a VBC 
with a periodicity not commensurate with our clusters, or other phase with topological 
order~\cite{Kumar2014}.

We next describe the main features of the principal VBC$_\rho$ and CSF phases. Generically, 
the CB Gutzwiller wave function for a particular VBC$_\rho$ depends exclusively on the hopping 
parameter $t$, and not on the chemical potential $\mu$. Moreover, it contains configurations with 
a definite number of hard-core bosons per cluster, $N_\square$. On the contrary, the CSF 
wave function, characterized by breaking of the global U(1) symmetry, i.e. onset of BEC, changes 
with both $\mu/V$ and $t/V$.

\paragraph{One-hole resonant state (VBC$_{8/9}$).---}
The VBC$_{8/9}$ phase is characterized by a fully stacked pattern of localized one-hole resonant 
states, which was shown to be one exact ground state of the XXZ Hamiltonian~\cite{Schulenburg2002} 
along the line defined in Eq. (\ref{eq:mufull_pos}). This state is exactly 
contained within the 9-HMFT Gutzwiller wave function and can be written as
\begin{equation} \hspace*{-0.3cm}
\ket{\Phi^{8/9}} = \frac{1}{\sqrt{6}} \left(
    \left\vert
    \begin{array}{c}
    \includegraphics[clip=true,scale=0.1]{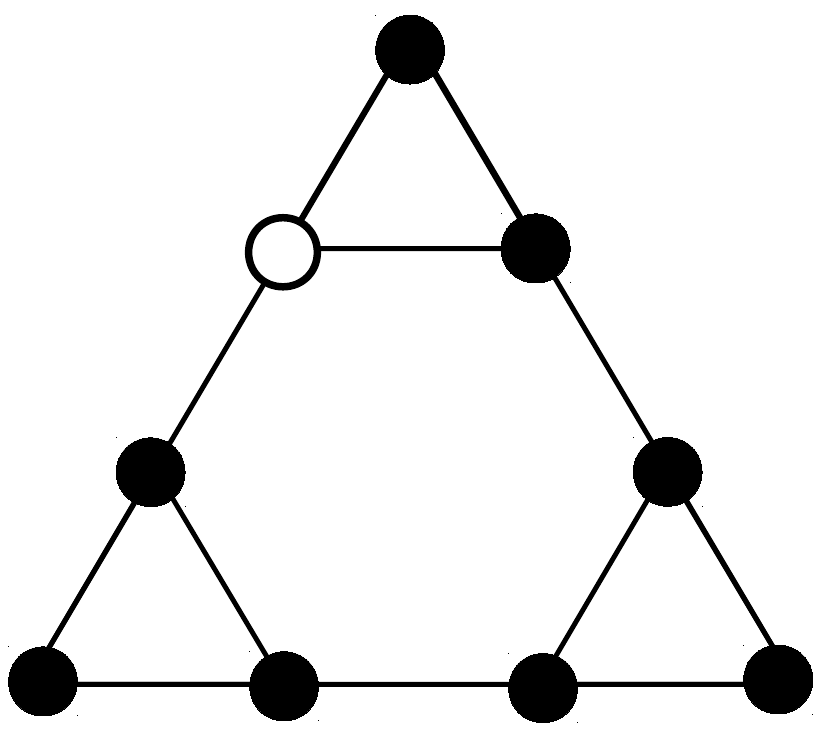}
    \end{array}
    \right\rangle
    -
    \left\vert
    \begin{array}{c}
    \includegraphics[clip=true,scale=0.1]{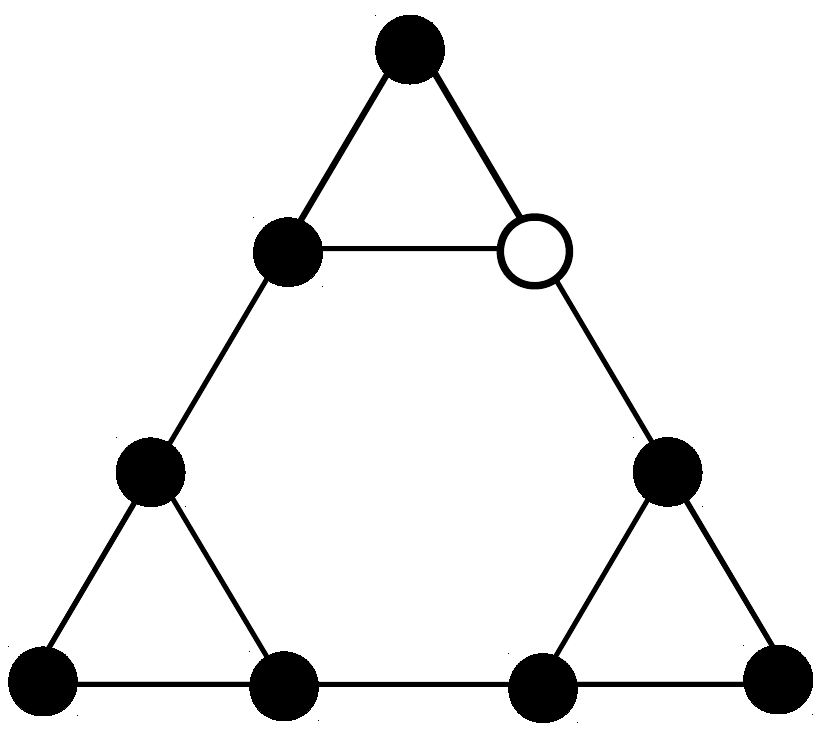}
    \end{array}
    \right\rangle+[C_3~\text{rots.}] \right) \label{exact_wf}.
\end{equation}
Equivalently, $\ket{\Phi^{8/9}}$ be written as a state where a two-spin singlet resonates in the 
hexagon of the 9-sites cluster.

\paragraph{Two-hole resonant state ($\pi$VBC$_{7/9}$).---}
The CB Gutzwiller wave function $\pi$VBC$_{7/9}$ is a localized two-hole resonant 
state that can be written in terms of a 9-sites clusters as
\begin{eqnarray}
\ket{\Phi^{7/9}_\pi} &=& \beta_1 \left(
    \left\vert
    \begin{array}{c}
    \includegraphics[clip=true,scale=0.1]{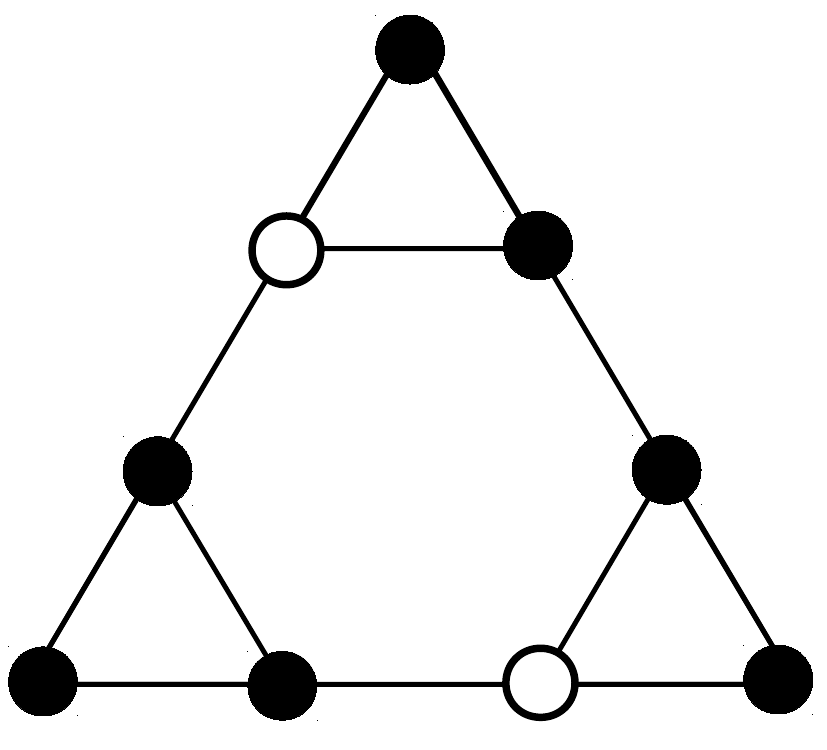}
    \end{array}
    \right\rangle
+ \left[C_3~\text{rots.}\right]
    \right)\notag\\
&&-\beta_2
    \left(
    \left\vert
    \begin{array}{c}
    \includegraphics[clip=true,scale=0.1]{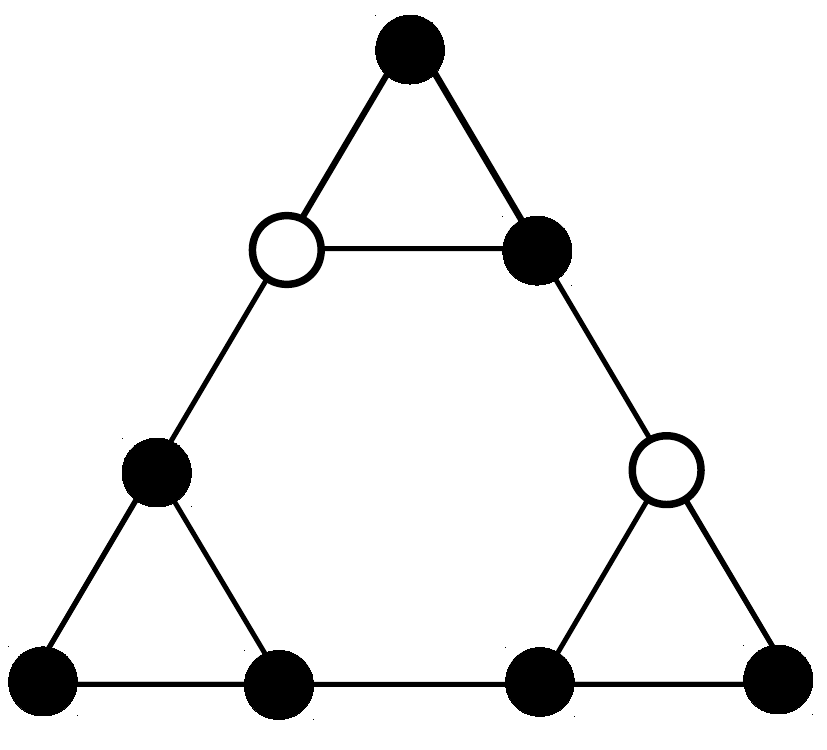}
    \end{array}
    \right\rangle
+
    \left\vert
    \begin{array}{c}
    \includegraphics[clip=true,scale=0.1]{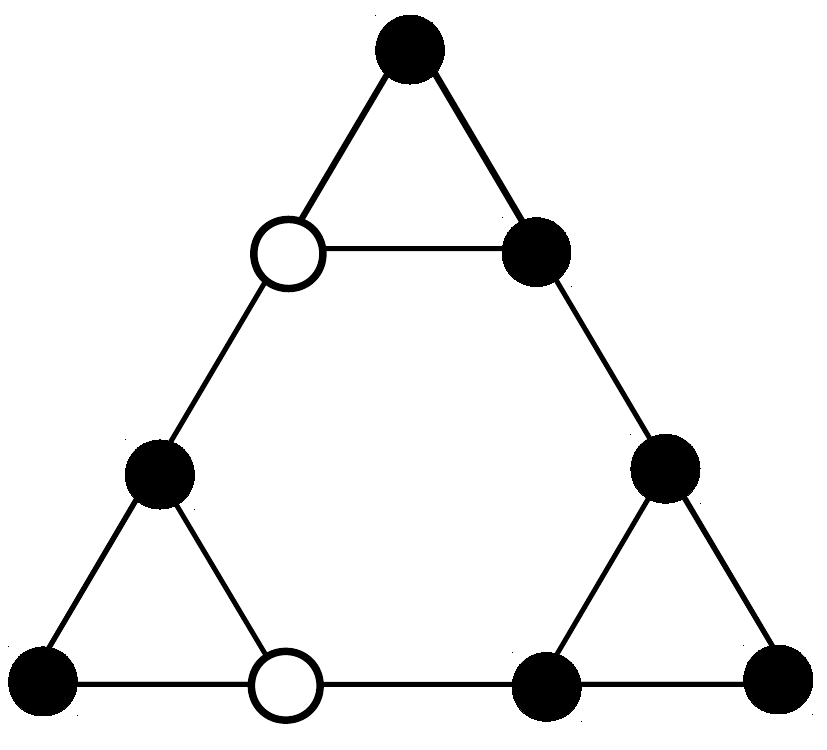}
    \end{array}
    \right\rangle
+ \left[C_3~\text{rots.}\right]\right)\notag\\
&&+\ldots  ,  \label{wf:79}
\end{eqnarray}
where $\beta_1$ and $\beta_2$ are positive real numbers, and the leading weights 
for the whole range up to the XY regime, i.e. $\beta_1^2=0.16$ and 
$\beta_1^2=0.09$ for $t=0.1V$, while $\beta_1^2=0.11$ and $\beta_1^2=0.08$ in the 
XY limit. 
This wave function, although not an exact eigenstate of the Hamiltonian (\ref{hardcore_ham}), 
is a good approximation to the ground state. It reasonably describes the Heisenberg and XY 
limits, as can be seen by computing the expectation value of the $B_{ij}$ 
operator (defined in Eq. \ref{bond_op}) shown in Table \ref{tab:bond_op}.

\begin{table}[t]
\begin{center}
\begin{tabular}{|c|c|c|c|}
\hline
~~$\rho$~~ & $M$ & Heisenberg & AFM XY \\
\hline
8/9 & 7/9 & 
 $\begin{array}{c}
 \includegraphics[width=0.15\textwidth]{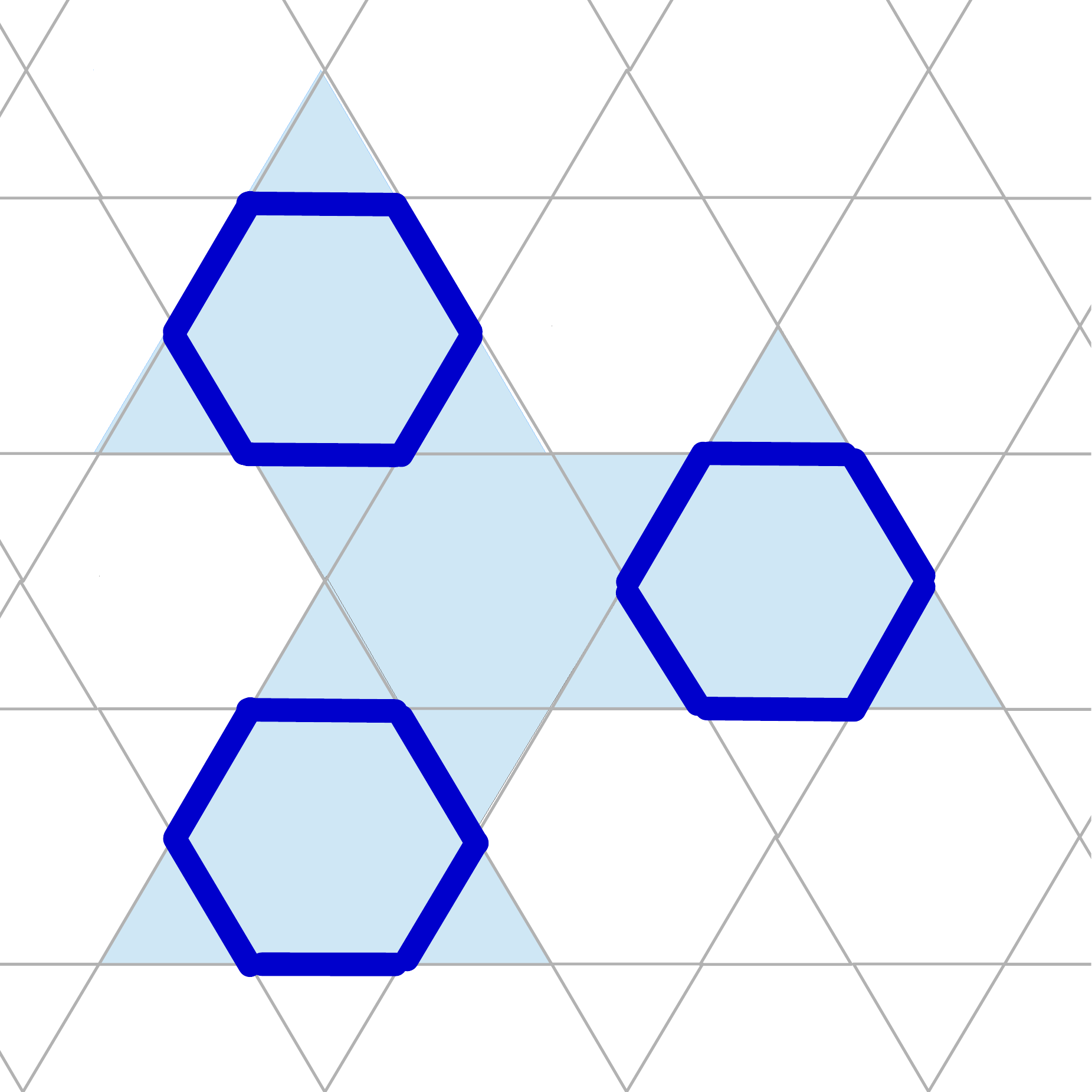}
 \end{array}$
& 
 $\begin{array}{c}
 \includegraphics[width=0.15\textwidth]{89.pdf}
 \end{array}$
\\
\hline
7/9 & 5/9 & 
$\begin{array}{c}
\includegraphics[width=0.15\textwidth]{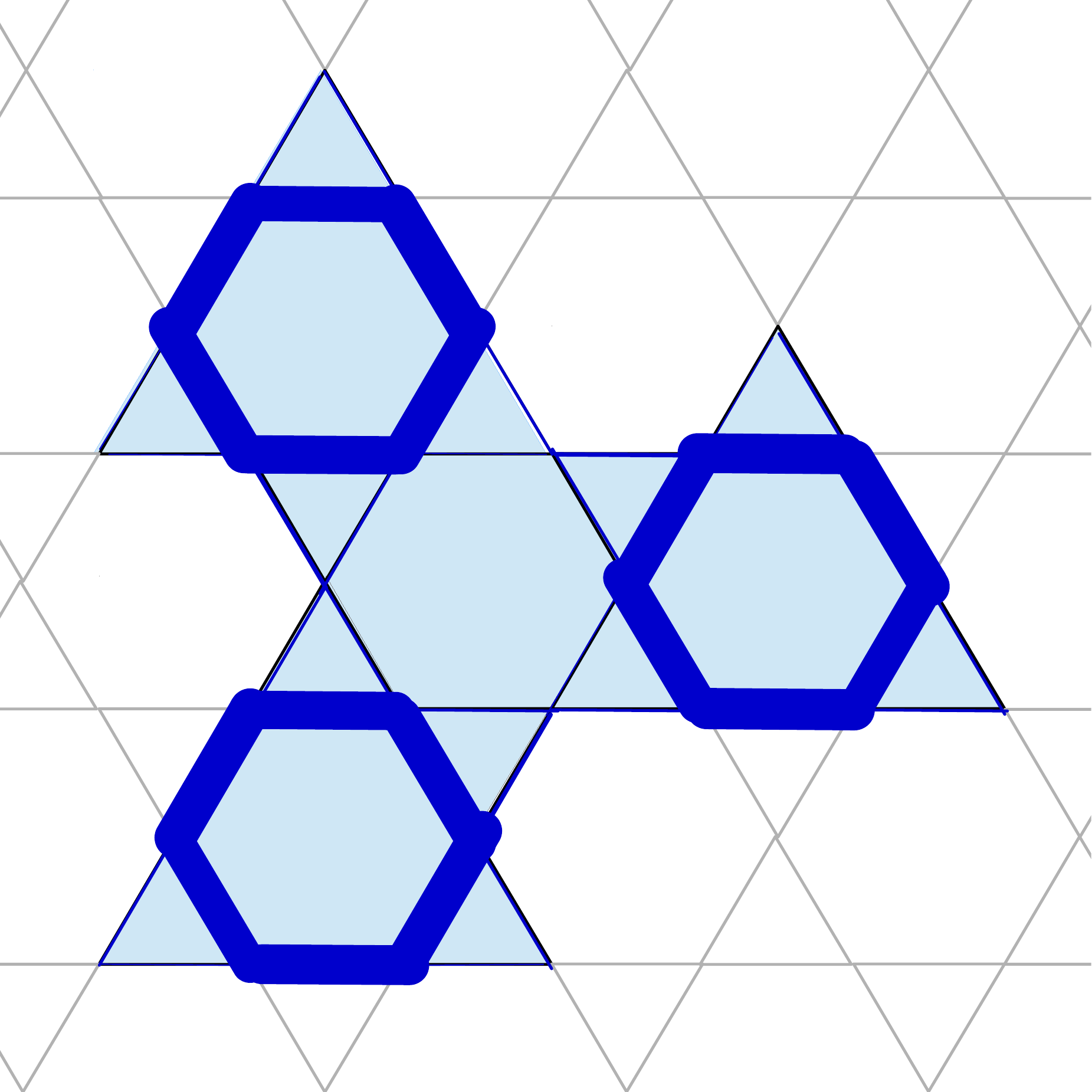}
\end{array}$& 
$\begin{array}{c}
\includegraphics[width=0.15\textwidth]{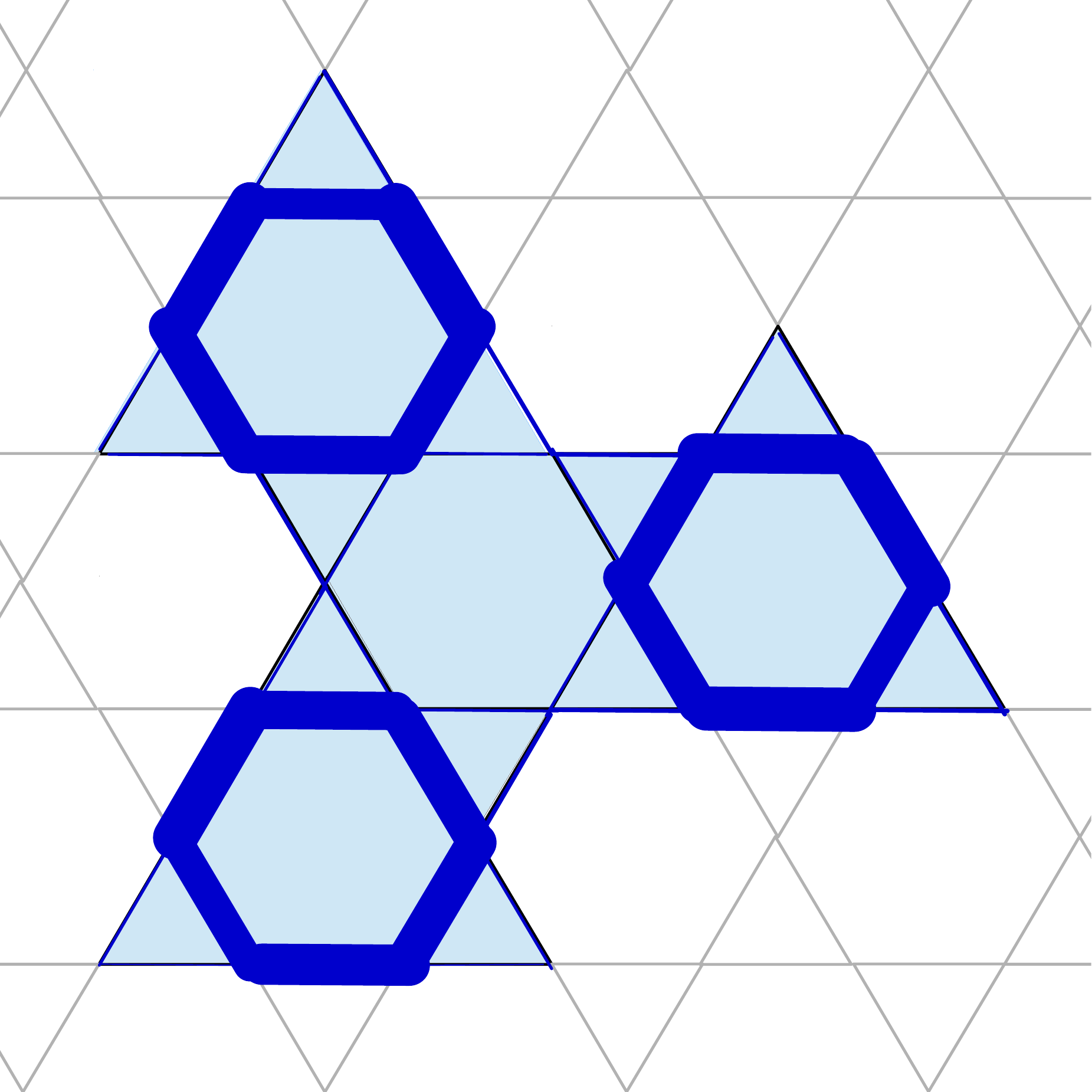}
\end{array}$
\\
\hline
2/3 & 1/3 &
$\begin{array}{c}
\includegraphics[width=0.15\textwidth]{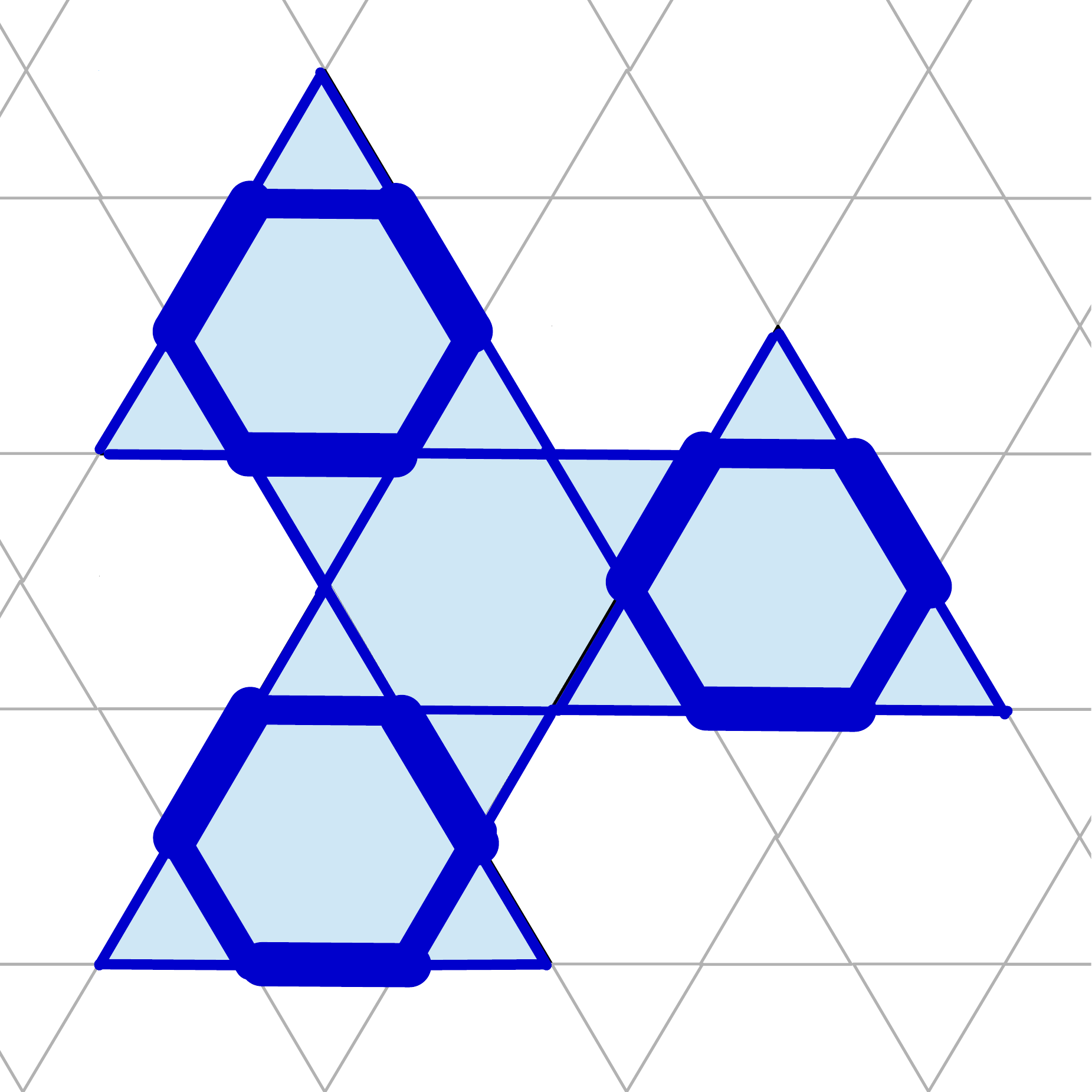}
\end{array}$& 
$\begin{array}{c}
\includegraphics[width=0.15\textwidth]{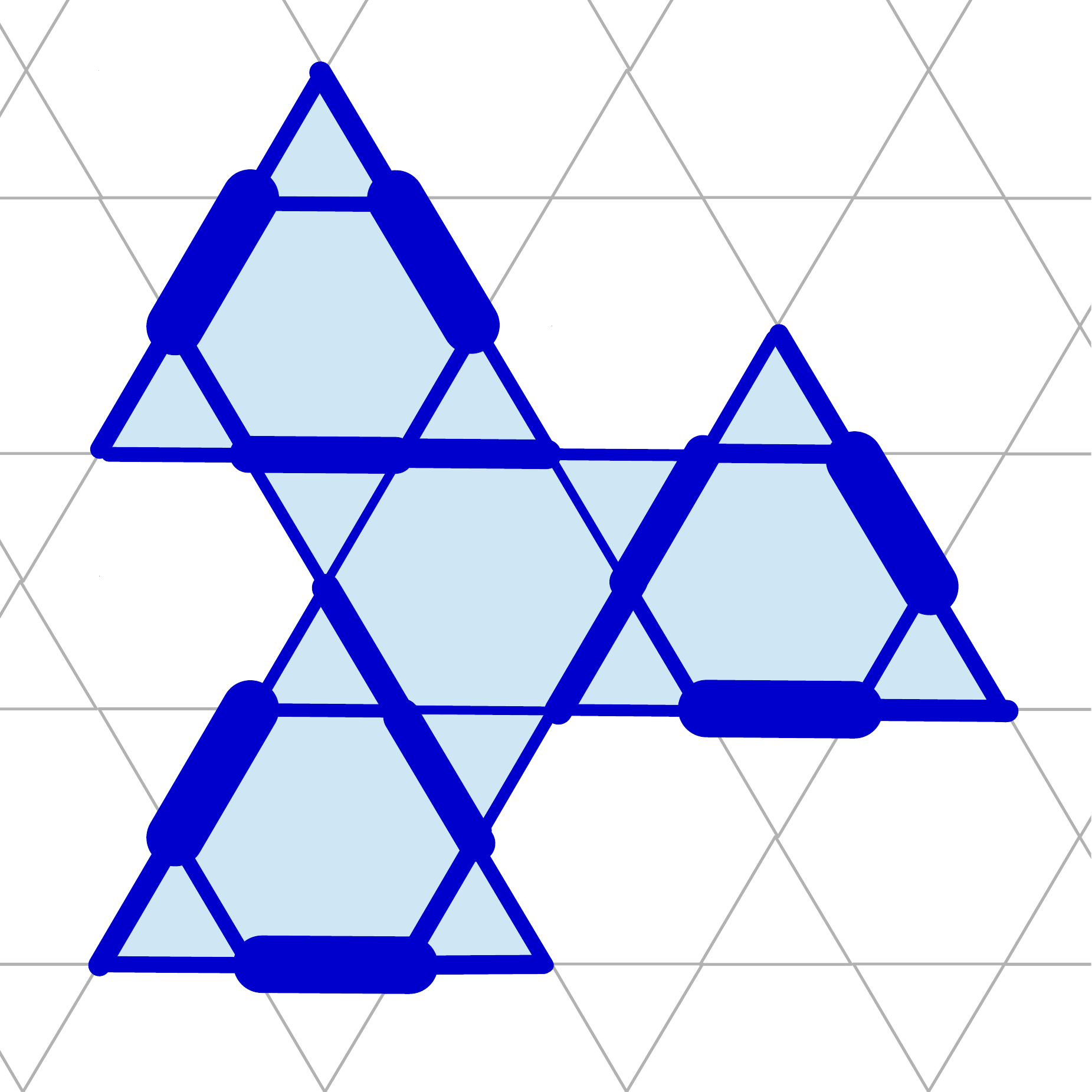}
\end{array}$
\\
\hline
5/9 & 1/9 &
$\begin{array}{c}
\includegraphics[width=0.14\textwidth]{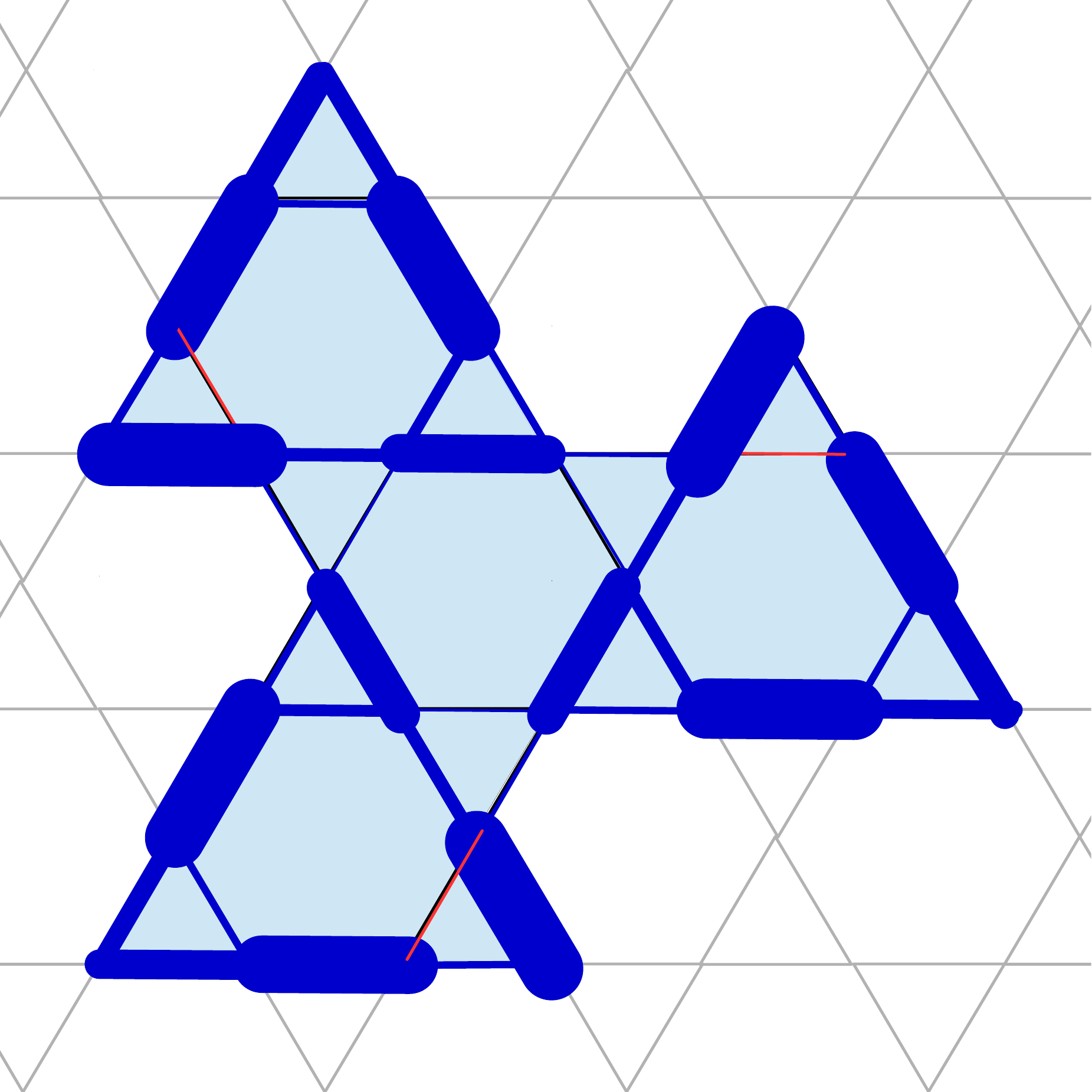}
\end{array}$& 
$\begin{array}{c}
\includegraphics[width=0.14\textwidth]{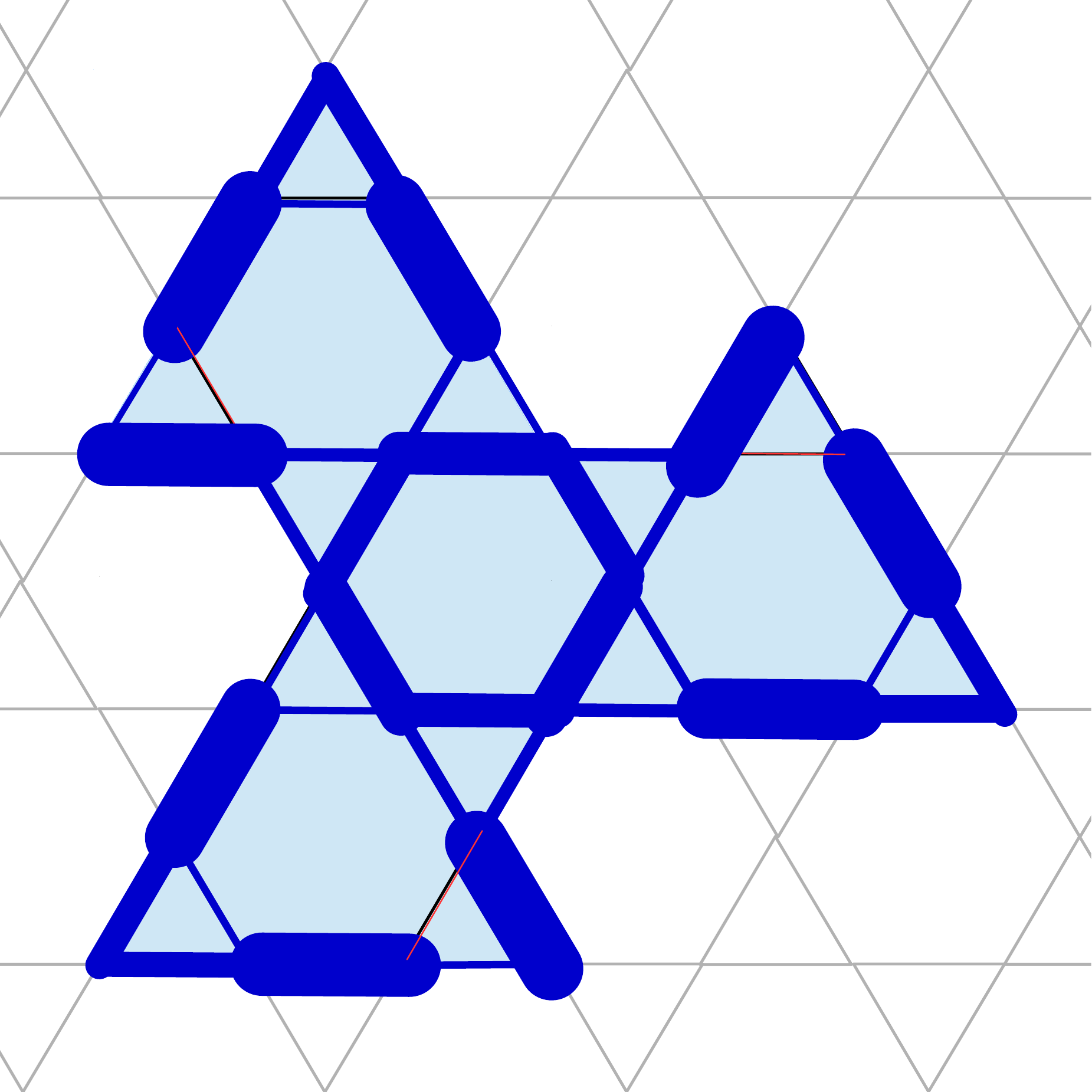}
\end{array}$
\\
\hline
\end{tabular}
\end{center}
\caption{\label{tab:bond_op}(Color online) Expectation value of the $B_{ij}$ operator (\ref{bond_op}) 
for the Heisenberg and XY limits, in the frustrated regime, within main 
VBC$_{\rho}$ plateaux. Thickness of the bonds is proportional to the absolute value of 
$\left\langle B_{ij}\right\rangle$. For the 27-HMFT, the maximum absolute value is 0.4 and 
blue (red) color refers to its negative (positive) sign. Notice that only the $\pi$VBC$_{5/9}$ 
exhibits some slightly positive  values for three of the bonds in the cluster.}
\end{table}

\begin{figure}[t]
\centering
\includegraphics[width=0.42\textwidth]{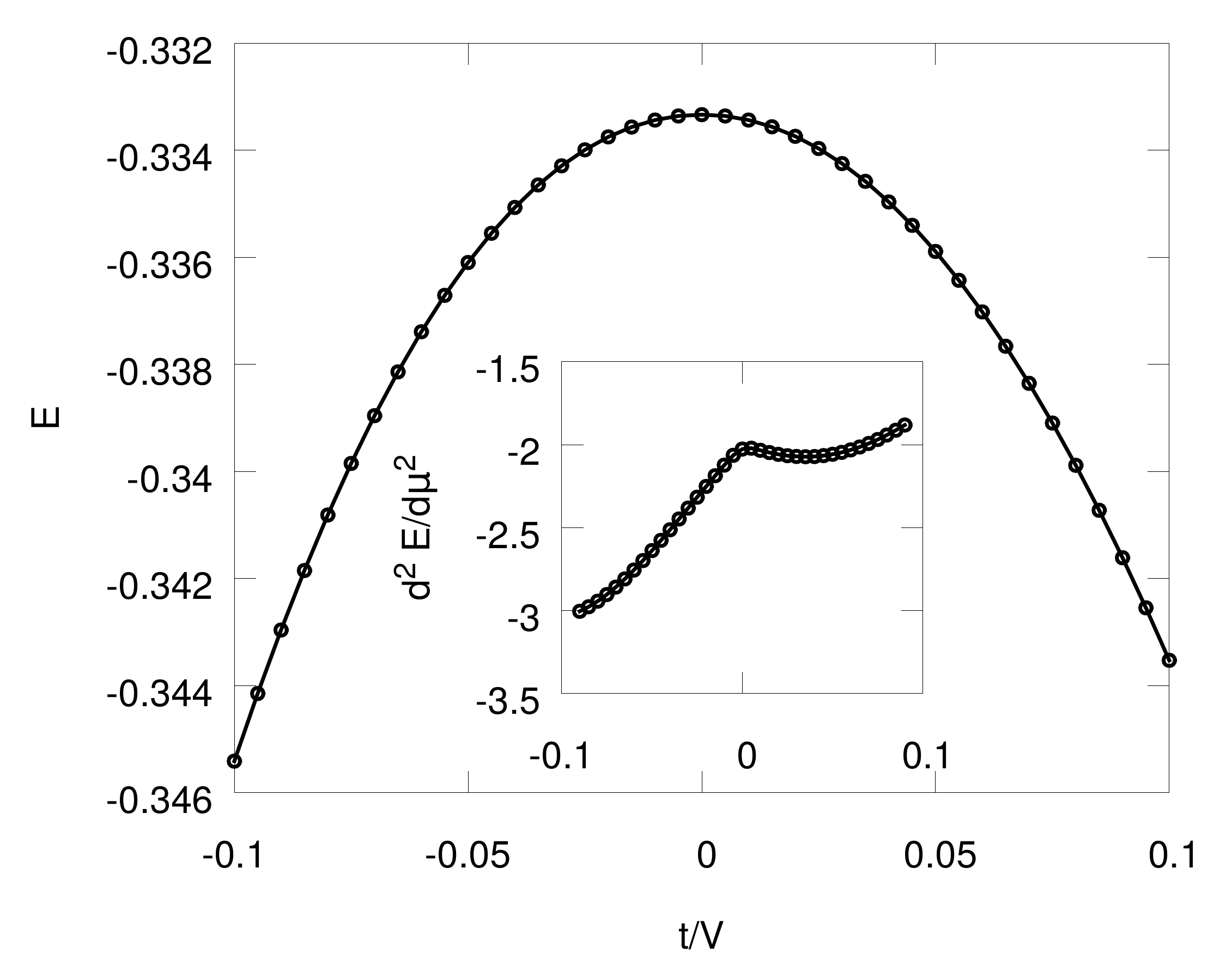}
\caption{\label{fig:ising23} Energy (in units of $V$) and second-order derivative (inset) across 
the $\pi$VBC$_{2/3}$ to VBC$_{2/3}$ transition for $\mu=3V$.}
\end{figure}

\paragraph{The $\pi$VBC$_{2/3}$ state.---}
Similarly to the FM regime, for small hopping amplitudes $(t<0.1V)$, the $\pi$VBC$_{2/3}$ can 
be written within 9-HMFT as
\begin{equation}
\ket{\Phi^{2/3}_\pi} = \gamma\left(
    \left\vert
    \begin{array}{c}
    \includegraphics[clip=true,scale=0.1]{up69.pdf}
    \end{array}
    \right\rangle-
    \left\vert
    \begin{array}{c}
    \includegraphics[clip=true,scale=0.1]{down69.pdf}
    \end{array}
    \right\rangle    \right)
    +\ldots \ , \label{wf:pi23}
\end{equation}
with amplitude $\gamma^2\sim 0.49$, and the next leading term being two orders of magnitude smaller. 
This situation changes smoothly with increasing $t/V$, until the XY limit, where $\gamma^2=0.095$ 
and the next leading term is of the same order of magnitude. Near the Ising limit 
-- and always at $\rho=2/3$ -- the small hopping amplitude generates an effective three-body 
ring-exchange hopping which lifts the Ising macroscopic degeneracy and stabilizes 
VBC order~\cite{Cabra2005}. Interestingly, the phase transition between the VBC$_{2/3}$ and $\pi$VBC$_{2/3}$ 
across the Ising line is not first order, but continuous, as can be 
seen by inspecting the energy and its second-order derivative in Fig. \ref{fig:ising23}.
This is probably due to the large number of quasi-degenerate low-lying $\pi$VBC$_{2/3}$ 
states present in the frustrated regime.
%

For $t>0.1V$, many other configurations with $N_\square=6$ start to have 
relevant weights, and thus the wave function cannot be approximated by a three-hole 
resonant state (\ref{wf:pi23}). This fact results in a complex VBC pattern, 
as can be seen when inspecting the expectation value of the $B_{ij}$ operator computed with 27-HMFT
(Table \ref{tab:bond_op}). 
Interestingly, in the XY limit, this phase was argued to be 
a fractional quantum Hall state characterized by non-trivial topological order, based 
on a Chern-Simons analysis~\cite{Kumar2014}.

\paragraph{The $\pi$VBC$_{5/9}$ state.---}
%
Leading contributions to this state cannot be interpreted 
in terms of resonant-holes over a background of fixed particles in any regime of the phase diagram. 
This is also evident in the related pattern of expectation values of the $B_{ij}$ 
operator in Table (\ref{tab:bond_op}). 
Interestingly, this phase was argued to be 
a translational invariant gapped phase with $\mathbb{Z}_3$ topological order, based on a 
DMRG study~\cite{Nishimoto2013},
and a VBC phase with eighteen-fold degeneracy, based on an iPEPS study~\cite{Picot2016}
Further analysis of the phase based on ED is presented below.

\begin{figure}[t]
\centering
\includegraphics[width=0.45\textwidth]{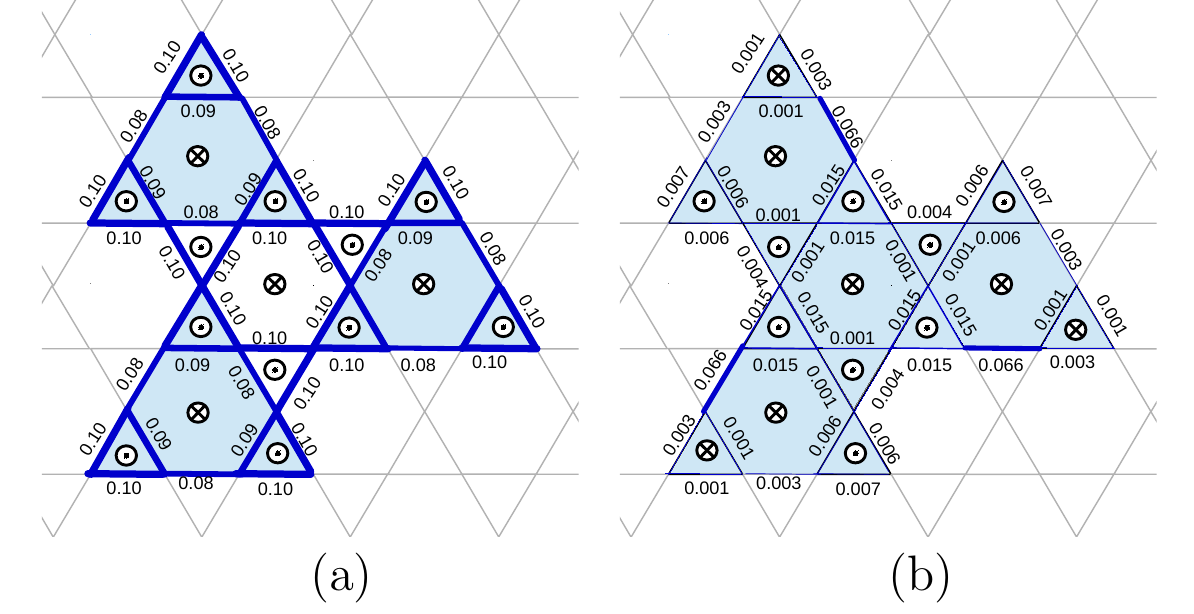}
\caption{\label{fig:csf_pattern} (Color online) Absolute value of the expectation of the 
$\mathcal{J}_{ij}$ operator defined in Eq. (\ref{eq:bond_current}) in the CSF phase as 
computed with (a) 9-HMFT, and  (b) 27-HMFT,  in the AFM XY limit  at $\mu/t=0.02,~V=0$. 
Thickness of the blue lines is proportional to  $\vert\langle\mathcal{J}_{ij}\rangle\vert$, and 
its maximum value is $\vert\langle\mathcal{J}_{ij}\rangle\vert=0.1$. The chirality of the 
loop currents is represented by a circled point (vortex) or a crossed circle (antivortex).}
\end{figure}

\paragraph{Chiral superfluid phase.---}
The doubly degenerate CSF is characterized by breakdown of the U(1) symmetry, signaled 
by a finite BEC at $\Gamma$, and nonvanishing bond-currents where triangles and 
hexagons  host vortices with opposite charge. Additionally, 
it possesses null total transverse magnetization per cluster, 
i.e. $\bra{\Psi} \sum_{j\in\square} \mathbf{S}_j^x \ket{\Psi}=0=
\bra{\Psi} \sum_{j\in\square} \mathbf{S}_j^y \ket{\Psi}$, a common feature with previous 
iPEPS studies in Ref.~\onlinecite{Picot2016}. Both the chirality (shown in Fig. \ref{fig:csf_pattern}) 
and condensate density are nevertheless supressed upon increasing the 
cluster size from 9-HMFT to 27-HMFT, and affected by finite-size effects, as three of the triangles 
in the 27-sites cluster change the sense of chirality. 

To assess the stability of the chiral order around half-filling, we have performed 18-HMFT 
computations~\footnote{The 18-sites cluster used is comprised of two vertically connected 9-sites 
clusters.} along the half-filling line, finding a SF in the AFM XY regime that transitions 
to a half-filled VBC$_{1/2}$ at $t\sim V$, both states without chiral order. 
The relation between the presence (absence) of chirality and the odd (even) number of sites of 
the cluster  is consistent with previous ED calculations of the KHAF,  
demonstrating the existence of low-lying S=1 states characterized by nontrivial (null) Chern 
numbers --related to breaking of time-reversal-- when using odd (even) clusters~\cite{Waldtmann1998}. 
Therefore, we conjecture that the nonvanishing chirality of the CSF found with 9- and 27-HMFT 
reflects the admixture of these nontrivial excited states when using clusters with an odd number of sites.

Moreover, from an energetics viewpoint, the CSF is competing with 
several VBC phases (the ones with odd fillings are doubly-degenerate due to particle-hole symmetry), 
indicating that it may eventually be replaced by a degenerate manifold of phases in the 
thermodynamic limit. In particular, at the KHAF point, the energies per site --i.e. the expectation 
value of (\ref{hardcore_ham}) for $t/V=1/2$ and $\mu/V=2$-- obtained with 9-, 18- and 27-HMFT and various cutoffs 
(27-HMFT) are 
\begin{eqnarray}
E_{9}&=&-0.4006,~\text{CSF},\notag\\
E_{27}&=&-0.4112,~\text{$\pi$VBC$_{5/9}$ and $\pi$VBC$_{4/9}$},\notag\\
E_{18}&=&-0.4135,~\text{VBC$_{1/2}$},\notag\\
E_{27}&=&-0.4148,~\text{CSF}~(13\leq N_\square\leq 14),\notag\\
E_{27}&=&-0.4169,~\text{VBC$_{13/27}$ and VBC$_{14/27}$},\notag\\
E_{27}&=&-0.4175,~\text{CSF}~(12\leq N_\square\leq 15),\notag
\end{eqnarray}
in units of $V$. These values are consistent with those obtained by the computationally 
demanding tensor network techniques, such as the projected 
entangled simplex states (PESS), where it was found an 
energy $E_{\text{PESS}}=-0.4364(1)$ for a 9-sites cell with bond dimension $D=13$, and a translationally 
invariant phase with no broken symmetry~\cite{Xie2014}.

\begin{table}[!tb]
\begin{center}
\begin{tabular}{|c|c|c|}
\hline
~~$\rho$~~ & near Ising & AFM XY \\
\hline
7/9 & 
$\begin{array}{c}
\includegraphics[width=0.19\textwidth]{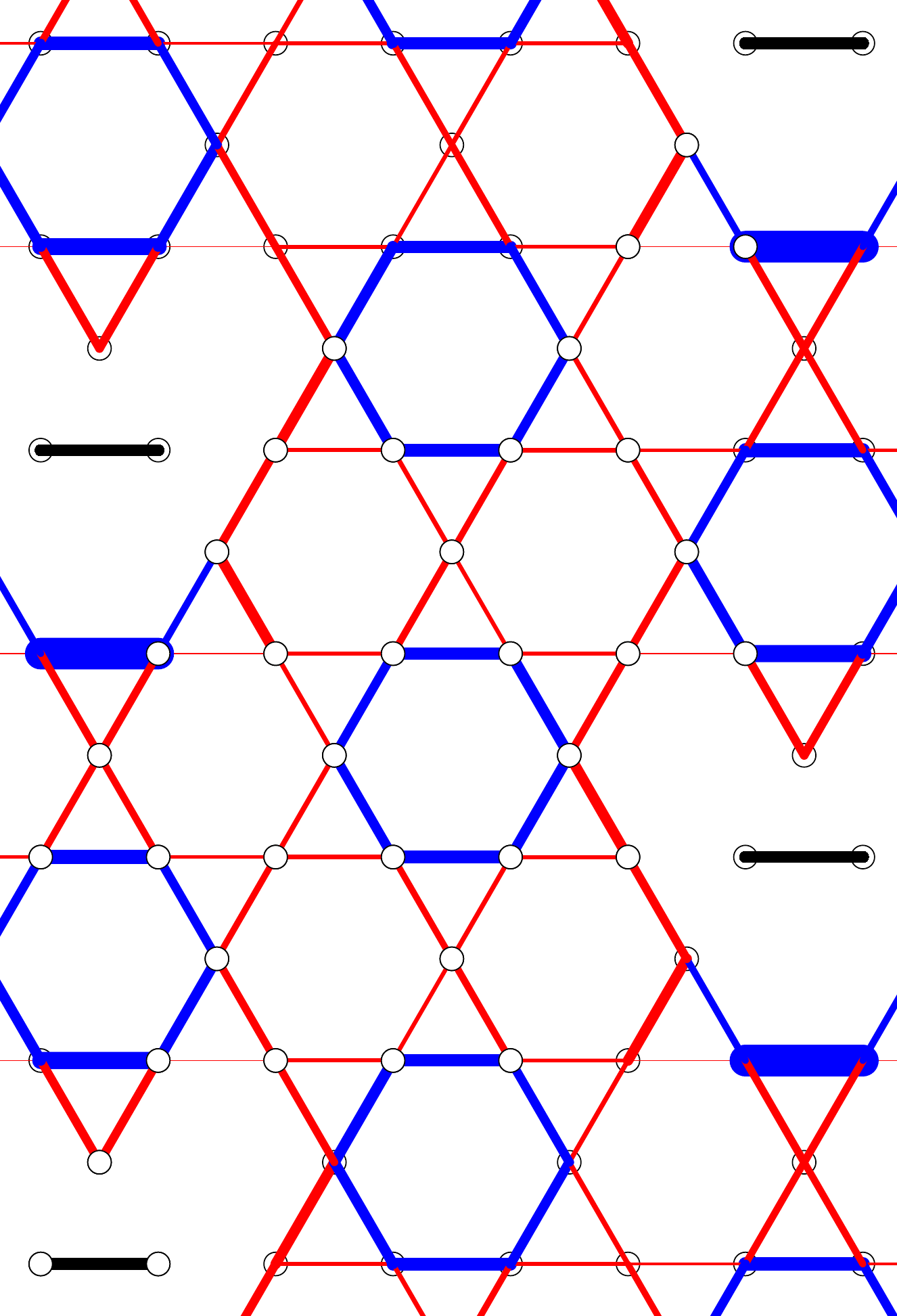}
\end{array}$& 
$\begin{array}{c}
\includegraphics[width=0.19\textwidth]{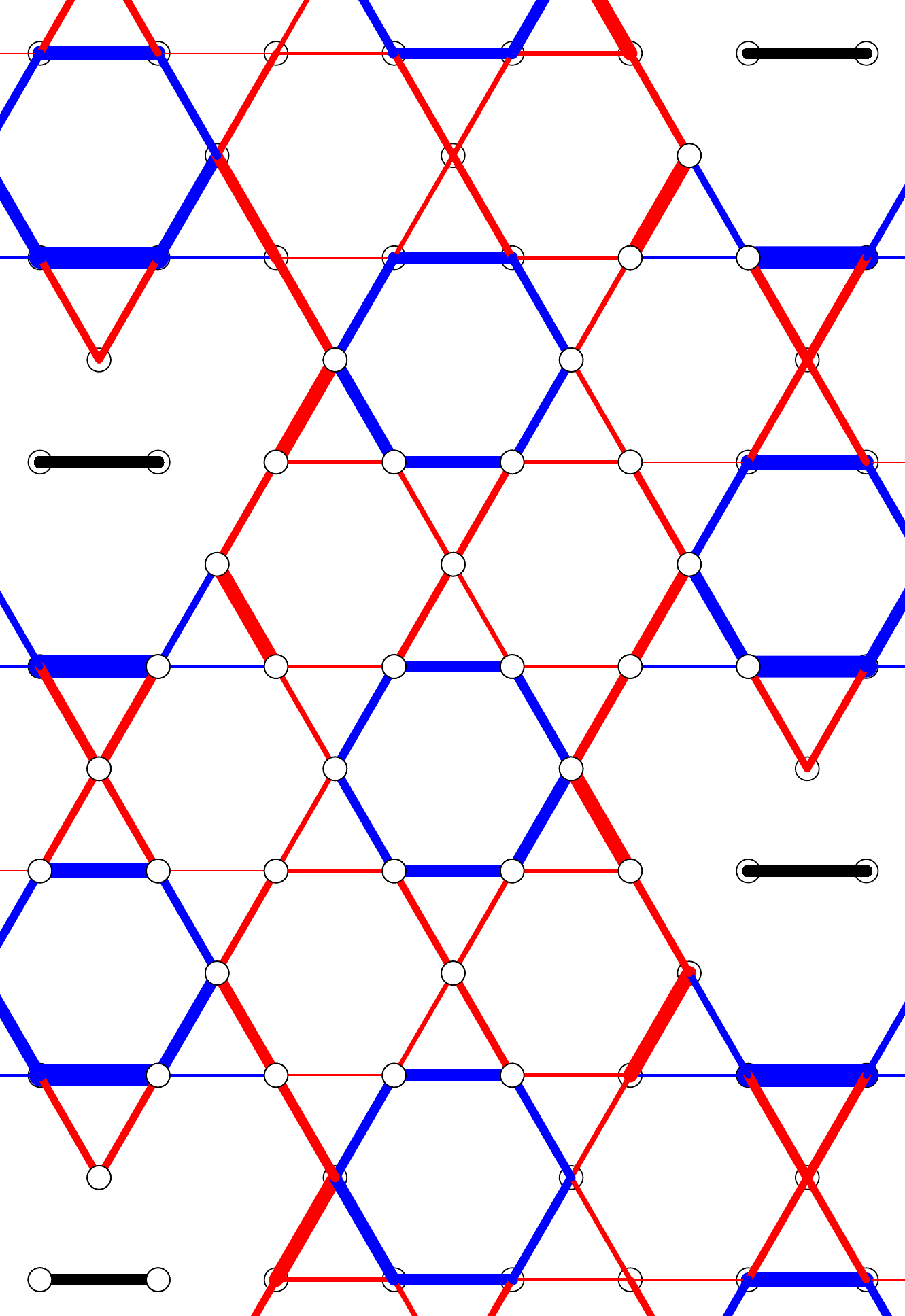}
\end{array}$
\\
\hline
2/3 &
$\begin{array}{c}
\includegraphics[width=0.19\textwidth]{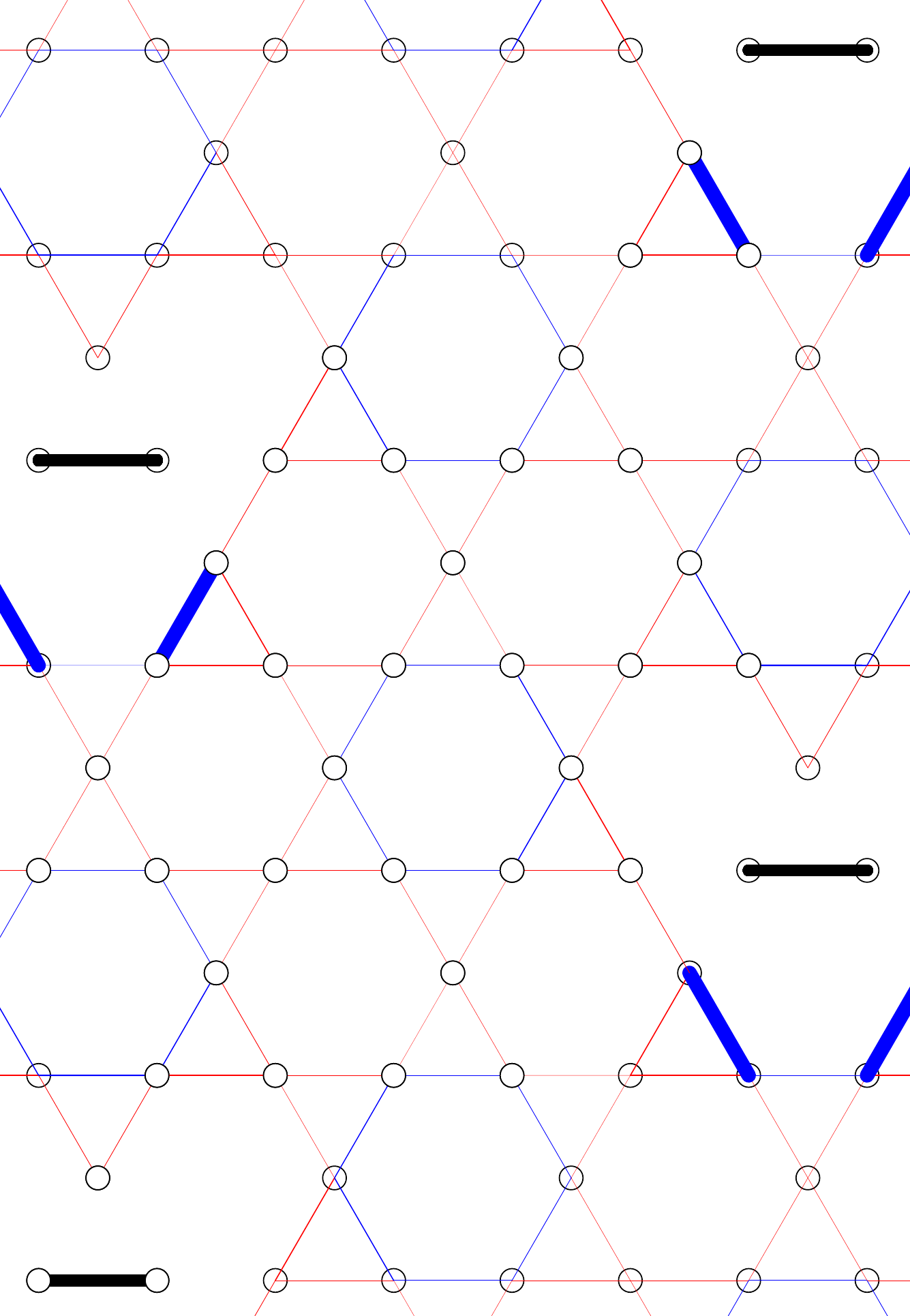}
\end{array}$& 
$\begin{array}{c}
\includegraphics[width=0.19\textwidth]{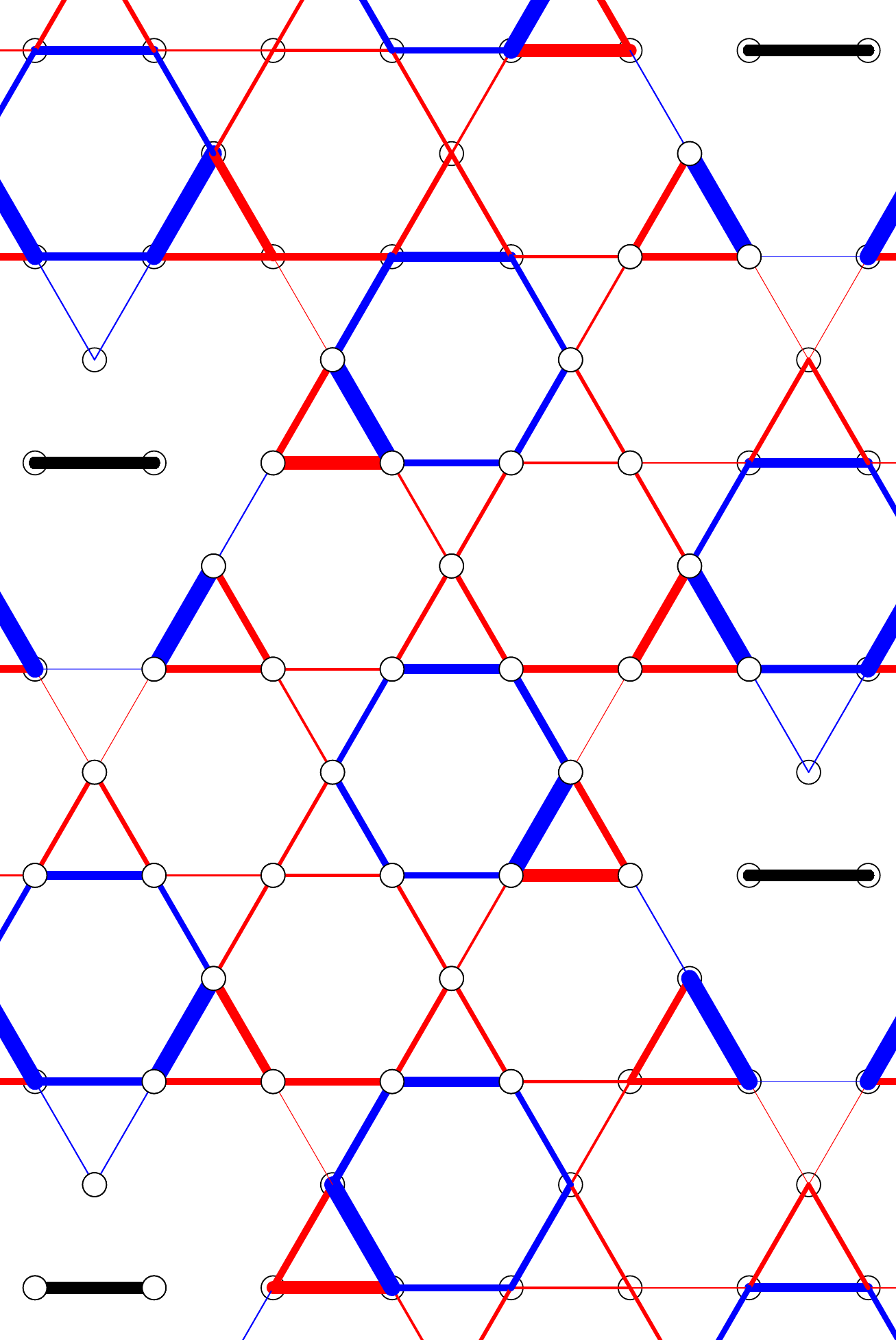}
\end{array}$
\\
\hline
\end{tabular}
\end{center}
\caption{\label{tab:ED_bond}(Color online) Kinetic bond-bond correlations (\ref{eq:dimerdimer}) 
computed by ED on a $N=36$ cluster at $t=0.1V$ (near Ising), and the 
AFM XY limit. Positive (negative) values are shown with blue (red) lines and their width is 
proportional to the data, the reference bond being shown in black.}
\end{table}

\begin{figure*}[ht]
\centering
\includegraphics[width=0.95\textwidth]{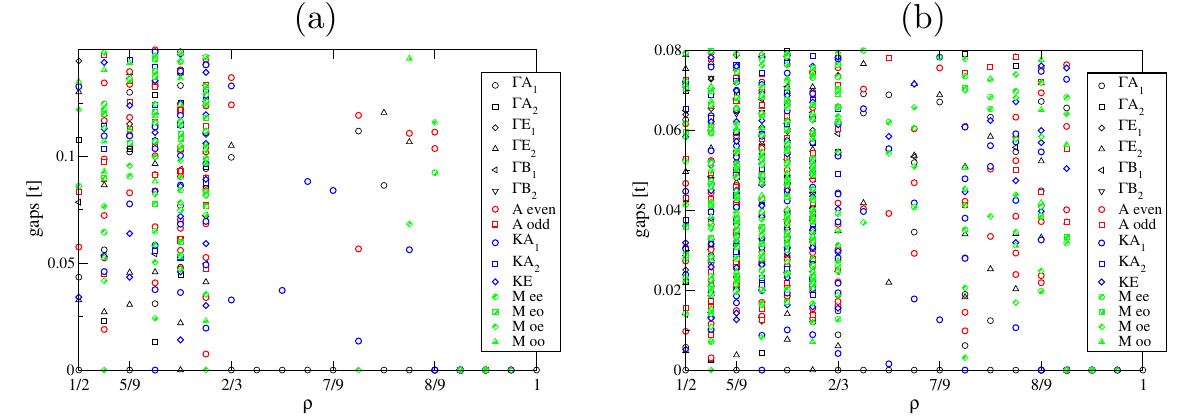}
\caption{\label{fig:ED_ToS}(Color online) Tower of states representing excitation energies 
(in units of $2t$) as a function of density $\rho$, obtained by ED on a $N=36$ cluster. States are 
labelled with respect to their quantum numbers associated to translation and point group 
symmetry (see Ref.~\onlinecite{Capponi2013} for more details): (a) $t=0.1V$ (near Ising regime), 
and (b) $V=0$ (AFM XY).}
\end{figure*}

\subsubsection{Exact Diagonalization approach}\label{sec:ED}
Using ground state energies for fixed densities and different cluster sizes and shapes, 
i.e. 27-, 36-, 36c- and 45-ED (see Fig. \ref{fig:clusters_ED}), we have computed the density 
as a function of the chemical potential using a Legendre transform for different values of the 
hopping amplitude. In particular, the widths of the main plateaux for $t/V=0.1,0.5,1.0$ and $1.5$ 
computed with a 36-ED cluster in the AFM regime are included in the general phase diagram 
(Fig.~\ref{fig:9site_diag}). Results on the density staircases as a function of the chemical 
potential are plotted in Fig.~\ref{densXY}(c) and (d) for both the FM and AFM XY regimes, 
respectively. Due to the low number of cluster sizes containing the exact VBC$_{8/9}$ 
available with current computational capabilities, performing an appropriate finite-size 
scaling analysis is difficult, thus we restrict ourselves to the analysis of the main plateaux 
$\rho=5/9,2/3,7/9,8/9$.

As a first attempt to characterizing the phase diagram, and specially to assess whether an 
adiabatic continuity from the non-interacting $V=0$ (XY) regime to the 
classical limit $t=0$ (Ising) exists, we computed bond-bond correlators (\ref{eq:dimerdimer}), shown 
in Table \ref{tab:ED_bond}, as well as the excitation energy spectrum as a function of the 
density in both regimes, shown in Fig.~\ref{fig:ED_ToS}.

\paragraph{One-hole resonant state (VBC$_{8/9}$).---}
At $\rho=8/9$, both in the Ising and XY regimes, there exists an exact three-fold degeneracy 
due to the existence of an exact localized magnon eigenstate~\cite{Schulenburg2002} for any 
regime of the interaction $V$. Due to the particular shape of the 36-ED cluster, small loops go around the cluster, 
and we find a greater degeneracy. However, on larger clusters (e.g. $63$-ED), 
the exact three-fold degeneracy with expected quantum numbers is recovered~\cite{Capponi2013}. 
From general considerations, the VBC$_{8/9}$ should possess gapped excitations and hence correspond to 
an extended region in the phase diagram.

\paragraph{Two-hole resonant state ($\pi$VBC$_{7/9}$).---}
At $\rho=7/9$, the spectra both at the Ising and XY regimes exhibit a two-fold degenerate state at 
momentum $K$A$_1$ very close to the ground state ($\Gamma$A$_1$), and then a rather large gap 
above them. Such data suggests that this first excited state may probably collapse to the ground state 
in the thermodynamic limit, stabilizing a three-fold degenerate VBC equivalent to the one found 
with HMFT at this density (see Table~\ref{tab:bond_op}). Further computations of the kinetic 
energy correlator (\ref{eq:dimerdimer}), shown in Table \ref{tab:ED_bond}, support this result. 

Moreover, the $\pi$VBC$_{7/9}$ found in this limit is equivalent to the one found at the Heisenberg limit in previous 
studies~\cite{Capponi2013,Nishimoto2013}, suggesting that it is stable in the whole AFM regime. 
On the contrary, we find no evidence of any additional degeneracy that would signal either an additional 
symmetry breaking, as proposed in Ref.~\onlinecite{Picot2016}, or the stabilization of a translational 
invariant gapped phase with topological order, as proposed in Ref.~\onlinecite{Kumar2014}. 

\paragraph{The $\pi$VBC$_{2/3}$ state.---}
At $\rho=2/3$ and in the Ising limit, there are clear signatures of the classical degeneracy 
expected for the $\pi$VBC$_{2/3}$ state. The first excited state over the ground state 
(at $\Gamma$A$_1$) is at momentum $K$A$_1$ (two-fold degenerate) while the next excited 
state is far in energy. Again, this suggests that this first excited state may collapse onto 
the ground state in the thermodynamic limit, giving a three-fold broken translational 
symmetry state, in agreement with the results obtained with HMFT.

However, in the XY regime, there is no clear separation of states in the low-energy spectrum. 
This may signal the stabilization in the thermodynamic limit of either a VBC state with larger 
unit cell and larger degeneracy (in case the first excited state $K$A$_1$ collapsed onto the 
ground state), or an eventual translational invariant gapped phase with manifold degeneracy 
(in case the $\Gamma$E$_2$ would collapse 
but not the $K$A$_1$), or even the breakdown of global U(1) (gapless phase) by BEC at some 
$\mathbf{k}$-point not commensurate with the clusters utilized in HMFT. 
Note, nevertheless, that the width of the plateau at this density increases when increasing the 
cluster size from 36 to 45 (see Fig. \ref{densXY} (d)).

\begin{figure}[b]
\centering
\includegraphics[width=0.45\textwidth]{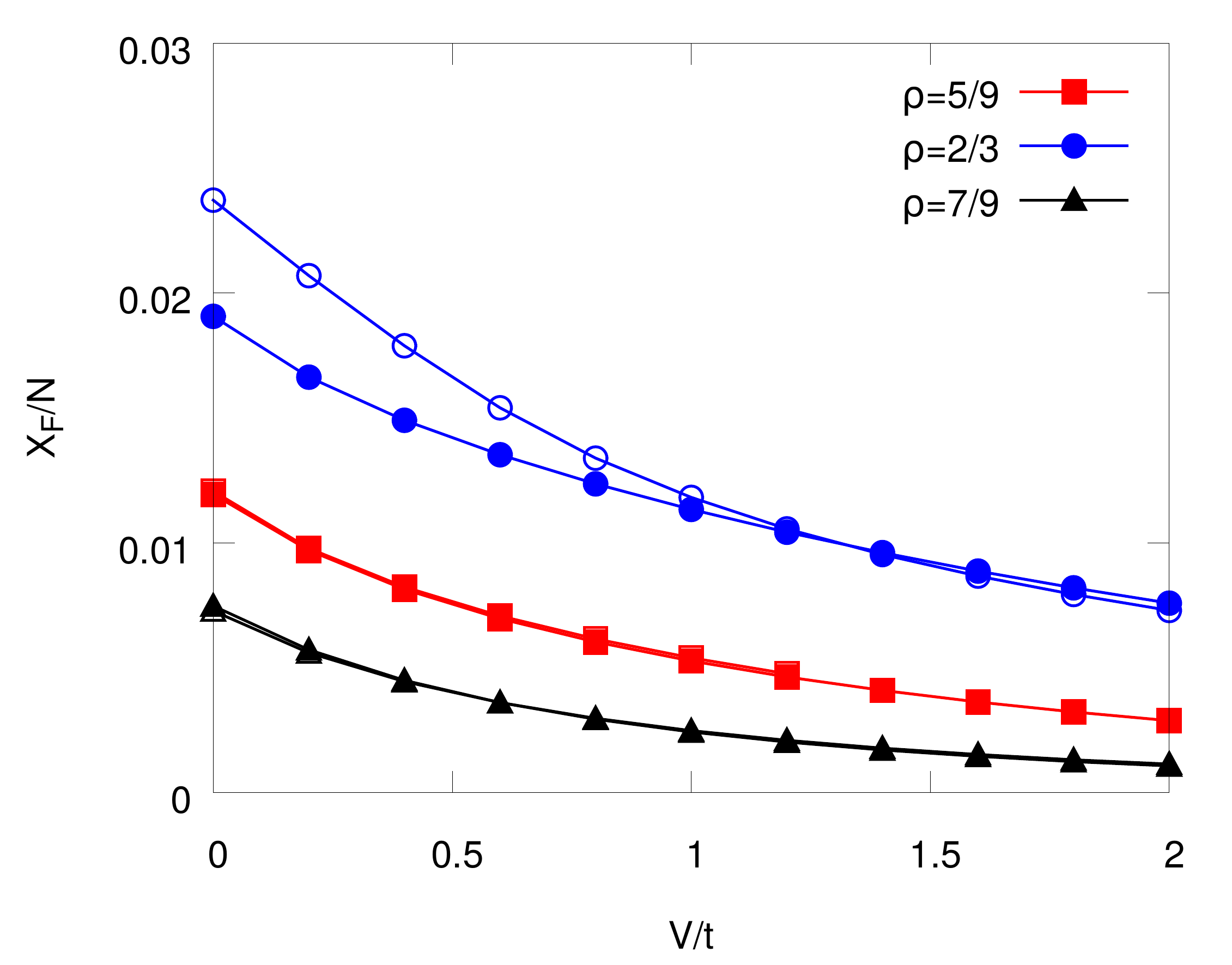}
\caption{\label{fig:ED_fid}(Color online) Fidelity susceptibility per site $\chi_F/N$ (\ref{eq:fidelity}) as a 
function of $V/t$ from the XY regime to the SU(2) point, computed by 27-ED (filled) and 36-ED (empty) 
for various densities. For $\rho=5/9~\text{and}~7/9$ filled and empty symbols are superimposed.}
\end{figure}
To further discuss the existence of an adiabatic continuity connecting the Ising, 
Heisenberg and XY regimes of the main plateaux in the AFM regime of the phase diagram, 
we show in Fig.~\ref{fig:ED_fid} the fidelity susceptibility $\chi_F$ per site defined in Eq. (\ref{eq:fidelity}). 
The size independence for $\rho=7/9$ is in agreement with the stability of the $\pi$VBC$_{7/9}$ 
over all regimes of interaction as found with HMFT. However, for $\rho=2/3$, the increase of $\chi_F$ with system 
size may signal a quantum phase transition in the vicinity of the AFM XY limit where, coincidently, the VBC 
pattern obtained with HMFT is no more of the localized resonant-magnon type. Larger clusters are 
required to reach a definite conclusion.

Similarly, in Table~\ref{tab:ED_bond} we show kinetic bond-bond correlations 
(\ref{eq:dimerdimer}) for $\rho=7/9$ and $\rho=2/3$ in both the AFM XY and Ising limits. 
Although the numerical data shows different amplitudes and short-range features, the 
overall pattern is compatible with the $\pi$VBC$_{7/9}$ and $\pi$VBC$_{2/3}$, respectively, 
known to be the ground states in the Ising~\cite{Cabra2005} and 
Heisenberg~\cite{Capponi2013,Nishimoto2013} limits.

In summary, it was claimed in previous ED studies~\cite{Cabra2005} that the 
ground state at $\rho=2/3$ is qualitatively similar for any $V>0$, 
i.e. of the same VBC nature as found in the Ising limit. From our ED data, we cannot 
conclusively establish the exact nature of the $\rho=2/3$ plateau in the XY limit, leaving 
the question of whether it is a more complex-pattern VBC (see Table~\ref{tab:bond_op}) 
or a gapped topological state~\cite{Kumar2014} for future studies.

\paragraph{The $\pi$VBC$_{5/9}$ state.---}
From the fidelity measurement (Fig.~\ref{fig:ED_fid}), we do not find any sign of a quantum 
phase transition taking place within the $\pi$VBC$_{5/9}$ plateau when varying the 
density-density interaction strength $V$. From our results, we conclude that the 
ground state  in the XY limit should be similar to the one encountered at the Ising and 
Heisenberg limits. 

Regarding the ED 
low-energy spectrum in the XY limit at this density (see Fig.~\ref{fig:ED_ToS}), we identify 
a two-fold energy state at the $\Gamma$ point ($\Gamma$E$_2$) very close to the ground state, 
and a tower of close states well above. This may result in an eventual collapse to the 
ground state in the thermodynamic limit, stabilizing a translational invariant three-fold 
degenerate ground state, which would be compatible with either a topological 
state~\cite{Nishimoto2013} or a peculiar VBC that would only break point-group 
symmetry but not translations.

Interestingly, in the Heisenberg limit, this $\pi$VBC$_{5/9}$ phase 
has been claimed to be a three-fold degenerate topological state, based on DMRG 
studies~\cite{Nishimoto2013}, and to be an eighteen-fold degenerate VBC, based 
on iPEPS computations with a 9-sites cell~\cite{Picot2016}. Our HMFT results 
also predicts a VBC with the same degree of degeneracy as the one of Ref.~\onlinecite{Picot2016}, 
but with a different bond pattern (see Table~\ref{tab:bond_op}). 

\section{SUMMARY and CONCLUSIONs\label{sec:conclusion}}
%
We have determined the quantum phase diagram of a system of repulsively interacting hard-core bosons 
on the Kagome lattice  by means of the hierarchical mean field theory (HMFT) and exact 
diagonalization (ED) techniques. The system is isomorphic to the spin S=1/2 XXZ model 
on the Kagome lattice in presence of an external magnetic field, a paradigmatic example of 
frustrated magnetism, and relevant to unveil magnetic properties of  Mott insulating materials. We have 
studied the non-frustrated (ferromagnetic),  and frustrated (antiferromagnetic) regimes, ranging from 
the purely repulsive (Ising) to the non-interacting (XY) limits. 

In the non-frustrated regime we found two valence bond crystal (VBC$_\rho$) lobes at 
densities $\rho=1/3$ and $\rho=2/3$ that melt into a standard superfluid characterized by the 
onset of Bose-Einstein condensation at momentum $\mathbf{k}=0$, in semi-quantitative agreement 
with previous quantum Monte Carlo (QMC) results.

In the frustrated regime, where QMC computations are impeded 
by the sign-problem, and away from half-filling, we have found a series of wide plateaux 
with VBC$_\rho$ order at densities $\rho$ commensurate with 1/9, in 
agreement with previous numerical studies carried out on the Heisenberg line  
(DMRG~\cite{Nishimoto2013}, ED~\cite{Capponi2013}, iPEPS~\cite{Picot2016}).  
These \textit{main plateaux} are separated by narrower plateaux at 
fillings commensurate  with 1/27, when using clusters with 27 sites. The physical 
mechanism responsible for the density (magnetization) plateaux is similar to the 
one found for the, in principle unrelated,  Shastry-Sutherland 
model~\cite{Shastry1981,Isaev2009SS}.   Moreover, near 
half-filling this {\it devil staircase} of crystal phases melts into a superfluid phase 
characterized by a Bose-Einstein condensate (BEC). Interestingly, we discovered a peculiar
``odd-even effect'': The superfluid is chiral (non-chiral) whenever the size of the 
simulated cluster is odd (even). Note, however, that several VBC$_\rho$ phases compete 
in energy near this half-filling region. The devil staircase physics and its melting mechanism 
prevail essentially up to the antiferromagnetic XY limit. This shows that 
correlations are adiabatically continued within a quantum phase as the 
hopping (or transverse exchange interaction in the spin language)  changes from the 
Ising to the XY limits.  

In particular, the plateaux at $\rho=8/9$ and $\rho=7/9$ (VBC$_{8/9}$ and $\pi$VBC$_{7/9}$, 
respectively) are characterized by a fully stacked pattern of
localized resonant magnons in all regimes.
The plateau at $\rho=2/3$ $(\pi$VBC$_{2/3})$ is also characterized by three localized resonant 
magnons, except for the antiferromagnetic XY limit,
where we found a more complex VBC pattern. 
Based on the analysis of the low-lying ED energy spectrum of 36-sites 
clusters, and the fidelity susceptibility for this density,
we cannot exclude that the XY limit may also be described by a gapped translational invariant 
phase in the thermodynamic limit, or even a gapless BEC.
Similarly, 
the plateau at $\rho=5/9$ is found to be a complex VBC$_\rho$ ($\pi$VBC$_{5/9}$) that changes 
smoothly from the Ising and Heisenberg regimes to the antiferromagnetic XY limit.
Further analysis on the low-lying ED energy spectrum of 36-sites clusters reveals the possibility 
of stabilization, in the thermodynamic limit, of a three-fold degenerate gapped translational 
invariant state, which would not be inconsistent with a proposed topological order~\cite{Nishimoto2013}.

We note that, based on the 36-ED tower of states (Fig. \ref{fig:ED_ToS}) and the HMFT analysis, 
one can clearly distinguish two regimes with respect to the density of low-lying excited states. For $\rho<2/3$
the density of low-lying excited states is much higher than for $\rho>2/3$.
This, together with the fact that the lower-$\mu$ $\pi$VBC$_{2/3}$ phase boundary computed with 
27-HMFT and with 36-ED coincide (see Fig. \ref{fig:9site_diag}), suggest that, in the 
thermodynamic limit, for densities 
$\rho\le2/3$ the staircase may be comprised of an infinite series of gapped phases.
It is clear that quantum fluctuations become more relevant as we move towards the half-filled and non-interacting limits of the quantum phase diagram.

Note Added: After submission of the current manuscript we noted the work of Ref.~\onlinecite{Kshetrimayum2016} that studied
the same model and finds an $M$=1/3 plateau in the entire frustrated regime. 

\begin{acknowledgements}
DH gratefully acknowledge the computing time granted on the supercomputer JURECA 
at J\"ulich Supercomputing Centre (JSC). DH and JD acknowledge support from the 
Spanish Ministry of Economy and Competitiveness through Grants FIS2012-34479 and 
FIS2015-63770-P (MINECO/FEDER). SC would like to acknowledge HPC resources from 
GENCI (grants x2015050225 and x2016050225) and CALMIP (grants 2015-P0677 and 
2016-P0677). The QMC SSE simulations were performed using the code from the 
ALPS libraries~\cite{ALPS2}.
\end{acknowledgements}

\bibliography{Kagome_crystal.bib}

\begin{thebibliography}{57}%
\makeatletter
\providecommand \@ifxundefined [1]{%
 \@ifx{#1\undefined}
}%
\providecommand \@ifnum [1]{%
 \ifnum #1\expandafter \@firstoftwo
 \else \expandafter \@secondoftwo
 \fi
}%
\providecommand \@ifx [1]{%
 \ifx #1\expandafter \@firstoftwo
 \else \expandafter \@secondoftwo
 \fi
}%
\providecommand \natexlab [1]{#1}%
\providecommand \enquote  [1]{``#1''}%
\providecommand \bibnamefont  [1]{#1}%
\providecommand \bibfnamefont [1]{#1}%
\providecommand \citenamefont [1]{#1}%
\providecommand \href@noop [0]{\@secondoftwo}%
\providecommand \href [0]{\begingroup \@sanitize@url \@href}%
\providecommand \@href[1]{\@@startlink{#1}\@@href}%
\providecommand \@@href[1]{\endgroup#1\@@endlink}%
\providecommand \@sanitize@url [0]{\catcode `\\12\catcode `\$12\catcode
  `\&12\catcode `\#12\catcode `\^12\catcode `\_12\catcode `\%12\relax}%
\providecommand \@@startlink[1]{}%
\providecommand \@@endlink[0]{}%
\providecommand \url  [0]{\begingroup\@sanitize@url \@url }%
\providecommand \@url [1]{\endgroup\@href {#1}{\urlprefix }}%
\providecommand \urlprefix  [0]{URL }%
\providecommand \Eprint [0]{\href }%
\providecommand \doibase [0]{http://dx.doi.org/}%
\providecommand \selectlanguage [0]{\@gobble}%
\providecommand \bibinfo  [0]{\@secondoftwo}%
\providecommand \bibfield  [0]{\@secondoftwo}%
\providecommand \translation [1]{[#1]}%
\providecommand \BibitemOpen [0]{}%
\providecommand \bibitemStop [0]{}%
\providecommand \bibitemNoStop [0]{.\EOS\space}%
\providecommand \EOS [0]{\spacefactor3000\relax}%
\providecommand \BibitemShut  [1]{\csname bibitem#1\endcsname}%
\let\auto@bib@innerbib\@empty
\bibitem [{\citenamefont {Lacroix}\ \emph {et~al.}(2011)\citenamefont
  {Lacroix}, \citenamefont {Mendels},\ and\ \citenamefont {Mila}}]{Lacroix}%
  \BibitemOpen
  \bibfield  {author} {\bibinfo {author} {\bibfnamefont {C.}~\bibnamefont
  {Lacroix}}, \bibinfo {author} {\bibfnamefont {P.}~\bibnamefont {Mendels}}, \
  and\ \bibinfo {author} {\bibfnamefont {F.}~\bibnamefont {Mila}},\ }\href@noop
  {} {\emph {\bibinfo {title} {{Introduction to Frustrated Magnetism:
  Materials, Experiments, Theory}}}}\ (\bibinfo  {publisher} {Springer
  Verlag},\ \bibinfo {year} {2011})\BibitemShut {NoStop}%
\bibitem [{\citenamefont {Takigawa}\ \emph {et~al.}(2013)\citenamefont
  {Takigawa}, \citenamefont {{Horvati\ifmmode \acute{c}\else {\'c}\fi{}}},
  \citenamefont {Waki}, \citenamefont {Kr{\"a}mer}, \citenamefont {Berthier},
  \citenamefont {L{\'e}vy-Bertrand}, \citenamefont {Sheikin}, \citenamefont
  {Kageyama}, \citenamefont {Ueda},\ and\ \citenamefont {Mila}}]{Takigawa2013}%
  \BibitemOpen
  \bibfield  {author} {\bibinfo {author} {\bibfnamefont {M.}~\bibnamefont
  {Takigawa}}, \bibinfo {author} {\bibfnamefont {M.}~\bibnamefont
  {{Horvati\ifmmode \acute{c}\else {\'c}\fi{}}}}, \bibinfo {author}
  {\bibfnamefont {T.}~\bibnamefont {Waki}}, \bibinfo {author} {\bibfnamefont
  {S.}~\bibnamefont {Kr{\"a}mer}}, \bibinfo {author} {\bibfnamefont
  {C.}~\bibnamefont {Berthier}}, \bibinfo {author} {\bibfnamefont
  {F.}~\bibnamefont {L{\'e}vy-Bertrand}}, \bibinfo {author} {\bibfnamefont
  {I.}~\bibnamefont {Sheikin}}, \bibinfo {author} {\bibfnamefont
  {H.}~\bibnamefont {Kageyama}}, \bibinfo {author} {\bibfnamefont
  {Y.}~\bibnamefont {Ueda}}, \ and\ \bibinfo {author} {\bibfnamefont
  {F.}~\bibnamefont {Mila}},\ }\href {\doibase 10.1103/PhysRevLett.110.067210}
  {\bibfield  {journal} {\bibinfo  {journal} {Phys. Rev. Lett.}\ }\textbf
  {\bibinfo {volume} {110}},\ \bibinfo {pages} {067210} (\bibinfo {year}
  {2013})}\BibitemShut {NoStop}%
\bibitem [{\citenamefont {Nishimori}\ and\ \citenamefont
  {Ortiz}(2011)}]{Nishimori-Ortiz-2011}%
  \BibitemOpen
  \bibfield  {author} {\bibinfo {author} {\bibfnamefont {H.}~\bibnamefont
  {Nishimori}}\ and\ \bibinfo {author} {\bibfnamefont {G.}~\bibnamefont
  {Ortiz}},\ }\href@noop {} {\emph {\bibinfo {title} {{Elements of Phase
  Transitions and Critical Phenomena}}}}\ (\bibinfo  {publisher} {Oxford
  University Press},\ \bibinfo {year} {2011})\BibitemShut {NoStop}%
\bibitem [{\citenamefont {Roger}\ \emph {et~al.}(1983)\citenamefont {Roger},
  \citenamefont {Hetherington},\ and\ \citenamefont {Delrieu}}]{Roger1983}%
  \BibitemOpen
  \bibfield  {author} {\bibinfo {author} {\bibfnamefont {M.}~\bibnamefont
  {Roger}}, \bibinfo {author} {\bibfnamefont {J.~H.}\ \bibnamefont
  {Hetherington}}, \ and\ \bibinfo {author} {\bibfnamefont {J.~M.}\
  \bibnamefont {Delrieu}},\ }\href {\doibase 10.1103/RevModPhys.55.1}
  {\bibfield  {journal} {\bibinfo  {journal} {Rev. Mod. Phys.}\ }\textbf
  {\bibinfo {volume} {55}},\ \bibinfo {pages} {1} (\bibinfo {year}
  {1983})}\BibitemShut {NoStop}%
\bibitem [{\citenamefont {Bak}(1982)}]{Bak1982}%
  \BibitemOpen
  \bibfield  {author} {\bibinfo {author} {\bibfnamefont {P.}~\bibnamefont
  {Bak}},\ }\href {http://stacks.iop.org/0034-4885/45/i=6/a=001} {\bibfield
  {journal} {\bibinfo  {journal} {Rep. Prog. Phys.}\ }\textbf {\bibinfo
  {volume} {45}},\ \bibinfo {pages} {587} (\bibinfo {year} {1982})}\BibitemShut
  {NoStop}%
\bibitem [{\citenamefont {Capponi}\ \emph
  {et~al.}(2013{\natexlab{a}})\citenamefont {Capponi}, \citenamefont {Derzhko},
  \citenamefont {Honecker}, \citenamefont {L{\"a}uchli},\ and\ \citenamefont
  {Richter}}]{Capponi2013}%
  \BibitemOpen
  \bibfield  {author} {\bibinfo {author} {\bibfnamefont {S.}~\bibnamefont
  {Capponi}}, \bibinfo {author} {\bibfnamefont {O.}~\bibnamefont {Derzhko}},
  \bibinfo {author} {\bibfnamefont {A.}~\bibnamefont {Honecker}}, \bibinfo
  {author} {\bibfnamefont {A.~M.}\ \bibnamefont {L{\"a}uchli}}, \ and\ \bibinfo
  {author} {\bibfnamefont {J.}~\bibnamefont {Richter}},\ }\href {\doibase
  10.1103/PhysRevB.88.144416} {\bibfield  {journal} {\bibinfo  {journal} {Phys.
  Rev. B}\ }\textbf {\bibinfo {volume} {88}},\ \bibinfo {pages} {144416}
  (\bibinfo {year} {2013}{\natexlab{a}})}\BibitemShut {NoStop}%
\bibitem [{\citenamefont {Nishimoto}\ \emph {et~al.}(2013)\citenamefont
  {Nishimoto}, \citenamefont {Shibata},\ and\ \citenamefont
  {Hotta}}]{Nishimoto2013}%
  \BibitemOpen
  \bibfield  {author} {\bibinfo {author} {\bibfnamefont {S.}~\bibnamefont
  {Nishimoto}}, \bibinfo {author} {\bibfnamefont {N.}~\bibnamefont {Shibata}},
  \ and\ \bibinfo {author} {\bibfnamefont {C.}~\bibnamefont {Hotta}},\ }\href
  {http://dx.doi.org/10.1038/ncomms3287} {\bibfield  {journal} {\bibinfo
  {journal} {Nat. Commun.}\ }\textbf {\bibinfo {volume} {4}},\ \bibinfo {pages}
  {3287} (\bibinfo {year} {2013})}\BibitemShut {NoStop}%
\bibitem [{\citenamefont {Picot}\ \emph {et~al.}(2016)\citenamefont {Picot},
  \citenamefont {Ziegler}, \citenamefont {Or{\'u}s},\ and\ \citenamefont
  {Poilblanc}}]{Picot2016}%
  \BibitemOpen
  \bibfield  {author} {\bibinfo {author} {\bibfnamefont {T.}~\bibnamefont
  {Picot}}, \bibinfo {author} {\bibfnamefont {M.}~\bibnamefont {Ziegler}},
  \bibinfo {author} {\bibfnamefont {R.}~\bibnamefont {Or{\'u}s}}, \ and\
  \bibinfo {author} {\bibfnamefont {D.}~\bibnamefont {Poilblanc}},\ }\href
  {\doibase 10.1103/PhysRevB.93.060407} {\bibfield  {journal} {\bibinfo
  {journal} {Phys. Rev. B}\ }\textbf {\bibinfo {volume} {93}},\ \bibinfo
  {pages} {060407} (\bibinfo {year} {2016})}\BibitemShut {NoStop}%
\bibitem [{\citenamefont {Schulenburg}\ \emph {et~al.}(2002)\citenamefont
  {Schulenburg}, \citenamefont {Honecker}, \citenamefont {Schnack},
  \citenamefont {Richter},\ and\ \citenamefont {Schmidt}}]{Schulenburg2002}%
  \BibitemOpen
  \bibfield  {author} {\bibinfo {author} {\bibfnamefont {J.}~\bibnamefont
  {Schulenburg}}, \bibinfo {author} {\bibfnamefont {A.}~\bibnamefont
  {Honecker}}, \bibinfo {author} {\bibfnamefont {J.}~\bibnamefont {Schnack}},
  \bibinfo {author} {\bibfnamefont {J.}~\bibnamefont {Richter}}, \ and\
  \bibinfo {author} {\bibfnamefont {H.-J.}\ \bibnamefont {Schmidt}},\ }\href
  {\doibase 10.1103/PhysRevLett.88.167207} {\bibfield  {journal} {\bibinfo
  {journal} {Phys. Rev. Lett.}\ }\textbf {\bibinfo {volume} {88}},\ \bibinfo
  {pages} {167207} (\bibinfo {year} {2002})}\BibitemShut {NoStop}%
\bibitem [{\citenamefont {Shores}\ \emph {et~al.}(2005)\citenamefont {Shores},
  \citenamefont {Nytko}, \citenamefont {Bartlett},\ and\ \citenamefont
  {Nocera}}]{Shores2005}%
  \BibitemOpen
  \bibfield  {author} {\bibinfo {author} {\bibfnamefont {M.~P.}\ \bibnamefont
  {Shores}}, \bibinfo {author} {\bibfnamefont {E.~A.}\ \bibnamefont {Nytko}},
  \bibinfo {author} {\bibfnamefont {B.~M.}\ \bibnamefont {Bartlett}}, \ and\
  \bibinfo {author} {\bibfnamefont {D.~G.}\ \bibnamefont {Nocera}},\ }\href
  {\doibase 10.1021/ja053891p} {\bibfield  {journal} {\bibinfo  {journal} {J.
  Am. Chem. Soc.}\ }\textbf {\bibinfo {volume} {127}},\ \bibinfo {pages}
  {13462} (\bibinfo {year} {2005})}\BibitemShut {NoStop}%
\bibitem [{\citenamefont {Helton}\ \emph {et~al.}(2007)\citenamefont {Helton},
  \citenamefont {Matan}, \citenamefont {Shores}, \citenamefont {Nytko},
  \citenamefont {Bartlett}, \citenamefont {Yoshida}, \citenamefont {Takano},
  \citenamefont {Suslov}, \citenamefont {Qiu}, \citenamefont {Chung},
  \citenamefont {Nocera},\ and\ \citenamefont {Lee}}]{Helton2007}%
  \BibitemOpen
  \bibfield  {author} {\bibinfo {author} {\bibfnamefont {J.~S.}\ \bibnamefont
  {Helton}}, \bibinfo {author} {\bibfnamefont {K.}~\bibnamefont {Matan}},
  \bibinfo {author} {\bibfnamefont {M.~P.}\ \bibnamefont {Shores}}, \bibinfo
  {author} {\bibfnamefont {E.~A.}\ \bibnamefont {Nytko}}, \bibinfo {author}
  {\bibfnamefont {B.~M.}\ \bibnamefont {Bartlett}}, \bibinfo {author}
  {\bibfnamefont {Y.}~\bibnamefont {Yoshida}}, \bibinfo {author} {\bibfnamefont
  {Y.}~\bibnamefont {Takano}}, \bibinfo {author} {\bibfnamefont
  {A.}~\bibnamefont {Suslov}}, \bibinfo {author} {\bibfnamefont
  {Y.}~\bibnamefont {Qiu}}, \bibinfo {author} {\bibfnamefont {J.-H.}\
  \bibnamefont {Chung}}, \bibinfo {author} {\bibfnamefont {D.~G.}\ \bibnamefont
  {Nocera}}, \ and\ \bibinfo {author} {\bibfnamefont {Y.~S.}\ \bibnamefont
  {Lee}},\ }\href {\doibase 10.1103/PhysRevLett.98.107204} {\bibfield
  {journal} {\bibinfo  {journal} {Phys. Rev. Lett.}\ }\textbf {\bibinfo
  {volume} {98}},\ \bibinfo {pages} {107204} (\bibinfo {year}
  {2007})}\BibitemShut {NoStop}%
\bibitem [{\citenamefont {Mendels}\ \emph {et~al.}(2007)\citenamefont
  {Mendels}, \citenamefont {Bert}, \citenamefont {de~Vries}, \citenamefont
  {Olariu}, \citenamefont {Harrison}, \citenamefont {Duc}, \citenamefont
  {Trombe}, \citenamefont {Lord}, \citenamefont {Amato},\ and\ \citenamefont
  {Baines}}]{Mendels2007}%
  \BibitemOpen
  \bibfield  {author} {\bibinfo {author} {\bibfnamefont {P.}~\bibnamefont
  {Mendels}}, \bibinfo {author} {\bibfnamefont {F.}~\bibnamefont {Bert}},
  \bibinfo {author} {\bibfnamefont {M.~A.}\ \bibnamefont {de~Vries}}, \bibinfo
  {author} {\bibfnamefont {A.}~\bibnamefont {Olariu}}, \bibinfo {author}
  {\bibfnamefont {A.}~\bibnamefont {Harrison}}, \bibinfo {author}
  {\bibfnamefont {F.}~\bibnamefont {Duc}}, \bibinfo {author} {\bibfnamefont
  {J.~C.}\ \bibnamefont {Trombe}}, \bibinfo {author} {\bibfnamefont {J.~S.}\
  \bibnamefont {Lord}}, \bibinfo {author} {\bibfnamefont {A.}~\bibnamefont
  {Amato}}, \ and\ \bibinfo {author} {\bibfnamefont {C.}~\bibnamefont
  {Baines}},\ }\href {http://link.aps.org/abstract/PRL/v98/e077204} {\bibfield
  {journal} {\bibinfo  {journal} {Phys. Rev. Lett.}\ }\textbf {\bibinfo
  {volume} {98}},\ \bibinfo {pages} {077204} (\bibinfo {year}
  {2007})}\BibitemShut {NoStop}%
\bibitem [{\citenamefont {Fu}\ \emph {et~al.}(2015)\citenamefont {Fu},
  \citenamefont {Imai}, \citenamefont {Han},\ and\ \citenamefont
  {Lee}}]{Fu2015}%
  \BibitemOpen
  \bibfield  {author} {\bibinfo {author} {\bibfnamefont {M.}~\bibnamefont
  {Fu}}, \bibinfo {author} {\bibfnamefont {T.}~\bibnamefont {Imai}}, \bibinfo
  {author} {\bibfnamefont {T.-H.}\ \bibnamefont {Han}}, \ and\ \bibinfo
  {author} {\bibfnamefont {Y.~S.}\ \bibnamefont {Lee}},\ }\href {\doibase
  10.1126/science.aab2120} {\bibfield  {journal} {\bibinfo  {journal}
  {Science}\ }\textbf {\bibinfo {volume} {350}},\ \bibinfo {pages} {655}
  (\bibinfo {year} {2015})}\BibitemShut {NoStop}%
\bibitem [{\citenamefont {Jo}\ \emph {et~al.}(2012)\citenamefont {Jo},
  \citenamefont {Guzman}, \citenamefont {Thomas}, \citenamefont {Hosur},
  \citenamefont {Vishwanath},\ and\ \citenamefont {Stamper-Kurn}}]{Jo2012}%
  \BibitemOpen
  \bibfield  {author} {\bibinfo {author} {\bibfnamefont {G.-B.}\ \bibnamefont
  {Jo}}, \bibinfo {author} {\bibfnamefont {J.}~\bibnamefont {Guzman}}, \bibinfo
  {author} {\bibfnamefont {C.~K.}\ \bibnamefont {Thomas}}, \bibinfo {author}
  {\bibfnamefont {P.}~\bibnamefont {Hosur}}, \bibinfo {author} {\bibfnamefont
  {A.}~\bibnamefont {Vishwanath}}, \ and\ \bibinfo {author} {\bibfnamefont
  {D.~M.}\ \bibnamefont {Stamper-Kurn}},\ }\href {\doibase
  10.1103/PhysRevLett.108.045305} {\bibfield  {journal} {\bibinfo  {journal}
  {Phys. Rev. Lett.}\ }\textbf {\bibinfo {volume} {108}},\ \bibinfo {pages}
  {045305} (\bibinfo {year} {2012})}\BibitemShut {NoStop}%
\bibitem [{\citenamefont {Cabra}\ \emph {et~al.}(2005)\citenamefont {Cabra},
  \citenamefont {Grynberg}, \citenamefont {Holdsworth}, \citenamefont
  {Honecker}, \citenamefont {Pujol}, \citenamefont {Richter}, \citenamefont
  {Schmalfu{\ss}},\ and\ \citenamefont {Schulenburg}}]{Cabra2005}%
  \BibitemOpen
  \bibfield  {author} {\bibinfo {author} {\bibfnamefont {D.~C.}\ \bibnamefont
  {Cabra}}, \bibinfo {author} {\bibfnamefont {M.~D.}\ \bibnamefont {Grynberg}},
  \bibinfo {author} {\bibfnamefont {P.~C.~W.}\ \bibnamefont {Holdsworth}},
  \bibinfo {author} {\bibfnamefont {A.}~\bibnamefont {Honecker}}, \bibinfo
  {author} {\bibfnamefont {P.}~\bibnamefont {Pujol}}, \bibinfo {author}
  {\bibfnamefont {J.}~\bibnamefont {Richter}}, \bibinfo {author} {\bibfnamefont
  {D.}~\bibnamefont {Schmalfu{\ss}}}, \ and\ \bibinfo {author} {\bibfnamefont
  {J.}~\bibnamefont {Schulenburg}},\ }\href {\doibase
  10.1103/PhysRevB.71.144420} {\bibfield  {journal} {\bibinfo  {journal} {Phys.
  Rev. B}\ }\textbf {\bibinfo {volume} {71}},\ \bibinfo {pages} {144420}
  (\bibinfo {year} {2005})}\BibitemShut {NoStop}%
\bibitem [{\citenamefont {Isakov}\ \emph {et~al.}(2006)\citenamefont {Isakov},
  \citenamefont {Wessel}, \citenamefont {Melko}, \citenamefont {Sengupta},\
  and\ \citenamefont {Kim}}]{Isakov2006}%
  \BibitemOpen
  \bibfield  {author} {\bibinfo {author} {\bibfnamefont {S.~V.}\ \bibnamefont
  {Isakov}}, \bibinfo {author} {\bibfnamefont {S.}~\bibnamefont {Wessel}},
  \bibinfo {author} {\bibfnamefont {R.~G.}\ \bibnamefont {Melko}}, \bibinfo
  {author} {\bibfnamefont {K.}~\bibnamefont {Sengupta}}, \ and\ \bibinfo
  {author} {\bibfnamefont {Y.~B.}\ \bibnamefont {Kim}},\ }\href
  {http://link.aps.org/doi/10.1103/PhysRevLett.97.147202} {\bibfield  {journal}
  {\bibinfo  {journal} {Phys. Rev. Lett.}\ }\textbf {\bibinfo {volume} {97}},\
  \bibinfo {pages} {147202} (\bibinfo {year} {2006})}\BibitemShut {NoStop}%
\bibitem [{\citenamefont {Mambrini}\ and\ \citenamefont
  {Mila}(2000)}]{Mambrini2000}%
  \BibitemOpen
  \bibfield  {author} {\bibinfo {author} {\bibfnamefont {M.}~\bibnamefont
  {Mambrini}}\ and\ \bibinfo {author} {\bibfnamefont {F.}~\bibnamefont
  {Mila}},\ }\href {\doibase 10.1007/PL00011071} {\bibfield  {journal}
  {\bibinfo  {journal} {Eur. Phys. J. B}\ }\textbf {\bibinfo {volume} {17}},\
  \bibinfo {pages} {651} (\bibinfo {year} {2000})}\BibitemShut {NoStop}%
\bibitem [{\citenamefont {Poilblanc}\ \emph {et~al.}(2010)\citenamefont
  {Poilblanc}, \citenamefont {Mambrini},\ and\ \citenamefont
  {Schwandt}}]{Poilblanc2010}%
  \BibitemOpen
  \bibfield  {author} {\bibinfo {author} {\bibfnamefont {D.}~\bibnamefont
  {Poilblanc}}, \bibinfo {author} {\bibfnamefont {M.}~\bibnamefont {Mambrini}},
  \ and\ \bibinfo {author} {\bibfnamefont {D.}~\bibnamefont {Schwandt}},\
  }\href {\doibase 10.1103/PhysRevB.81.180402} {\bibfield  {journal} {\bibinfo
  {journal} {Phys. Rev. B}\ }\textbf {\bibinfo {volume} {81}},\ \bibinfo
  {pages} {180402} (\bibinfo {year} {2010})}\BibitemShut {NoStop}%
\bibitem [{\citenamefont {Waldtmann}\ \emph {et~al.}(1998)\citenamefont
  {Waldtmann}, \citenamefont {Everts}, \citenamefont {Bernu}, \citenamefont
  {Lhuillier}, \citenamefont {Sindzingre}, \citenamefont {Lecheminant},\ and\
  \citenamefont {Pierre}}]{Waldtmann1998}%
  \BibitemOpen
  \bibfield  {author} {\bibinfo {author} {\bibfnamefont {C.}~\bibnamefont
  {Waldtmann}}, \bibinfo {author} {\bibfnamefont {H.-U.}\ \bibnamefont
  {Everts}}, \bibinfo {author} {\bibfnamefont {B.}~\bibnamefont {Bernu}},
  \bibinfo {author} {\bibfnamefont {C.}~\bibnamefont {Lhuillier}}, \bibinfo
  {author} {\bibfnamefont {P.}~\bibnamefont {Sindzingre}}, \bibinfo {author}
  {\bibfnamefont {P.}~\bibnamefont {Lecheminant}}, \ and\ \bibinfo {author}
  {\bibfnamefont {L.}~\bibnamefont {Pierre}},\ }\href {\doibase
  10.1007/s100510050274} {\bibfield  {journal} {\bibinfo  {journal} {Eur. Phys.
  J. B}\ }\textbf {\bibinfo {volume} {2}},\ \bibinfo {pages} {501} (\bibinfo
  {year} {1998})}\BibitemShut {NoStop}%
\bibitem [{\citenamefont {Singh}\ and\ \citenamefont
  {Huse}(2007)}]{Singh-H-07}%
  \BibitemOpen
  \bibfield  {author} {\bibinfo {author} {\bibfnamefont {R.~R.~P.}\
  \bibnamefont {Singh}}\ and\ \bibinfo {author} {\bibfnamefont {D.~A.}\
  \bibnamefont {Huse}},\ }\href {http://link.aps.org/abstract/PRB/v76/e180407}
  {\bibfield  {journal} {\bibinfo  {journal} {Phys. Rev. B}\ }\textbf {\bibinfo
  {volume} {76}},\ \bibinfo {pages} {180407} (\bibinfo {year}
  {2007})}\BibitemShut {NoStop}%
\bibitem [{\citenamefont {Yan}\ \emph {et~al.}(2011)\citenamefont {Yan},
  \citenamefont {Huse},\ and\ \citenamefont {White}}]{Yan2011}%
  \BibitemOpen
  \bibfield  {author} {\bibinfo {author} {\bibfnamefont {S.}~\bibnamefont
  {Yan}}, \bibinfo {author} {\bibfnamefont {D.~A.}\ \bibnamefont {Huse}}, \
  and\ \bibinfo {author} {\bibfnamefont {S.~R.}\ \bibnamefont {White}},\ }\href
  {\doibase 10.1126/science.1201080} {\bibfield  {journal} {\bibinfo  {journal}
  {Science}\ }\textbf {\bibinfo {volume} {332}},\ \bibinfo {pages} {1173}
  (\bibinfo {year} {2011})}\BibitemShut {NoStop}%
\bibitem [{\citenamefont {Depenbrock}\ \emph {et~al.}(2012)\citenamefont
  {Depenbrock}, \citenamefont {McCulloch},\ and\ \citenamefont
  {Schollw{\"o}ck}}]{Depenbrock2012}%
  \BibitemOpen
  \bibfield  {author} {\bibinfo {author} {\bibfnamefont {S.}~\bibnamefont
  {Depenbrock}}, \bibinfo {author} {\bibfnamefont {I.~P.}\ \bibnamefont
  {McCulloch}}, \ and\ \bibinfo {author} {\bibfnamefont {U.}~\bibnamefont
  {Schollw{\"o}ck}},\ }\href {\doibase 10.1103/PhysRevLett.109.067201}
  {\bibfield  {journal} {\bibinfo  {journal} {Phys. Rev. Lett.}\ }\textbf
  {\bibinfo {volume} {109}},\ \bibinfo {pages} {067201} (\bibinfo {year}
  {2012})}\BibitemShut {NoStop}%
\bibitem [{\citenamefont {Jiang}\ \emph {et~al.}(2012)\citenamefont {Jiang},
  \citenamefont {Wang},\ and\ \citenamefont {Balents}}]{Jiang2012}%
  \BibitemOpen
  \bibfield  {author} {\bibinfo {author} {\bibfnamefont {H.-C.}\ \bibnamefont
  {Jiang}}, \bibinfo {author} {\bibfnamefont {Z.}~\bibnamefont {Wang}}, \ and\
  \bibinfo {author} {\bibfnamefont {L.}~\bibnamefont {Balents}},\ }\href
  {http://dx.doi.org/10.1038/nphys2465} {\bibfield  {journal} {\bibinfo
  {journal} {Nat. Phys.}\ }\textbf {\bibinfo {volume} {8}},\ \bibinfo {pages}
  {902} (\bibinfo {year} {2012})}\BibitemShut {NoStop}%
\bibitem [{\citenamefont {Messio}\ \emph {et~al.}(2012)\citenamefont {Messio},
  \citenamefont {Bernu},\ and\ \citenamefont {Lhuillier}}]{Messio2012}%
  \BibitemOpen
  \bibfield  {author} {\bibinfo {author} {\bibfnamefont {L.}~\bibnamefont
  {Messio}}, \bibinfo {author} {\bibfnamefont {B.}~\bibnamefont {Bernu}}, \
  and\ \bibinfo {author} {\bibfnamefont {C.}~\bibnamefont {Lhuillier}},\ }\href
  {\doibase 10.1103/PhysRevLett.108.207204} {\bibfield  {journal} {\bibinfo
  {journal} {Phys. Rev. Lett.}\ }\textbf {\bibinfo {volume} {108}},\ \bibinfo
  {pages} {207204} (\bibinfo {year} {2012})}\BibitemShut {NoStop}%
\bibitem [{\citenamefont {Capponi}\ \emph
  {et~al.}(2013{\natexlab{b}})\citenamefont {Capponi}, \citenamefont {Chandra},
  \citenamefont {Auerbach},\ and\ \citenamefont {Weinstein}}]{Capponi2013a}%
  \BibitemOpen
  \bibfield  {author} {\bibinfo {author} {\bibfnamefont {S.}~\bibnamefont
  {Capponi}}, \bibinfo {author} {\bibfnamefont {V.~R.}\ \bibnamefont
  {Chandra}}, \bibinfo {author} {\bibfnamefont {A.}~\bibnamefont {Auerbach}}, \
  and\ \bibinfo {author} {\bibfnamefont {M.}~\bibnamefont {Weinstein}},\ }\href
  {\doibase 10.1103/PhysRevB.87.161118} {\bibfield  {journal} {\bibinfo
  {journal} {Phys. Rev. B}\ }\textbf {\bibinfo {volume} {87}},\ \bibinfo
  {pages} {161118} (\bibinfo {year} {2013}{\natexlab{b}})}\BibitemShut
  {NoStop}%
\bibitem [{\citenamefont {Iqbal}\ \emph {et~al.}(2013)\citenamefont {Iqbal},
  \citenamefont {Becca}, \citenamefont {Sorella},\ and\ \citenamefont
  {Poilblanc}}]{Iqbal2013}%
  \BibitemOpen
  \bibfield  {author} {\bibinfo {author} {\bibfnamefont {Y.}~\bibnamefont
  {Iqbal}}, \bibinfo {author} {\bibfnamefont {F.}~\bibnamefont {Becca}},
  \bibinfo {author} {\bibfnamefont {S.}~\bibnamefont {Sorella}}, \ and\
  \bibinfo {author} {\bibfnamefont {D.}~\bibnamefont {Poilblanc}},\ }\href
  {\doibase 10.1103/PhysRevB.87.060405} {\bibfield  {journal} {\bibinfo
  {journal} {Phys. Rev. B}\ }\textbf {\bibinfo {volume} {87}},\ \bibinfo
  {pages} {060405} (\bibinfo {year} {2013})}\BibitemShut {NoStop}%
\bibitem [{\citenamefont {Matsubara}\ and\ \citenamefont
  {Matsuda}(1956)}]{Matsubara1956}%
  \BibitemOpen
  \bibfield  {author} {\bibinfo {author} {\bibfnamefont {T.}~\bibnamefont
  {Matsubara}}\ and\ \bibinfo {author} {\bibfnamefont {H.}~\bibnamefont
  {Matsuda}},\ }\href@noop {} {\bibfield  {journal} {\bibinfo  {journal}
  {Progr. Theoret. Phys. (Kyoto)}\ }\textbf {\bibinfo {volume} {16}},\ \bibinfo
  {pages} {569} (\bibinfo {year} {1956})}\BibitemShut {NoStop}%
\bibitem [{Note1()}]{Note1}%
  \BibitemOpen
  \bibinfo {note} {This terminology is often used since the model is amenable
  to sign-free QMC simulations but we should emphasize that the density-density
  interaction is always frustrated.}\BibitemShut {Stop}%
\bibitem [{\citenamefont {Ortiz}\ and\ \citenamefont
  {Batista}(2003)}]{Ortiz2003}%
  \BibitemOpen
  \bibfield  {author} {\bibinfo {author} {\bibfnamefont {G.}~\bibnamefont
  {Ortiz}}\ and\ \bibinfo {author} {\bibfnamefont {C.~D.}\ \bibnamefont
  {Batista}},\ }\href {\doibase 10.1103/PhysRevB.67.134301} {\bibfield
  {journal} {\bibinfo  {journal} {Phys. Rev. B}\ }\textbf {\bibinfo {volume}
  {67}},\ \bibinfo {pages} {134301} (\bibinfo {year} {2003})}\BibitemShut
  {NoStop}%
\bibitem [{\citenamefont {Batista}\ and\ \citenamefont
  {Ortiz}(2004)}]{Batista2004}%
  \BibitemOpen
  \bibfield  {author} {\bibinfo {author} {\bibfnamefont {C.~D.}\ \bibnamefont
  {Batista}}\ and\ \bibinfo {author} {\bibfnamefont {G.}~\bibnamefont
  {Ortiz}},\ }\href {\doibase 10.1080/00018730310001642086} {\bibfield
  {journal} {\bibinfo  {journal} {Adv. Phys.}\ }\textbf {\bibinfo {volume}
  {53}},\ \bibinfo {pages} {1} (\bibinfo {year} {2004})}\BibitemShut {NoStop}%
\bibitem [{\citenamefont {Isaev}\ \emph
  {et~al.}(2009{\natexlab{a}})\citenamefont {Isaev}, \citenamefont {Ortiz},\
  and\ \citenamefont {Dukelsky}}]{Isaev2009}%
  \BibitemOpen
  \bibfield  {author} {\bibinfo {author} {\bibfnamefont {L.}~\bibnamefont
  {Isaev}}, \bibinfo {author} {\bibfnamefont {G.}~\bibnamefont {Ortiz}}, \ and\
  \bibinfo {author} {\bibfnamefont {J.}~\bibnamefont {Dukelsky}},\ }\href
  {\doibase 10.1103/PhysRevB.79.024409} {\bibfield  {journal} {\bibinfo
  {journal} {Phys. Rev. B}\ }\textbf {\bibinfo {volume} {79}},\ \bibinfo
  {pages} {024409} (\bibinfo {year} {2009}{\natexlab{a}})}\BibitemShut
  {NoStop}%
\bibitem [{\citenamefont {Isaev}\ and\ \citenamefont
  {Ortiz}(2012)}]{Isaev2012}%
  \BibitemOpen
  \bibfield  {author} {\bibinfo {author} {\bibfnamefont {L.}~\bibnamefont
  {Isaev}}\ and\ \bibinfo {author} {\bibfnamefont {G.}~\bibnamefont {Ortiz}},\
  }\href {\doibase 10.1103/PhysRevB.86.100402} {\bibfield  {journal} {\bibinfo
  {journal} {Phys. Rev. B}\ }\textbf {\bibinfo {volume} {86}},\ \bibinfo
  {pages} {100402(R)} (\bibinfo {year} {2012})}\BibitemShut {NoStop}%
\bibitem [{\citenamefont {Huerga}\ \emph {et~al.}(2014)\citenamefont {Huerga},
  \citenamefont {Dukelsky}, \citenamefont {Laflorencie},\ and\ \citenamefont
  {Ortiz}}]{Huerga2014}%
  \BibitemOpen
  \bibfield  {author} {\bibinfo {author} {\bibfnamefont {D.}~\bibnamefont
  {Huerga}}, \bibinfo {author} {\bibfnamefont {J.}~\bibnamefont {Dukelsky}},
  \bibinfo {author} {\bibfnamefont {N.}~\bibnamefont {Laflorencie}}, \ and\
  \bibinfo {author} {\bibfnamefont {G.}~\bibnamefont {Ortiz}},\ }\href
  {\doibase 10.1103/PhysRevB.89.094401} {\bibfield  {journal} {\bibinfo
  {journal} {Phys. Rev. B}\ }\textbf {\bibinfo {volume} {89}},\ \bibinfo
  {pages} {094401} (\bibinfo {year} {2014})}\BibitemShut {NoStop}%
\bibitem [{\citenamefont {Sandvik}\ and\ \citenamefont
  {Kurkij{\"a}rvi}(1991)}]{Sandvik1991}%
  \BibitemOpen
  \bibfield  {author} {\bibinfo {author} {\bibfnamefont {A.~W.}\ \bibnamefont
  {Sandvik}}\ and\ \bibinfo {author} {\bibfnamefont {J.}~\bibnamefont
  {Kurkij{\"a}rvi}},\ }\href {\doibase 10.1103/PhysRevB.43.5950} {\bibfield
  {journal} {\bibinfo  {journal} {Phys. Rev. B}\ }\textbf {\bibinfo {volume}
  {43}},\ \bibinfo {pages} {5950} (\bibinfo {year} {1991})}\BibitemShut
  {NoStop}%
\bibitem [{\citenamefont {Sylju{\aa}sen}\ and\ \citenamefont
  {Sandvik}(2002)}]{Syljuasen2002}%
  \BibitemOpen
  \bibfield  {author} {\bibinfo {author} {\bibfnamefont {O.~F.}\ \bibnamefont
  {Sylju{\aa}sen}}\ and\ \bibinfo {author} {\bibfnamefont {A.~W.}\ \bibnamefont
  {Sandvik}},\ }\href {\doibase 10.1103/PhysRevE.66.046701} {\bibfield
  {journal} {\bibinfo  {journal} {Phys. Rev. E}\ }\textbf {\bibinfo {volume}
  {66}},\ \bibinfo {pages} {046701} (\bibinfo {year} {2002})}\BibitemShut
  {NoStop}%
\bibitem [{\citenamefont {Bauer}\ \emph {et~al.}(2011)\citenamefont {Bauer},
  \citenamefont {Carr}, \citenamefont {Evertz}, \citenamefont {Feiguin},
  \citenamefont {Freire}, \citenamefont {Fuchs}, \citenamefont {Gamper},
  \citenamefont {Gukelberger}, \citenamefont {Gull}, \citenamefont {Guertler},
  \citenamefont {Hehn}, \citenamefont {Igarashi}, \citenamefont {Isakov},
  \citenamefont {Koop}, \citenamefont {Ma}, \citenamefont {Mates},
  \citenamefont {Matsuo}, \citenamefont {Parcollet}, \citenamefont
  {Paw{\l}owski}, \citenamefont {Picon}, \citenamefont {Pollet}, \citenamefont
  {Santos}, \citenamefont {Scarola}, \citenamefont {Schollw{\"o}ck},
  \citenamefont {Silva}, \citenamefont {Surer}, \citenamefont {Todo},
  \citenamefont {Trebst}, \citenamefont {Troyer}, \citenamefont {Wall},
  \citenamefont {Werner},\ and\ \citenamefont {Wessel}}]{ALPS2}%
  \BibitemOpen
  \bibfield  {author} {\bibinfo {author} {\bibfnamefont {B.}~\bibnamefont
  {Bauer}}, \bibinfo {author} {\bibfnamefont {L.~D.}\ \bibnamefont {Carr}},
  \bibinfo {author} {\bibfnamefont {H.~G.}\ \bibnamefont {Evertz}}, \bibinfo
  {author} {\bibfnamefont {A.}~\bibnamefont {Feiguin}}, \bibinfo {author}
  {\bibfnamefont {J.}~\bibnamefont {Freire}}, \bibinfo {author} {\bibfnamefont
  {S.}~\bibnamefont {Fuchs}}, \bibinfo {author} {\bibfnamefont
  {L.}~\bibnamefont {Gamper}}, \bibinfo {author} {\bibfnamefont
  {J.}~\bibnamefont {Gukelberger}}, \bibinfo {author} {\bibfnamefont
  {E.}~\bibnamefont {Gull}}, \bibinfo {author} {\bibfnamefont {S.}~\bibnamefont
  {Guertler}}, \bibinfo {author} {\bibfnamefont {A.}~\bibnamefont {Hehn}},
  \bibinfo {author} {\bibfnamefont {R.}~\bibnamefont {Igarashi}}, \bibinfo
  {author} {\bibfnamefont {S.~V.}\ \bibnamefont {Isakov}}, \bibinfo {author}
  {\bibfnamefont {D.}~\bibnamefont {Koop}}, \bibinfo {author} {\bibfnamefont
  {P.~N.}\ \bibnamefont {Ma}}, \bibinfo {author} {\bibfnamefont
  {P.}~\bibnamefont {Mates}}, \bibinfo {author} {\bibfnamefont
  {H.}~\bibnamefont {Matsuo}}, \bibinfo {author} {\bibfnamefont
  {O.}~\bibnamefont {Parcollet}}, \bibinfo {author} {\bibfnamefont
  {G.}~\bibnamefont {Paw{\l}owski}}, \bibinfo {author} {\bibfnamefont {J.~D.}\
  \bibnamefont {Picon}}, \bibinfo {author} {\bibfnamefont {L.}~\bibnamefont
  {Pollet}}, \bibinfo {author} {\bibfnamefont {E.}~\bibnamefont {Santos}},
  \bibinfo {author} {\bibfnamefont {V.~W.}\ \bibnamefont {Scarola}}, \bibinfo
  {author} {\bibfnamefont {U.}~\bibnamefont {Schollw{\"o}ck}}, \bibinfo
  {author} {\bibfnamefont {C.}~\bibnamefont {Silva}}, \bibinfo {author}
  {\bibfnamefont {B.}~\bibnamefont {Surer}}, \bibinfo {author} {\bibfnamefont
  {S.}~\bibnamefont {Todo}}, \bibinfo {author} {\bibfnamefont {S.}~\bibnamefont
  {Trebst}}, \bibinfo {author} {\bibfnamefont {M.}~\bibnamefont {Troyer}},
  \bibinfo {author} {\bibfnamefont {M.~L.}\ \bibnamefont {Wall}}, \bibinfo
  {author} {\bibfnamefont {P.}~\bibnamefont {Werner}}, \ and\ \bibinfo {author}
  {\bibfnamefont {S.}~\bibnamefont {Wessel}},\ }\href
  {http://stacks.iop.org/1742-5468/2011/i=05/a=P05001} {\bibfield  {journal}
  {\bibinfo  {journal} {J. Stat. Mech.}\ }\textbf {\bibinfo {volume} {2011}},\
  \bibinfo {pages} {P05001} (\bibinfo {year} {2011})}\BibitemShut {NoStop}%
\bibitem [{\citenamefont {Rousseau}(2014)}]{Rousseau2014}%
  \BibitemOpen
  \bibfield  {author} {\bibinfo {author} {\bibfnamefont {V.~G.}\ \bibnamefont
  {Rousseau}},\ }\href {\doibase 10.1103/PhysRevB.90.134503} {\bibfield
  {journal} {\bibinfo  {journal} {Phys. Rev. B}\ }\textbf {\bibinfo {volume}
  {90}},\ \bibinfo {pages} {134503} (\bibinfo {year} {2014})}\BibitemShut
  {NoStop}%
\bibitem [{\citenamefont {Gong}\ \emph {et~al.}(2014)\citenamefont {Gong},
  \citenamefont {Zhu},\ and\ \citenamefont {Sheng}}]{Gong2014}%
  \BibitemOpen
  \bibfield  {author} {\bibinfo {author} {\bibfnamefont {S.-S.}\ \bibnamefont
  {Gong}}, \bibinfo {author} {\bibfnamefont {W.}~\bibnamefont {Zhu}}, \ and\
  \bibinfo {author} {\bibfnamefont {D.~N.}\ \bibnamefont {Sheng}},\ }\href
  {http://dx.doi.org/10.1038/srep06317} {\bibfield  {journal} {\bibinfo
  {journal} {Sci. Rep.}\ }\textbf {\bibinfo {volume} {4}},\ \bibinfo {pages}
  {6317 EP} (\bibinfo {year} {2014})}\BibitemShut {NoStop}%
\bibitem [{\citenamefont {Bauer}\ \emph {et~al.}(2014)\citenamefont {Bauer},
  \citenamefont {Cincio}, \citenamefont {Keller}, \citenamefont {Dolfi},
  \citenamefont {Vidal}, \citenamefont {Trebst},\ and\ \citenamefont
  {Ludwig}}]{Bauer2014}%
  \BibitemOpen
  \bibfield  {author} {\bibinfo {author} {\bibfnamefont {B.}~\bibnamefont
  {Bauer}}, \bibinfo {author} {\bibfnamefont {L.}~\bibnamefont {Cincio}},
  \bibinfo {author} {\bibfnamefont {B.~P.}\ \bibnamefont {Keller}}, \bibinfo
  {author} {\bibfnamefont {M.}~\bibnamefont {Dolfi}}, \bibinfo {author}
  {\bibfnamefont {G.}~\bibnamefont {Vidal}}, \bibinfo {author} {\bibfnamefont
  {S.}~\bibnamefont {Trebst}}, \ and\ \bibinfo {author} {\bibfnamefont
  {A.~W.~W.}\ \bibnamefont {Ludwig}},\ }\href
  {http://dx.doi.org/10.1038/ncomms6137} {\bibfield  {journal} {\bibinfo
  {journal} {Nat. Commun.}\ }\textbf {\bibinfo {volume} {5}} (\bibinfo {year}
  {2014})}\BibitemShut {NoStop}%
\bibitem [{\citenamefont {He}\ \emph {et~al.}(2014)\citenamefont {He},
  \citenamefont {Sheng},\ and\ \citenamefont {Chen}}]{He2014}%
  \BibitemOpen
  \bibfield  {author} {\bibinfo {author} {\bibfnamefont {Y.-C.}\ \bibnamefont
  {He}}, \bibinfo {author} {\bibfnamefont {D.~N.}\ \bibnamefont {Sheng}}, \
  and\ \bibinfo {author} {\bibfnamefont {Y.}~\bibnamefont {Chen}},\ }\href
  {\doibase 10.1103/PhysRevLett.112.137202} {\bibfield  {journal} {\bibinfo
  {journal} {Phys. Rev. Lett.}\ }\textbf {\bibinfo {volume} {112}},\ \bibinfo
  {pages} {137202} (\bibinfo {year} {2014})}\BibitemShut {NoStop}%
\bibitem [{\citenamefont {{L{\"a}uchli}}\ and\ \citenamefont
  {{Moessner}}(2015)}]{LauchliMoessner}%
  \BibitemOpen
  \bibfield  {author} {\bibinfo {author} {\bibfnamefont {A.~M.}\ \bibnamefont
  {{L{\"a}uchli}}}\ and\ \bibinfo {author} {\bibfnamefont {R.}~\bibnamefont
  {{Moessner}}},\ }\href@noop {} {\bibfield  {journal} {\bibinfo  {journal}
  {ArXiv e-prints}\ } (\bibinfo {year} {2015})},\ \Eprint
  {http://arxiv.org/abs/1504.04380} {arXiv:1504.04380 [cond-mat.quant-gas]}
  \BibitemShut {NoStop}%
\bibitem [{\citenamefont {He}\ and\ \citenamefont {Chen}(2015)}]{He2015}%
  \BibitemOpen
  \bibfield  {author} {\bibinfo {author} {\bibfnamefont {Y.-C.}\ \bibnamefont
  {He}}\ and\ \bibinfo {author} {\bibfnamefont {Y.}~\bibnamefont {Chen}},\
  }\href {\doibase 10.1103/PhysRevLett.114.037201} {\bibfield  {journal}
  {\bibinfo  {journal} {Phys. Rev. Lett.}\ }\textbf {\bibinfo {volume} {114}},\
  \bibinfo {pages} {037201} (\bibinfo {year} {2015})}\BibitemShut {NoStop}%
\bibitem [{\citenamefont {Kumar}\ \emph {et~al.}(2014)\citenamefont {Kumar},
  \citenamefont {Sun},\ and\ \citenamefont {Fradkin}}]{Kumar2014}%
  \BibitemOpen
  \bibfield  {author} {\bibinfo {author} {\bibfnamefont {K.}~\bibnamefont
  {Kumar}}, \bibinfo {author} {\bibfnamefont {K.}~\bibnamefont {Sun}}, \ and\
  \bibinfo {author} {\bibfnamefont {E.}~\bibnamefont {Fradkin}},\ }\href
  {\doibase 10.1103/PhysRevB.90.174409} {\bibfield  {journal} {\bibinfo
  {journal} {Phys. Rev. B}\ }\textbf {\bibinfo {volume} {90}},\ \bibinfo
  {pages} {174409} (\bibinfo {year} {2014})}\BibitemShut {NoStop}%
\bibitem [{\citenamefont {Greschner}\ \emph {et~al.}(2015)\citenamefont
  {Greschner}, \citenamefont {Huerga}, \citenamefont {Sun}, \citenamefont
  {Poletti},\ and\ \citenamefont {Santos}}]{Greschner2015}%
  \BibitemOpen
  \bibfield  {author} {\bibinfo {author} {\bibfnamefont {S.}~\bibnamefont
  {Greschner}}, \bibinfo {author} {\bibfnamefont {D.}~\bibnamefont {Huerga}},
  \bibinfo {author} {\bibfnamefont {G.}~\bibnamefont {Sun}}, \bibinfo {author}
  {\bibfnamefont {D.}~\bibnamefont {Poletti}}, \ and\ \bibinfo {author}
  {\bibfnamefont {L.}~\bibnamefont {Santos}},\ }\href {\doibase
  10.1103/PhysRevB.92.115120} {\bibfield  {journal} {\bibinfo  {journal} {Phys.
  Rev. B}\ }\textbf {\bibinfo {volume} {92}},\ \bibinfo {pages} {115120}
  (\bibinfo {year} {2015})}\BibitemShut {NoStop}%
\bibitem [{\citenamefont {Huerga}\ \emph {et~al.}(2013)\citenamefont {Huerga},
  \citenamefont {Dukelsky},\ and\ \citenamefont {Scuseria}}]{Huerga2013}%
  \BibitemOpen
  \bibfield  {author} {\bibinfo {author} {\bibfnamefont {D.}~\bibnamefont
  {Huerga}}, \bibinfo {author} {\bibfnamefont {J.}~\bibnamefont {Dukelsky}}, \
  and\ \bibinfo {author} {\bibfnamefont {G.~E.}\ \bibnamefont {Scuseria}},\
  }\href {\doibase 10.1103/PhysRevLett.111.045701} {\bibfield  {journal}
  {\bibinfo  {journal} {Phys. Rev. Lett.}\ }\textbf {\bibinfo {volume} {111}},\
  \bibinfo {pages} {045701} (\bibinfo {year} {2013})}\BibitemShut {NoStop}%
\bibitem [{\citenamefont {Leggett}(2001)}]{Leggett2001}%
  \BibitemOpen
  \bibfield  {author} {\bibinfo {author} {\bibfnamefont {A.~J.}\ \bibnamefont
  {Leggett}},\ }\href {\doibase 10.1103/RevModPhys.73.307} {\bibfield
  {journal} {\bibinfo  {journal} {Rev. Mod. Phys.}\ }\textbf {\bibinfo {volume}
  {73}},\ \bibinfo {pages} {307} (\bibinfo {year} {2001})}\BibitemShut
  {NoStop}%
\bibitem [{\citenamefont {Al-Hassanieh}\ \emph {et~al.}(2009)\citenamefont
  {Al-Hassanieh}, \citenamefont {Batista}, \citenamefont {Ortiz},\ and\
  \citenamefont {Bulaevskii}}]{Hassanieh2009}%
  \BibitemOpen
  \bibfield  {author} {\bibinfo {author} {\bibfnamefont {K.~A.}\ \bibnamefont
  {Al-Hassanieh}}, \bibinfo {author} {\bibfnamefont {C.~D.}\ \bibnamefont
  {Batista}}, \bibinfo {author} {\bibfnamefont {G.}~\bibnamefont {Ortiz}}, \
  and\ \bibinfo {author} {\bibfnamefont {L.~N.}\ \bibnamefont {Bulaevskii}},\
  }\href {\doibase 10.1103/PhysRevLett.103.216402} {\bibfield  {journal}
  {\bibinfo  {journal} {Phys. Rev. Lett.}\ }\textbf {\bibinfo {volume} {103}},\
  \bibinfo {pages} {216402} (\bibinfo {year} {2009})}\BibitemShut {NoStop}%
\bibitem [{\citenamefont {Gu}(2010)}]{Gu2010}%
  \BibitemOpen
  \bibfield  {author} {\bibinfo {author} {\bibfnamefont {S.-J.}\ \bibnamefont
  {Gu}},\ }\href {\doibase 10.1142/S0217979210056335} {\bibfield  {journal}
  {\bibinfo  {journal} {Int. J. Mod. Phys. B}\ }\textbf {\bibinfo {volume}
  {24}},\ \bibinfo {pages} {4371} (\bibinfo {year} {2010})}\BibitemShut
  {NoStop}%
\bibitem [{\citenamefont {Albuquerque}\ \emph {et~al.}(2010)\citenamefont
  {Albuquerque}, \citenamefont {Alet}, \citenamefont {Sire},\ and\
  \citenamefont {Capponi}}]{Albuquerque2010}%
  \BibitemOpen
  \bibfield  {author} {\bibinfo {author} {\bibfnamefont {A.~F.}\ \bibnamefont
  {Albuquerque}}, \bibinfo {author} {\bibfnamefont {F.}~\bibnamefont {Alet}},
  \bibinfo {author} {\bibfnamefont {C.}~\bibnamefont {Sire}}, \ and\ \bibinfo
  {author} {\bibfnamefont {S.}~\bibnamefont {Capponi}},\ }\href {\doibase
  10.1103/PhysRevB.81.064418} {\bibfield  {journal} {\bibinfo  {journal} {Phys.
  Rev. B}\ }\textbf {\bibinfo {volume} {81}},\ \bibinfo {pages} {064418}
  (\bibinfo {year} {2010})}\BibitemShut {NoStop}%
\bibitem [{\citenamefont {Jiang}\ and\ \citenamefont {Ye}(2006)}]{Jiang2006}%
  \BibitemOpen
  \bibfield  {author} {\bibinfo {author} {\bibfnamefont {L.}~\bibnamefont
  {Jiang}}\ and\ \bibinfo {author} {\bibfnamefont {J.}~\bibnamefont {Ye}},\
  }\href {http://stacks.iop.org/0953-8984/18/i=29/a=028} {\bibfield  {journal}
  {\bibinfo  {journal} {J. Phys. Condens. Matter}\ }\textbf {\bibinfo {volume}
  {18}},\ \bibinfo {pages} {6907} (\bibinfo {year} {2006})}\BibitemShut
  {NoStop}%
\bibitem [{\citenamefont {Sengupta}\ \emph {et~al.}(2006)\citenamefont
  {Sengupta}, \citenamefont {Isakov},\ and\ \citenamefont
  {Kim}}]{Sengupta2006}%
  \BibitemOpen
  \bibfield  {author} {\bibinfo {author} {\bibfnamefont {K.}~\bibnamefont
  {Sengupta}}, \bibinfo {author} {\bibfnamefont {S.~V.}\ \bibnamefont
  {Isakov}}, \ and\ \bibinfo {author} {\bibfnamefont {Y.~B.}\ \bibnamefont
  {Kim}},\ }\href {\doibase 10.1103/PhysRevB.73.245103} {\bibfield  {journal}
  {\bibinfo  {journal} {Phys. Rev. B}\ }\textbf {\bibinfo {volume} {73}},\
  \bibinfo {pages} {245103} (\bibinfo {year} {2006})}\BibitemShut {NoStop}%
\bibitem [{\citenamefont {Damle}\ and\ \citenamefont
  {Senthil}(2006)}]{Damle2006}%
  \BibitemOpen
  \bibfield  {author} {\bibinfo {author} {\bibfnamefont {K.}~\bibnamefont
  {Damle}}\ and\ \bibinfo {author} {\bibfnamefont {T.}~\bibnamefont
  {Senthil}},\ }\href {\doibase 10.1103/PhysRevLett.97.067202} {\bibfield
  {journal} {\bibinfo  {journal} {Phys. Rev. Lett.}\ }\textbf {\bibinfo
  {volume} {97}},\ \bibinfo {pages} {067202} (\bibinfo {year}
  {2006})}\BibitemShut {NoStop}%
\bibitem [{Note2()}]{Note2}%
  \BibitemOpen
  \bibinfo {note} {The 18-sites cluster used is comprised of two vertically
  connected 9-sites clusters.}\BibitemShut {Stop}%
\bibitem [{\citenamefont {Xie}\ \emph {et~al.}(2014)\citenamefont {Xie},
  \citenamefont {Chen}, \citenamefont {Yu}, \citenamefont {Kong}, \citenamefont
  {Normand},\ and\ \citenamefont {Xiang}}]{Xie2014}%
  \BibitemOpen
  \bibfield  {author} {\bibinfo {author} {\bibfnamefont {Z.~Y.}\ \bibnamefont
  {Xie}}, \bibinfo {author} {\bibfnamefont {J.}~\bibnamefont {Chen}}, \bibinfo
  {author} {\bibfnamefont {J.~F.}\ \bibnamefont {Yu}}, \bibinfo {author}
  {\bibfnamefont {X.}~\bibnamefont {Kong}}, \bibinfo {author} {\bibfnamefont
  {B.}~\bibnamefont {Normand}}, \ and\ \bibinfo {author} {\bibfnamefont
  {T.}~\bibnamefont {Xiang}},\ }\href {\doibase 10.1103/PhysRevX.4.011025}
  {\bibfield  {journal} {\bibinfo  {journal} {Phys. Rev. X}\ }\textbf {\bibinfo
  {volume} {4}},\ \bibinfo {pages} {011025} (\bibinfo {year}
  {2014})}\BibitemShut {NoStop}%
\bibitem [{\citenamefont {Shastry}\ and\ \citenamefont
  {Sutherland}(1981)}]{Shastry1981}%
  \BibitemOpen
  \bibfield  {author} {\bibinfo {author} {\bibfnamefont {B.~S.}\ \bibnamefont
  {Shastry}}\ and\ \bibinfo {author} {\bibfnamefont {B.}~\bibnamefont
  {Sutherland}},\ }\href {\doibase 10.1016/0378-4363(81)90838-X} {\bibfield
  {journal} {\bibinfo  {journal} {Physica B+C}\ }\textbf {\bibinfo {volume}
  {108}},\ \bibinfo {pages} {1069} (\bibinfo {year} {1981})}\BibitemShut
  {NoStop}%
\bibitem [{\citenamefont {Isaev}\ \emph
  {et~al.}(2009{\natexlab{b}})\citenamefont {Isaev}, \citenamefont {Ortiz},\
  and\ \citenamefont {Dukelsky}}]{Isaev2009SS}%
  \BibitemOpen
  \bibfield  {author} {\bibinfo {author} {\bibfnamefont {L.}~\bibnamefont
  {Isaev}}, \bibinfo {author} {\bibfnamefont {G.}~\bibnamefont {Ortiz}}, \ and\
  \bibinfo {author} {\bibfnamefont {J.}~\bibnamefont {Dukelsky}},\ }\href
  {\doibase 10.1103/PhysRevLett.103.177201} {\bibfield  {journal} {\bibinfo
  {journal} {Phys. Rev. Lett.}\ }\textbf {\bibinfo {volume} {103}},\ \bibinfo
  {pages} {177201} (\bibinfo {year} {2009}{\natexlab{b}})}\BibitemShut
  {NoStop}%
\bibitem [{\citenamefont {Kshetrimayum}\ \emph {et~al.}(2016)\citenamefont
  {Kshetrimayum}, \citenamefont {Picot}, \citenamefont {Orus},\ and\
  \citenamefont {Poilblanc}}]{Kshetrimayum2016}%
  \BibitemOpen
  \bibfield  {author} {\bibinfo {author} {\bibfnamefont {A.}~\bibnamefont
  {Kshetrimayum}}, \bibinfo {author} {\bibfnamefont {T.}~\bibnamefont {Picot}},
  \bibinfo {author} {\bibfnamefont {R.}~\bibnamefont {Orus}}, \ and\ \bibinfo
  {author} {\bibfnamefont {D.}~\bibnamefont {Poilblanc}},\ }\href@noop {} {\
  (\bibinfo {year} {2016})},\ \Eprint {http://arxiv.org/abs/1608.00437}
  {arXiv:1608.00437 [cond-mat.str-el]} \BibitemShut {NoStop}%
\end{thebibliography}%

\end{document}